\documentclass[12pt,showpacs,amsmath,amssymb]{revtex4}%
\usepackage{color}
\usepackage{epic}
\usepackage{eepic}
\usepackage{graphicx}
\usepackage{bm}

\usepackage{graphicx}
\usepackage{dcolumn}
\usepackage{bm}
\usepackage[hyperfootnotes=false]{hyperref}
\hypersetup{
  colorlinks   = true, 
  urlcolor     = blue, 
  linkcolor    = blue, 
  citecolor   = red 
}

\begin{document}
\pagestyle{plain}
\title
{New Concept for Studying the Classical and Quantum Three-Body Problem:
Fundamental Irreversibility and Time's Arrow of Dynamical Systems  }

\author{\firstname{A.~S.}~\surname{Gevorkyan}}
\email{g\_ashot@sci.am}
\address{\it Institute for Informatics and Automation Problems, NAS of Armenia}
\address{\it Institute of Chemical Physics, NAS of Armenia}

\begin{abstract}
The article formulates the classical three-body problem in conformal-Euclidean space
(Riemannian manifold), and its equivalence to the Newton three-body problem is
mathematically rigorously proved. It is shown that a curved space with a local coordinate
system allows us to detect new hidden symmetries of the internal motion of a dynamical
system, which allows us to reduce the three-body problem to the 6\emph{th} order system.
A new approach makes the system of geodesic equations with respect to the evolution
parameter of a dynamical system (\emph{internal time}) \emph{fundamentally irreversible}.
To describe the motion of three-body system in different random environments, the
corresponding stochastic differential equations (SDEs) are obtained. Using these
SDEs, Fokker-Planck-type equations are obtained that describe the joint probability
distributions of geodesic flows in phase and configuration spaces.

The paper also formulates the quantum three-body problem in conformal-Euclidean space. In
particular, the corresponding wave equations have been obtained for studying the three-body
bound states, as well as for investigating multichannel quantum scattering in the framework
of the concept of \emph{internal time}. This allows us to solve the extremely important
\emph{quantum-classical correspondence problem} for  dynamical \emph{Poincar\'{e} systems}.

\end{abstract}


\maketitle

\section{I\lowercase{ntroduction}} \label{sec:INTRO}
\quad\,\,\,\qquad\qquad\qquad \qquad\qquad \qquad\qquad \qquad\qquad {\emph{One geometry
cannot be more accurate than}}

\quad\qquad\qquad\qquad \qquad\qquad \qquad\qquad \qquad\qquad \emph{ another, it may
only be more convenient ...}

\quad \qquad\qquad\qquad \qquad\qquad \qquad\qquad \qquad\qquad \emph{ A. Poincar\'{e}}

The general three-body classical problem is one of the oldest and most complex
problems in classical mechanics \cite{HP,Whitt,Chen,Valt,Lin,Lema}. Briefly,
the meaning of the task is to study the motion of three bodies in space under
the influence of pairwise interactions of bodies in accordance with Newton's
theory of gravitation.

As Bruns \cite{Brun} showed, the problem under consideration is described in
an 18 - dimensional phase space and has 10 integrals of motion. Note that this
property does not allow to solve the problem in the same way as it does for
two bodies, and therefore it is believed that it belongs to the class of
non-integrable classical systems or the so-called \emph{Poincar\'{e} systems}.
Recall that the three-body problem in Euclidean space has well-defined symmetries,
which in general case generate only 10 integrals of motion. The procedure
for reducing the number of equations of a dynamical system is based on the use
of these integrals of motion, which allows us to reduce the three-body problem
to the system of 8\emph{th} order. Recall that the latter means that the evolution
of a dynamical system in phase space is described using 8\emph{th} ordinary
differential equations of 1\emph{st} order.

It is important to note that the three-body problem has served as the most important
source for the development of scientific thought in many areas of mathematics,
mechanics and physics since Newton. However, it was Poincar\'{e} who
opened a new era, developing
geometric, topological and probabilistic methods for studying a nontrivial and highly
complex behavior of this dynamical problem. The three-body problem arising from
celestial mechanics \cite{AKN,Marchal,Bruno}, remains extremely urgent even
now in connection with the search for stable new  periodic trajectories that cannot
be calculated by analytical methods \cite{Suv, Li,Orlov, Xi}. Note that analysis of
current trends in technology development indicates that there is increasing need
for accurate data on elementary atomic-molecular collisions occurring in various
physicochemical processes \cite{Hersch, Levine, Cross, Guichardet, Iwai, Lin1}.
This fact additionally motivates a comprehensive theoretical and algorithmic studies
of this problem. It is important to note that  significant number of elementary
atomic-molecular processes, including chemical reactions that take into account
external effects, are described and can be described in the framework of this
seemingly simple classical model.

Thus, new mathematical studies are fundamentally important for the creation
of effective algorithms allowing to calculate complex multichannel processes
from the first principles of classical mechanics. It should be noted that
the problems of atomic-molecular collisions have their own quit subtle features,
which can stimulate the development of fundamentally new ideas in the theory
of dynamical systems. In particular, one of the important and insufficiently
studied problems of the theory of collisions is the accurate account of the
contribution of multichannel scattering to a specific elementary atomic-molecular
process.

Another unsolved problem, which is of great importance for modern chemistry,
is to take into account the regular and stochastic effects of the medium on
the dynamics of elementary atomic-molecular processes, the ultimate goal
of which is to control these processes.

When solving complex dynamical problems, it is important not only to perform
convenient coordinate transformations, but also to choose the appropriate
geometry for solving a specific problem. In this sense, Krylov made
one of the first successful attempts to study the dynamics of $N$ classical
bodies on a Riemannian manifold, which is the hypersurface of the energy
of the system of bodies \cite{Kry}. Recall that the main goal of the study was
to substantiate statistical mechanics based on the first principles of
classical mechanics. Note that later this method was successfully used to
study the statistical properties of the non-Abelian Yang-Mills gauge fields
\cite{Sav1} and the relaxation properties of stellar systems \cite{Gurz1,Gurz2}.

In this work we significantly develop the above geometric and other ideas for
studying the classical and quantum three-body problem in order to find new
theoretical and algorithmic possibilities for the effective solution of these
problems. Unlike previous authors, we solved the complex problem of mapping
Euclidean geometry to  Riemann geometry, which allowed us to make the theory
consistent and mathematically rigorous  \cite{gev}. In other words, we prove
the equivalence of the original Newton three-body problem to the problem of
geodesic flows on a Riemannian manifold.

As shown in a series of works \cite{AshGev, gev, gev1,gev0}, a representation
developed on the basis of Riemannian geometry allows one to detect new hidden
internal symmetries of dynamical systems. The latter allows one to realize a more
complete integration of the three-body problem, which in the general case in the
sense of Poincar\'{e} is a \emph{non-integrable dynamical system}. However, more
importantly, this formulation of the problem allows us to answer the following
fundamental question concerning the foundations of quantum physics, namely:
\emph{is the irreversibility fundamental for describing the classical world} \cite{Brig}?
In particular, the proof of the \emph{irreversibility} of the general three-body
problem with respect to the \emph{internal time} of the system allows us to
solve the fundamentally important problem of \emph{quantum-classical correspondence}
for dynamical \emph{Poincar\'{e} systems}.

In the work, classical and quantum three-body problems are considered in a more
general formulation. In particular, in addition to the potentials of two- and
three-particle interactions, the contribution of external regular and random
forces to elementary processes is also taken into account. The latter creates
new opportunities and prospects for studying the three-body problem,
taking into account its wide application in various applied problems of
physics, chemistry and material science.

The manuscript is organized as follows:

Section II briefly describes  the general classical three-body problem and
proves that it reduces to the problem of the motion of an \emph{imaginary point}
with effective mass $\mu_0$  in the configuration space $6D$ under the influence
of an external field.

In Section III, the classical three-body problem is formulated as the problem of
geodesic flows on a $6D$ Riemannian manifold. A system of six geodesic
equations is obtained, three of which are exactly solved. As a result of this,
the problem was reduced to the system of order $6th$, and in the case of fixed
energy, to the system of $5th$ order. In this section, the reduced Hamiltonian
of the three-body system is obtained, which is defined in the $6D$ phase space.
This Hamiltonian is later used to formulate the quantum three-body problem
in the framework of the concept of  \emph{internal time} in  section 10.

In Section IV, the proposition on homeomorphism between the subspace
$\mathbb{E}^6\in \mathbb{R}^6$
and the $6D$ \emph{Riemannian} manifold $\mathcal{M}$ in detail is proved,
which plays a key role in proving the equivalence of the developed
representation with the Newtonian three-body problem. This section analyzes
the connection of the above proposition  with the well-known \emph{Poincar\'{e}
conjecture}  (see \emph{Millennium Challenges} \cite{kly}).

In Section V,  transformations between the global and local coordinate systems in
differential form are obtained. The peculiarities of  \emph{internal time} are discussed in
detail, as a result of which its key role in the occurrence of \emph{irreversibility} even in
a closed classical three-body system is revealed, contrary to the well-known
\emph{Poincar\'{e}'s return theorem}.

In Section VI, the restricted classical three-body problems with holonomic connections
 are studied. The possibility of finding all families of stable solutions by algebraic and
geometrical  methods is proved.

In Section VII,  an equation  for deviation of the geodesic trajectories of one
family is obtained, which makes it possible to study the important characteristics
of the motion of a dynamical system.

In Section VIII, the three-body problem in a random environment is considered,
taking into account various conditions. Various equations of the Fokker- Planck
type are obtained, which describe the evolution of  geodesic trajectories
flows in the phase and configuration spaces.

In Section IX, a new criterion for assessing chaos in the classical statistical
system is substantiated using the Kullback - Leibler idea of the
distance of two continuous distributions (in considered case, between two
tubes of probabilistic currents). An expression is constructed for the deviation
of two different tubes of probability currents in phase space. The mathematical
expectation of the transition between two asymptotic states $(in)$ and $(out)$ is
constructed using rigorous probabilistic reasoning.

In Section X, the quantum problem is formulated for the case of a
three-particle bound state and scattering with rearrangement of particles.
The corresponding equations are obtained that describe the evolution of
the wave state of a quantum system with the possibility of occurrence
\emph{quantum-wave chaos} both for a coupled system and for a scattering
one. To describe the scattering process with rearrangement of particles,
$\mathbf{S}$ - matrix  elements of transitions  are constructed. The necessity of
additional averaging of $\mathbf{S}$ - matrix elements in connection with the
quantum-chaotic behavior of the system in the case of multichannel
scattering is substantiated.

In Section XI, the obtained results are discussed in detail and further ways
of development of the problems under consideration are indicated.

In Section XIII which includes appendices \textbf{A}, {\bf  B}, {\bf  C}, {\bf D}, {\bf  E},
{\bf F} and {\bf G}, provides important proof supporting the mathematical
rigor of the developed approaches.

\section{T\lowercase{he classical three-body problem}}
As already mentioned, the classical three-body problem is still rather associated
with the problems of celestial mechanics, the purpose of which studying the relative
motion of three bodies interacting according to Newton's law  (for example, the Sun,
Earth and the Moon) \cite{HP}. Recall that for celestial mechanics, the solutions
 that lead to the appearance of periodic or spatially bounded trajectories are
especially interesting and important, and are currently and being intensively
studied (see \cite{Xi}).

However, if we consider the three-body problem for
an atomic-molecular collision, then this is a typical problem of multichannel
scattering, where interactions between particles can be arbitrary. On this basis,
the three-body collision  in the most general case, taking into account a number
of possible asymptotic results, can be represented schematically as:
\vskip-3mm
$$
1\,\,+\,(23)\quad\longrightarrow  \quad \begin{cases} 1\,\,+\,(23),
\\ 1+2+3,\\(12)\,\,+\,3,\\(13)\,\,+\,2,\\
\quad (123)^\star\quad
\longrightarrow\quad
\begin{cases}1\,\,+\,(23),
\\ 1+2+3,\\(12)\,\,+\,3,\\(13)\,\,+\,
2,\\
(123)^{\star\star}\to\begin{cases}...\end{cases},
\end{cases}
\end{cases}
$$
\vskip6mm
\textbf{Scheme 1.} \emph{Where 1, 2 and 3 indicate single bodies, the bracket
$(\cdot\cdot\cdot)$ denotes the two-body bound state, while $^{"\star"}$ and
$^{"\star\star"}$ denote, respectively,  some short-lived bound states  of
three bodies, which in the chemical literature are also called transition states.}

\textbf{Definition 1.} \emph{The classical three-body dynamics in the laboratory
coordinate system is described by the Hamiltonian of the form:}
\begin{equation}
 H \bigl(\{\emph{\textbf{r}}\};\{\emph{\textbf{p}}\}\bigr)=
\sum_{i=1}^3\frac{||\emph{\textbf{p}}_i||^2}{2m_i}
+V\bigl(\{\emph{\textbf{r}}\}\bigr),
\label{01a}
\end{equation}
\emph{where}  $\{\emph{\textbf{r}}\}=(\emph{\textbf{r}}_1,\emph{\textbf{r}}_2,\emph{\textbf{r}}_3)\in
 \mathbb{R}^3\times\mathbb{R}^3\times\mathbb{R}^3$ and
$\{\emph{\textbf{p}}\}=(\emph{\textbf{p}}_1,\emph{\textbf{p}}_2,\emph{\textbf{p}}_3)\in
\mathbb{R^\ast}^3\times \mathbb{R^\ast}^3\times\mathbb{R^\ast}^3$ \emph{are the
sets of radius vectors and  momenta of bodies with masses} $m_1, m_2 $ \emph{and}
$ m_3,$ \,\emph{respectively,} \emph{here the sign above the symbol $\,\,\,"^\ast "$ denotes the
transposed space,} $||\cdot\cdot\cdot||$ \emph{is the Euclidean norm, and
$\,\,"\times" $ denotes a direct product of subspaces.}

We will consider the most general form of the total interaction potential, depending on
the relative distances between the bodies:
\begin{equation}
 V(\{\emph{\textbf{r}}\})= \bar{V}\bigl(||\emph{\textbf{r}}_{12}||,||\emph{\textbf{r}}_{13}||,
 ||\emph{\textbf{r}}_{23}||\bigr),
\label{01b}
\end{equation}
where $\emph{\textbf{r}}_{12}=\emph{\textbf{r}}_{1}-\emph{\textbf{r}}_{2}$,
$\emph{\textbf{r}}_{13}=\emph{\textbf{r}}_{1}-\emph{\textbf{r}}_{3}$, and
$\emph{\textbf{r}}_{23}=\emph{\textbf{r}}_{2}-\emph{\textbf{r}}_{3}$ are relative
displacements between the bodies,  in addition, the set of radius vectors
$(\emph{\textbf{r}}_{12},\emph{\textbf{r}}_{13},
\emph{\textbf{r}}_{23})\in \mathbb{R}^3\times\mathbb{R}^3\times\mathbb{R}^3 \setminus \oslash$
(\emph{where  $\oslash$ denotes an empty set}),
which means the impossibility of a situation where two bodies occupy the same position.
Note that  the potential (\ref{01b}), in addition to two-particle interactions, can
also taking into account the contribution of three-particle interactions and  as
well as the influence of external fields. The latter circumstance significantly
expands the range of  problems studied related to the classical three-body problem.
Obviously, the configuration space for describing the dynamics of three bodies
without any restrictions should be $\mathbb{R}^9$.
In this regard, it is important to note that; $V:\mathbb{R}^9\to\mathbb{R}^1$
 and $\bar{V}:\mathbb{R}^3\to\mathbb{R}^1$, in addition,
$H:\mathbb{R}^{18}\to\mathbb{R}^1$. Recall that the not reduced Hamiltonian of
three-body problem (\ref{01a}) is a function of the 18\,-dimensional phase space
$\mathbb{R}^{18}$.
\begin{figure}
\includegraphics[width=95mm]{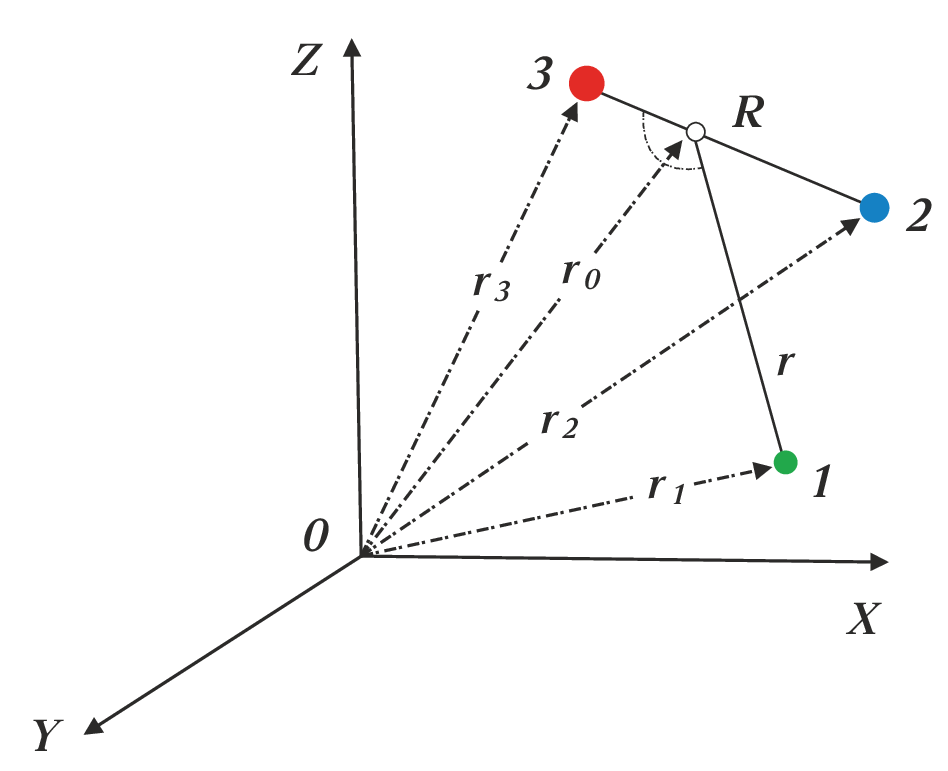}
 \caption{\emph{The Cartesian coordinate system where the set of radius vectors}
$\textbf{\emph{r}}_1,\textbf{\emph{r}}_2$ and $\textbf{\emph{r}}_3$
\emph{denote  positions of the} 1, 2 \emph{and} 3 \emph{bodies, respectively.
The circle} $"\circ"$ \emph{denotes the center of mass of pair }(23)
\emph{which in the Cartesian system is denoted by}
$\textbf{\emph{r}}_0$. \emph{The radius vectors} $\textbf{{\emph{R}}}$ \emph{and
}$\textbf{\emph{r}}$ \emph{determine the Jacobi coordinate system, and}  $\theta$
 \emph{denotes the scattering angle.}}
\label{Fig.1}
\end{figure}

The three-body Hamiltonian (\ref{01a}), after the Jacobi coordinate  transformations \cite{Delves1}
acquires the form:
 \begin{equation}
\breve{H} =\sum_{i=1}^3\frac{\textbf{P}^2_i}{2\mu_i}
+\breve{V}\bigl(||\emph{\textbf{r}}-\lambda_{-}\emph{\textbf{R}}||,||\emph{\textbf{R}}||,
 ||\emph{\textbf{r}}+\lambda_{+}\emph{\textbf{R}}||\bigr),
\label{01}
\end{equation}
where the radius vector ${\textbf{\emph{R}}}$ denotes the relative displacement
between 2 and 3 bodies (see FIG. 1),
${\textbf{\emph{r}}}=\emph{\textbf{r}}_1-\emph{\textbf{r}}_0$ is the relative
displacement between the particle 1 and center of mass of the pair of particles (2,\,3), while
$\emph{\textbf{r}}_0=(m_2\emph{\textbf{r}}_2+m_3\emph{\textbf{r}}_3)/(m_1+m_2)$
is the radius vector of the center of mass of the pair  (2,\,3). In addition, the following
notations are made in the equation (\ref{01}) (see also \cite{AshGev}):
$$
\textbf{P}_1=\emph{\textbf{p}}_1+\emph{\textbf{p}}_2+\emph{\textbf{p}}_3,\quad
\textbf{P}_2=\frac{m_3\emph{\textbf{p}}_2-
m_2\emph{\textbf{p}}_3}{m_2+m_3},\quad \textbf{P}_3=\frac{(m_2+m_3)\emph{\textbf{p}}_1-
m_1(\emph{\textbf{p}}_2+\emph{\textbf{p}}_3)}{\mu_1},
$$
$$
\mu_1=m_1+m_2+m_3, \quad \mu_2=\frac{m_2m_3}{m_2+m_3},\quad \mu_3=\frac{m_1(m_2+m_3)}{\mu_1},
\quad \lambda_{-}=\frac{\mu_2}{m_2},\quad \lambda_{+}=\frac{\mu_2}{m_3}.
$$
Removing the motion of the center of mass of the three-body system, that is equivalent to
the condition  $\textbf{P}_1=0$, leads the  equation (\ref{01}) to the form (see \cite{gev1}):
 \begin{equation}
 \tilde{H} =\frac{1}{2\mu_0}\Bigl( \tilde{\textbf{P}}^2_2+\tilde{\textbf{P}}^2_3\Bigr)
+\breve{V}\bigl(||\emph{\textbf{r}}-\lambda_{-}\emph{\textbf{R}}||,||\emph{\textbf{R}}||,
 ||\emph{\textbf{r}}+\lambda_{+}\emph{\textbf{R}}||\bigr).
 \label{1c}
\end{equation}
In the equation (\ref{1c}) the following notations are made:
 $$
 \mu_0=\Bigl(\frac{m_1m_2m_3}{\mu_1}\Bigr)^{1/2},\qquad \tilde{\textbf{P}}_2=
 \sqrt{\mu_2\mu_0}\dot{\emph{\textbf{R}}},
 \qquad \tilde{\textbf{P}}_3=\sqrt{\mu_3\mu_0}\dot{\emph{\textbf{r}}},
 $$
where $\dot{\emph{\textbf{x}}}=d\emph{\textbf{x}}/dt$ and $\emph{\textbf{x}}=
(\emph{\textbf{R}},\emph{\textbf{r}}).$

Finally, the Hamiltonian (\ref{1c}) can be written as:
\begin{equation}
\mathbb{H}(\textbf{r},\textbf{p}) =\frac{1}{2\mu_0}{\textbf{p}}^2
+\mathbb{V}(\textbf{r}),
 \label{01c}
\end{equation}
where $ \mathbb{V}(\textbf{r})= \breve{V}\bigl(||\emph{\textbf{r}}-
\lambda_{-}\emph{\textbf{R}}||,||\emph{\textbf{R}}||,
 ||\emph{\textbf{r}}+\lambda_{+}\emph{\textbf{R}}||\bigr)$.\\
Note that (\ref{01c}) can be interpreted as a single-particle Hamiltonian
with effective mass $\mu_0$ in a $12D$ phase space.
In addition (\ref{01c}) the following notations are made:
\begin{equation}
\textbf{r}=\textbf{\emph{r}}\oplus{\textbf{\emph{R}}}\in
{\mathbb{R}}^6,\qquad \textbf{p}=\tilde{\textbf{\emph{P}}}_2\oplus
\tilde{{\textbf{\emph{P}}}}_3\in {\mathbb{R}^\ast}^6,
\label{02t}
\end{equation}
where $"\oplus"$ denotes the direct sum of the $3D$ vectors and, accordingly,
$\textbf{r}$ and $\textbf{p} $ are the radius vector and the momentum of an
 imaginary point  in the $6D$  configuration space. It is obvious that; $\mathbb{V}:\mathbb{R}^{3}\to
\mathbb{R}^{1}$ and $\mathbb{H}:\mathbb{R}^{12}\to \mathbb{R}^1$.

Let us consider the following system of hyper-spherical coordinates:
\begin{eqnarray}
\rho_1=r=||\textbf{\emph{r}}||,\quad\rho_2=R=||\textbf{\emph{R}}||,\quad
 \rho_3=\theta,\quad \rho_4=\Theta,\quad \rho_5=\Phi,\quad
\rho_6=\Omega,
\label{02}
\end{eqnarray}
where the first set of three coordinates (coordinates of the \emph{internal space} or
 \emph{the internal coordinates}) $\{\bar{\rho}\}=(\rho_1,\rho_2,\rho_3)$
determines the position of the effective mass $\mu_0$ (\emph{imaginary point})
on the plane formed by three bodies.  Note that the domain of definition of these coordinates,
 respectively,  are $(\rho_1,\rho_2)\in[0,\infty]$ and $\theta\in[0,\pi].$
The  set of coordinates $\{\underline{\rho}\}=(\Theta,\Phi,\Omega)$  will be
called \emph{external coordinates}. The domain of definition of these coordinates,
respectively, are   $\Theta\in(-\pi,+\pi]$,\, $\Phi=(-\pi,+\pi]$ and $\Omega\in[0,\pi]$.
Note that the \emph{external coordinates} are the \emph{Euler} angles describing the
rotation of the plane in $3D$ space.

As was shown \cite{Klar,Johonson,Johonson1,Smor,Kup,Schatz,Vinit,Gusev}, it is
convenient to represent the motion of a three-body system as translational and
rotational motion of a three-body triangle  $ \triangle(1,2,3)$, and also deformation
of sides of the same triangle \cite{gev,gev1,gev0}. In particular,
the kinetic energy in this case can be written in the form \cite{Fiz}:
\begin{eqnarray}
T=\frac{\mu_0}{2}
\bigl\{\dot{{\textbf{\emph{R}}}}^2+\dot{{\textbf{\emph{r}}}}^2\bigr\}=\frac{\mu_0}{2 }
\Bigl\{\dot{\emph{R}}^{2}+{\emph{R}}^{2}\bigl[\bm \omega\times\textbf{k}\bigr]^2+
\bigl(\dot{{\textbf{\emph{r}}}}+[\bm\omega\times
{\textbf{\emph{r}}}]\bigr)^2\Bigr\},
 \label{16a}
\end{eqnarray}
where the direction of the unit vector $\textbf{k}$ in the moving reference
frame $\{\rho\}$ is determined by the expression
${\emph{\textbf{R}}}{||\emph{\textbf{R}}||}^{-1}=\pm \textbf{k}$. Below we will
assume that the vector $\textbf{k}=(0,0,1)$ is directed toward the positive
direction of the axis $OZ$ (below will be designated as the axis $z$ ), and the
angular velocity  $\bm\omega$ describes the rotation of the frame
$\{\bar{\rho}\}$ relative to the laboratory system.

Having carried out simple calculations in the expression (\ref{16a}) it is easy to find:
\begin{equation}
T=\frac{\mu_0}{2}\Bigl\{\dot{{{\emph{R}}}}^{2}+\dot{{{\emph{r}}}}^{\,2}+
r^{2}\dot{\theta}^2+A{{{\emph{R}}}}^{2}+ B{{{\emph{r}}}}^{\,2}\Bigr\},  \label{17a}
\end{equation}
where the following notations are made:
 $$
A=\omega_x^2+\omega_y^2,
\qquad
 B=\omega^2_y+\bigl(\omega_x\cos\theta-\omega_z\sin\theta\bigr)^2.
$$
Note that when deriving the expression (\ref{17a}) we  used the definition of a
moving system $\{\bar{\varrho}\}$, suggesting that the unit vector
$\bm\gamma=\textbf{\emph{r}}||\textbf{\emph{r}}||^{-1}$ lies on the plane $OXZ$ at the
angle $\theta$ relative to the axis $OZ$, that is; $\bm\gamma=(\sin\theta,0,\cos\theta)$.
As for angular velocity projections, they satisfy the following equations:
\begin{eqnarray}
\omega_x=\dot{\Phi}\sin\Theta\sin\Omega+\dot{\Theta}\cos\Omega,
\nonumber\\
\omega_y=\dot{\Phi}\sin\Theta\cos\Omega-\dot{\Theta}\sin\Omega,
\nonumber\\
\omega_z=\dot{\Phi}\cos\Theta-\dot{\Omega}. \label{18a}
\end{eqnarray}
Taking into account (\ref{17a}) and (\ref{18a}), the kinetic energy of the three-body
 system in Euclidean space can be written in the tensor form:
$$
T=\frac{\mu_0}{2}\gamma^{\alpha\beta}\frac{d\rho_\alpha}{dt}\frac{d\rho_\beta}{dt},
\qquad \alpha,\beta=(1,2,...,6)=\overline{1,6},
$$
where $\gamma^{\alpha\beta}$ is the metric tensor, which has the form:
\begin{equation}
\gamma^{\alpha\beta}=\left(
  \begin{array}{cccccc}
    \gamma^{11}\, &\,  0\, & 0 \, & 0\,  & \, 0\, &\,  0 \\
    0\, & \gamma^{22} &\,  0 \, & 0\,  &\,  0\, &\,  0 \\
    0\, & \,0\, &  \gamma^{33} & 0\, & \,0 \,&\, 0 \\
    0\,  &\,  0\,  & \, 0\,  &\gamma^{44} &  \gamma^{45} &  \gamma^{46} \\
    0\,  & \, 0\, &\,  0\,  & \gamma^{54} & \gamma^{55} &  \gamma^{56}\\
    0\, &\, 0\,  &\,  0\, & \gamma^{64} &  \gamma^{65} &  \gamma^{66} \\
\end{array}
\right),
 \label{19a}
\end{equation}
in addition, the following notations are made (see Appendix {\bf A}):
$$
\gamma^{11}=\gamma^{22}=1,\quad  \gamma^{33}=r^2,\quad
\gamma^{44}= R^2 +\,r^2\bigl(1\,-
\sin^2\theta\cos^2\Omega\bigr),\quad   \gamma^{55}\,= R^2
 \sin^2\Theta \,\,+
$$
$$
r^2 \bigl\{\sin^2\Theta(1- \sin^2\theta\sin^2\Omega)+\sin^2\theta\cos^2\Theta+
(1/2)\sin2\theta\sin2\Theta\sin\Omega\bigr\},
\quad \gamma^{66}=
r^2\sin^2\theta,
$$
$$
\gamma^{45}=\gamma^{54}\,=\,-(1/ 2)
r^2\, \bigl(\sin^2\theta\sin\Theta\sin2\Omega+\sin2\theta\cos\Theta\cos\Omega\bigr),
\quad \gamma^{46}=\gamma^{64}= (1/ 2)\,\times
$$
$$
r^2\sin2\theta\cos\Omega,
\qquad
\gamma^{56}=\gamma^{65}= -(1/ 2)r^2
\bigl(\sin2\theta\sin\Theta\sin\Omega-2\sin^2\theta\cos\Theta\bigr).
$$
Using the metric tensor (\ref{19a}), one can write a linear infinitesimal element
of Euclidean space in hyperspherical coordinates:
\begin{equation}
(ds)^2=\gamma^{\alpha\beta}(\{\rho\})d\rho_\alpha d\rho_\beta,\qquad \alpha,\beta=\overline{1,6}.
 \label{19b}
\end{equation}

\textbf{Definition 2.} \emph{Let}  $(F,G):\mathbb{R}^{12}\to \mathbb{R}^{1}$
 \emph{be functions of 12 variables}  $(r_\alpha,p_\alpha),  $\emph{where}
  $\alpha=\overline{1,6}.$ \emph{The Poisson bracket on
 the phase space}  $\mathcal{P}\cong\mathbb{R}^{12}$ \emph{is defined by the following form:}
\begin{equation}
\{F,G\}=\sum_{\alpha=1}^6\biggl(\frac{\partial F}{\partial r_\alpha}\frac{\partial G}{\partial p_\alpha}-
\frac{\partial F}{\partial p_\alpha}\frac{\partial G}{\partial r_\alpha}\biggr).
 \label{19c}
\end{equation}
\emph{Note that the variables}  $r_\alpha$ and $p_\alpha$ \emph{denote the projections of  6D
radius vector} $\textbf{r}\in \mathbb{R}^6$ \emph{and the momentum} $\textbf{p}\in
\mathbb{R}^{\ast6},$ \emph{respectively (see equation (\ref{02t}), and also the \bf{Definition 1}).}

\textbf{Definition 3.} \emph{Let} $\mathbb{H}:\mathbb{R}^{12}\to \mathbb{R}^{1}$ \emph{be the Hamiltonian
of the imaginary point with the mass} $\mu_0$ \emph{in the 12-dimensional phase space.} The \emph{Hamiltonian
vector field} $\textbf{X}_{\mathbb{H}}:\mathbb{R}^{12}\to \mathbb{R}^{12}$ \emph{satisfies the equation:}
\begin{equation}
\textbf{X}_{\mathbb{H}}(\textbf{z})= \{\textbf{z},\mathbb{H}\}, \qquad \textbf{z}\in\mathbb{R}^{12}.
 \label{19d}
\end{equation}

\textbf{Definition 4.} \emph{The  Hamiltonian  equations in the phase space}
$\mathcal{P}\cong\mathbb{R}^{12}$ \emph{will be defined as follows:}
\begin{equation}
  \dot{\textbf{z}} =\textbf{X}_{\mathbb{H}}, \qquad \dot{\textbf{z}}=\frac{d\textbf{z}}{dt} \in\mathbb{R}^{12},
 \label{19e}
\end{equation}
\emph{or, equivalently:}
\begin{equation}
  \dot{r}_\alpha =\frac{\partial\mathbb{H} }{\partial p_\alpha},\qquad
  \dot{p}_\alpha =-\frac{\partial\mathbb{H} }{\partial r_\alpha}.
 \label{19f}
\end{equation}

Without going into well-known details, we note that the problem under consideration, having
 in the general case 10 independent integrals of motion, reduces to the system of 8\emph{th} order.
In the case when the total energy is fixed, the reduction of the problem leads to the system
of  7\emph{th} order system (see \cite{Whitt}, and also \cite{Chen}).

Note that only in very few specific cases, the problem of the gravity of three bodies is exactly integrated.

\section{T\lowercase{hree-body problem as a problem of geodesic flows on} R\lowercase{iemannian manifold}}

The classical three-body system  moving in the Euclidean  $3D$ space
  continuously forms a triangle, and, therefore, Newton's  equations
describe a dynamical system on the space of such triangles \cite{Fiz}.
The latter means that we can formally divide the motion into two parts, the
first of which is the rotational motion of the triangle of bodies in $3D$ Euclidean
space, and the second is the internal motion of bodies in the plane of the triangle.

As well-known, the configuration space of the solid body $\mathbb{R}^6$ can
 be represented as a direct product of two subspaces \cite{Arnold1}:
\begin{equation}
 \mathbb{R}^6:\Leftrightarrow \mathbb{R}^3\times \mathbb{S}^3,
\label{19ab}
\end{equation}
where $:\Leftrightarrow$ by definition denotes equivalence, ${\mathbb{R}}^3$ is a
manifold that is defined as the orthonormal space of relative distances between bodies,
and $\mathbb{S}^3$ denotes the space of the rotation group $SO(3)$. Note that in
the considered problem the connections between the bodies are not holonomic, and
therefore the representation (\ref{19ab}) for the configuration space is incorrect.

\textbf{Definition 5.} \emph{Let} $\mathcal{M}$ \emph{be a 6D Riemannian
manifold on which the local coordinate system is defined:}
\begin{equation}
\overline{x^1,x^6}=\{x\}=(x^1,...,x^6)\in \mathcal{M},
\label{03}
\end{equation}
\emph{where the set $\{\bar{x}\}=(x^1, x^2, x^3) $ will be called the internal coordinates,
and the set $\{\underline{x}\} = (x^4, x^5, x^6) $, respectively, the external coordinates.}\\
\emph{It is assumed that $\mathcal{M} $ is a conformal-Euclidean manifold or Weyl space
(see \cite{Nord})
immersed in the Euclidean space $\mathbb{R}^6$, which is determined by the metric tensor:}
\begin{equation}
g_{\mu\nu}(\{\bar{x}\})=g(\{\bar{x}\})\delta_{\mu\nu},\qquad
 g(\{\bar{x}\})=\bigl[\mathrm{E}-U(\{\bar{x}\})\bigr]U^{-1}_0\neq0,\qquad
\mu,\nu=\overline{1,6},
\label{03ab}
\end{equation}
\emph{where}  $\delta_{\mu\nu}$ \emph{denotes the Kronecker symbol}, $\mathrm{E}$
\emph{is the total energy of three-body system,} $U(\{\bar{x}\})$
\emph{is the total interaction potential between bodies and } $U_0=max|U(\{\bar{x}\})|$.

\textbf{Proposition 1.} \emph{If 6D manifold $\mathcal{M}$ is described by the metric tensor
(\ref{03ab}), then it can be represented as a direct product of two subspaces:
\begin{equation}
\mathcal{M}:\Leftrightarrow  \mathcal{M}^{(3)}\times \mathcal{S}^3_{M_k}.
\label{09az}
\end{equation}
where  $\mathcal{M}^{(3)}$ denotes $3D$ \emph{Riemannian} manifold defined as follows:
 $$\mathcal{M}^{(3)}=\bigl[\{\bar{x}\}=(x^1,x^2,x^3)\in
\mathcal{M}^{(3)}_t;\,g_{ij}(\{\bar{x}\})=g(\{\bar{x}\})\delta_{ij};\,\,g(\{\bar{x}\})
\neq 0\bigr]. $$
In addition, $\mathcal{M}^{(3)}_t\cong \bigcup_{k}M_k$ denotes the atlas of the manifold
$\mathcal{M}^{(3)}$ (internal space) and $M_k \ni (x^1,x^2,x^3)_k$ is the $k$-$th$ card.
Note that the atlas $\mathcal{M}^{(3)}_t$, immersed in the manifold $\mathcal{M}$,
is invariant under the local rotations group $SO(3)_{M_k}$ (external space
$\mathcal{S}_{M_k}^3\ni(x^4,x^5,x^6)_{M_k}$).}\\
\textbf{Proof.}

Using the \emph{Maupertuis' variational principle}, one can derive equations for geodesic
trajectories on the Riemannian manifold $\mathcal{M}$ (see \cite{Arnold1,BubrNovFom}):
\begin{eqnarray}
\ddot{x}^{\mu}+\Gamma^\mu_{\nu\gamma}(\{\bar{x}\}) \dot{x}^{\,\nu}
\dot{x}^{\,\gamma} =0,\qquad \mu,\nu,\gamma=\overline{1,6},
\label{05}
\end{eqnarray}
where
\begin{equation}
\dot{x}^{\,\mu}=\frac{dx^\mu}{ds},\qquad
\ddot{x}^{\,\mu}=\frac{d^{\,2}x^\mu}{ds^2},\qquad  s=\int
 \sqrt{g_{\mu\nu}(\{\bar{x}\})dx^\mu dx^\nu}.
\label{05t}
\end{equation}
Recall that $\dot{x}^\mu$ and $\ddot{x}^{\,\mu}$ denote the geodesic velocity
and acceleration, respectively, while $``s"$ denotes the length of a curve along
a geodesic trajectory. Note that it plays the role of a chronological parameter,
which orders the sequence of stages of the dynamical system motion, and in
further will be called \emph{internal time} of the system. In the equations (\ref{05})
$\Gamma^\mu_{\nu\gamma}(\{x\})$ denotes the Christoffel symbol:
$$
\Gamma^\mu_{\nu\gamma}(\{\bar{x}\})=\frac{1}{2}g^{\mu
\mu}\bigl({\partial_\gamma g_{\mu \nu}}+\partial_\nu g_{\gamma
\mu}-\partial_\mu g_{\nu \gamma}\bigr),\qquad \partial_\mu\equiv\partial_{x^\mu}.
$$
Taking into account (\ref{03ab}) and (\ref{05}), one can obtain the following
  equations for geodesic trajectories:
$$
\ddot{x}^{1} =a_1\Bigl\{\bigl(\dot{x}^{1}\bigr)^2-\sum_{\mu\neq
1,\,\mu=2}^6\bigl(\dot{x}^{\mu}\bigr)^2\Bigr\}
+2\dot{x}^{1}\Bigl\{a_2\dot{x}^{2}+a_3
\dot{x}^{3} \Bigr\},
$$
$$
\ddot{x}^{2} =a_2\Bigl\{\bigl(\dot{x}^{\,2}\bigr)^2-\sum_{\mu=1,\, \mu\neq
2}^6\bigl(\dot{x}^\mu\bigr)^2\Bigr\}
+2\dot{x}^{2}\Bigl\{a_3\dot{x}^{3}+a_1\dot{x}^{1} \Bigr\},
$$
$$
\ddot{x}^{3}
=a_3\Bigl\{\bigl(\dot{x}^{3}\bigr)^2-\sum_{\mu=1,\,\mu\neq3}^6\bigl(\dot{x}^{\mu}\bigr)^{2}\Bigr\}
+2\dot{x}^{3}\Bigl\{a_1 \dot{x}^{1}+a_2\dot{x}^{2}\Bigr\},
$$
\begin{eqnarray}
\qquad\qquad\qquad\qquad\quad\,
\ddot{x}^{4}=2\dot{x}^{4}\Bigl\{a_1\dot{x}^{1}+a_2 \dot{x}^{2}+a_3\dot{x}^{3}\Bigr\},
\nonumber\\
\ddot{x}^{5}=2\dot{x}^{5}\Bigl\{a_1
\dot{x}^{1}+a_2 \dot{x}^{2}+a_3\dot{x}^{3}\Bigr\},
\nonumber\\
\ddot{x}^{6}=2\dot{x}^{6}\Bigl\{a_1 \dot{x}^{1}+a_2
\dot{x}^{2}+a_3\dot{x}^{3}\Bigr\},
\label{06}
\end{eqnarray}
where
\begin{equation}
a_i(\{\bar{x}\})=-\partial_{x^i}\ln
\sqrt{g(\{\bar{x}\})}, \qquad \partial_{x^i}\equiv\partial/\partial x^i,
\label{06a}
\end{equation}
in addition,   the metric $g_{\mu\nu}$ is the conformal-Euclidean and, therefore,
$g(\{\bar{x}\})=g_{11}(\{\bar{x}\})=...=g_{66}(\{\bar{x}\})$.

It is easy to show  that in the system (\ref{06}) the last three equations can be exactly integrated:
\begin{equation}
\dot{x}^{\mu}=J_{\mu-3}/g(\{\bar{x}\}),\quad J_{\mu-3}=const_{\mu-3},
\label{07}
\end{equation}
where $\mu=\overline{4,6 }.$

Note that $J_1,J_2$ and $J_3$ are integrals of the motion of the problem.
They can be interpreted as projections of the total angular momentum of the
three-body system $J=\sqrt{\sum_{i=1}^3J^2_i}=const$ on the corresponding
three orthogonal local axes $\bigl(x^1,x^2,x^3\bigr)$. Recall that for the
classical problem these projections can continuously change and take arbitrary values.

Substituting (\ref{07}) into the equations (\ref{06}), we obtain the following
system of second-order nonlinear ordinary differential equations:
\begin{eqnarray}
\ddot{x}^1 =a_1\bigl\{(\dot{x}^1)^2-(\dot{x}^{2})^2-(\dot{x}^{3})^2-
\Lambda^2\bigr\}+ 2\dot{x}^1\bigl\{a_2\dot{x}^{2}+a_3
\dot{x}^{3} \bigr\},
\nonumber\\
\ddot{x}^2 =a_2\bigl\{(\dot{x}^2)^2-(\dot{x}^3)^2-(\dot{x}^1)^2-
\Lambda^2\bigr\}+ 2\dot{x}^{2}\bigl\{a_3 \dot{x}^{3}+a_1
\dot{x}^{1}\bigr\},
\nonumber\\
\ddot{x}^{3} =a_3
\bigl\{(\dot{x}^{3})^2-(\dot{x}^{1})^2-(\dot{x}^{2})^2-
\Lambda^2\bigr\}+2\dot{x}^{3}\bigl\{a_1 \dot{x}^{1}+a_2\dot{x}^{2}\bigr\},\,
 \label{08}
\end{eqnarray}
where $a_i\equiv a_i(\{\bar{x}\})$ and $\Lambda^2\equiv\Lambda^2(\{\bar{x}\})=
\bigl(J/g(\{\bar{x}\})\bigr)^2.$

The system of equations (\ref{08}) describes  motion of geodesic flows on an oriented
$3D$ submanifold $\mathcal{M}^{(3)}_{\{\bar{J}\}}\,$(\emph{the set of  projections
$\{\bar{J}\}=(J_1,J_2,J_3)$ defines the submanifold orientation}), which is
 immersed in the $6D$ manifold (space) $\mathcal{M}$.

The system of equations (\ref{08}) can be represented as a 6\emph{th order system},
that is, a system consisting of six first order differential equations:
\begin{eqnarray}
\dot{\xi^1} =a_1\bigl\{(\xi^1)^2-(\xi^{2})^2-(\xi^{3})^2-
\Lambda^2\bigr\}+ 2\xi^1\bigl\{a_2\xi^{2}+a_3
\xi^{3} \bigr\},\qquad \xi^1=\dot{x}^1,
\nonumber\\
\dot{\xi}^2 =a_2\bigl\{(\xi^2)^2-(\xi^3)^2-(\xi^1)^2-
\Lambda^2\bigr\}+ 2\xi^{2}\bigl\{a_3 \xi^{3}+a_1
\xi^{1}\bigr\},\qquad \xi^2=\dot{x}^2,
\nonumber\\
\dot{\xi}^{3} =a_3
\bigl\{(\xi^{3})^2-(\xi^{1})^2-(\xi^{2})^2-
\Lambda^2\bigr\}+2\xi^{3}\bigl\{a_1 \xi^{1}+a_2\xi^{2}\bigr\},\,\qquad \xi^3=\dot{x}^3.
\label{07a}
\end{eqnarray}

Thus, we proved that the last three equations in  (\ref{06}) describing the external
three coordinates $\{\underline{x}\}$ are exactly integrated and form a local rotation
group $SO(3)_{M_i}$.  The latter means that the 6$D$ manifold $\mathcal{M}$ can be
continuously filled with the submanifold $\mathcal{M}^{(3)}_{\{\bar{J}\}}$, rotating it
 according to the law of the local symmetry group $SO(3)_{M_i} $ and therefore the
representation (\ref{09az}) is true.

\textbf{Proposition 1} is proved.

\subsection{Reduced Hamiltonian in the \emph{ internal space} $\mathbb{E}^3\subset \mathbb{R}^3$}

Taking into account (\ref{03ab}) and (\ref{07}),  we can
reduce the Hamiltonian and obtain the following representation for it:
\begin{eqnarray}
\mathcal{H}\bigl(\{\bar{x}\};\{\bar{p}\}\bigr)=\frac{1}{2\mu_0}
g^{\mu\nu}(\{\bar{x}\}p_{\mu}p_{\nu}= \frac{1 }{2\mu_0g(\{\bar{x}\})}
p_{\mu}p_{\nu}\delta^{\mu\nu}\,\, \,\,
\nonumber\\
 =\frac{1}{2}\mu_0g(\{\bar{x}\}) \Biggl\{\sum_{i=1}^3\bigl(\dot{x}^{i}\bigr)^2\,+\,
 \biggl(\frac{J}{g(\{\bar{x}\})}\biggr)^2 \Biggr\},
\label{09}
\end{eqnarray}
where $\{\bar{p}\}=(p_{\,1},p_{\,2},p_{\,3})$ and $\mu,\nu=\overline{1,6}.$\\
Note that the reduced Hamiltonian (\ref{09}) is clearly independent of the
mass of the bodies. If we analyze the stages of obtaining the expression
(\ref{09}), we will see that the representation contains a dependence on
the masses, however it is hidden in coordinate transformations (see
transformations above (\ref{01})). The system of geodesic equations
(\ref{08}) can be obtained using the Hamilton equations:
\begin{eqnarray}
 \dot{x}^i =\frac{\partial \mathcal{H}}{\partial p_i}=g^{ik}(\{\bar{x}\})p_k,\qquad
\dot{p}_i =-\frac{\partial \mathcal{H}}{\partial x^i}=-\frac{1}{2\mu_0}
\frac{\partial g^{kl}(\{\bar{x}\})}{\partial x^i}p_k p_l,
\label{09a}
\end{eqnarray}
where  $ i,k,l=\overline{1,3}$.

Finally, assuming that in the three-body system the total energy is fixed:
\begin{equation}
 \mathrm{E} =\mathcal{H}\bigl(\{\bar{x}\};\{\bar{p}\}\bigr)= const,
\label{09at}
\end{equation}
the problem can be reduced  to the 5\emph{th} order system.

Thus, the system of equations (\ref{07a}) is the 6\emph{th} order system,
which describes the dynamics of an \emph{imaginary point} with an effective mass
$\mu_0$ on the 3\emph{D} Riemannian  manifold $\mathcal{M}^{(3)}_{\{\bar{J}\}}$.
Note that the system of equations (\ref{07a}) can also be obtained from
the Hamilton equations (\ref{09a})  using the reduced Hamiltonian (\ref{09}).
Using the system of equations (\ref{07a}), we can study in detail
the behavior of geodesic flows of various elementary atom-molecular
processes in the \emph{internal space} $\mathbb{E}^3\subset\mathbb{R}^3$.

\section{T\lowercase{he mappings between} 6\emph{D} E\lowercase{uclidean and} 6\emph{D}
\lowercase{conformal}-E\lowercase{uclidean subspaces}}
Now the main problem is to prove that the 6\emph{th} order system (\ref{07a})
is equivalent to the original three-particle Newtonian problem (\ref{19f}).
Recall, that both representations will be equivalent, if we prove that there
exists continuous one-to-one mappings between the two following  manifolds
$\mathbb{E}^6$ and $\mathcal{M}$, where $\mathbb{E}^6\subset \mathbb{R}^6$
is a subspace allocated from the Euclidean space  $\mathbb{R}^6$
taking into account the condition:
\begin{equation}
\breve{g}(\{\bar{\rho}\})=\mathrm{E}-\mathbb{V}(\{\bar{\rho}\})\neq0.
 \label{11a}
\end{equation}
In other words, we  must prove that between two sets of coordinates
$\overline{\rho^1,\rho^6}=\{\rho\}\in \mathbb{E}^6$ and
$\overline{x^1,x^6}=\{x\}\in \mathcal{M}$,
there are continuous direct and inverse one-to-one mappings.

In this regard, it makes sense to consider three cases:

a. When  $\breve{g}(\{\bar{\rho}\})<0$,  the system of equations (\ref{07a})
 obviously describes a restricted three-body problem.

b. When $\breve{g}(\{\bar{\rho}\})>0$, we are dealing with a typical scattering
problem in a three-body system.

c. When $\breve{g}(\{\bar{\rho}\})=0$. This is a special and very important
 case, which, generally speaking,  requires an extension  of  the
\emph{Maupertuis-Hamilton principle of least action}
on the case of complex-classical trajectories. In this article, we will touch
upon this problem problem when considering a restricted three-body problem.

\subsection{On a homeomorphism between the subspace $\mathbb{E}^{6}\subset\mathbb{R}^6$
and the manifold $\mathcal{M}$}

\textbf{Proposition 2.} \emph{If the interaction potential between the three bodies has
the form (\ref{01b}) and, moreover, it belongs to the class $\mathbb{V}(\{\bar{\rho}\})
\in\mathbb{C}^1(\mathbb{R}^6)$, then the Euclidean subspace $\mathbb{E}^{6}\subset\mathbb{R}^6$
 is homeomorphic to the manifold $\mathcal{M}$.}\\
\textbf{Proof.}

Let us consider a linear infinitesimal element $(ds)$ in both coordinate systems
$\{\rho\}\in \mathbb{E}^6$ and $\{x\}\in \mathcal{M}$. Equating them, we can write:
\begin{eqnarray}
(ds)^2=
\gamma^{\alpha\beta}(\{\mathrm{\rho}\})d\mathrm{\rho}_{\alpha}
d\mathrm{\rho}_\beta=
g_{\mu\nu}(\{\bar{x}\})dx^\mu{dx^\nu},\quad
\alpha,\beta,\mu,\nu=\overline{1,6},
\label{11}
\end{eqnarray}
from which one can obtain the following system of algebraic equations:
\begin{equation}
\gamma^{\alpha\beta}(\{\mathrm{\rho}\})\rho_{\alpha,\mu}\rho_{\beta,\nu}=
g_{\mu\nu}(\{\bar{x}\})= g(\{\bar{x}\})\delta_{\mu\nu},
 \label{12}
\end{equation}
where it is necessary to prove that the coefficients
$\mathrm{\rho}_{\alpha,\mu}(\{x\})=\partial\mathrm{\rho}_\alpha/\partial{x}^\mu$
have the meaning of derivatives. In this regard, we must prove that the function
$\mathrm{\rho}_{\alpha}(\{x\})$ is twice differentiable and continuous in  the
whole domain of its definition and  satisfy the symmetry condition:
\begin{equation}
\mathrm{\rho}_{\alpha,\mu\nu}(\{x\})=\mathrm{\rho}_{\alpha,\nu\mu}(\{x\}),
\qquad\qquad\forall\,\, \mu,\nu= \overline{1,6},
 \label{12w}
\end{equation}
\emph{(Schwartz's theorem on the symmetry of second derivatives)}.

Recall that the set of coefficients $\mathrm{\rho}_{\alpha,\mu}(\{x\})$
allows us to perform coordinate transformations $\{\rho\}\mapsto\{x\}$,
which we shall call \emph{direct transformations}.

Similarly, from (\ref{11}), one can obtain a system of algebraic equations
defining inverse transformations:
\begin{eqnarray}
\gamma_{\alpha\beta}(\{\rho\})
 g^{-1}(\{\bar{x}\})=x^{\mu}_{\,\,,\,\alpha}x^{\nu}_{\,\,,\,\beta}\,\delta_{\mu\nu},
  \label{13}
\end{eqnarray}
where
$x^{\mu}_{\,\,,\,\alpha}(\{\rho\})=\partial x^\mu/\partial{\rho}^\alpha$ and
$\gamma_{\alpha\beta}(\{{\rho}\})=\gamma_{\alpha\bar{\alpha}}(\{{\rho}\})\,\gamma_{\beta\bar{\beta}}
(\{{\rho}\})\,\gamma^{\bar{\alpha}\bar{\beta}}(\{{\rho}\})$.

At first we consider the system of equations (\ref{12}), which is related to direct coordinate
transformations. It is not difficult to see that the system of algebraic equations (\ref{12}) is
underdetermined with respect to the variables $ \mathrm{\rho}_ {\alpha,\mu}(\{x\}) $,
since it consists of 21 equations, while the number of unknown variables is 36. Obviously,
when these equations are compatible, then the system of equations (\ref{12}) has an
infinite number of real and complex solutions.
Note that for the classical three-body problem, the real solutions of the system (\ref{12})
are important, which  form a 15\,-dimensional manifold. Since the system of equations (\ref{13})
is still defined in a rather arbitrary way we can impose additional conditions on it in order
to find the minimal dimension of the manifold allowing a separation of the base $\mathcal{M}^{(3)}_{\{\bar{J}\}}$
from the layer $\bigcup_i\mathcal{S}^3_{M_i}$ (see  expression (\ref{09az})).

Let us make a new notations:
\begin{eqnarray}
\alpha_\mu=\rho_{1,\mu},\quad \beta_\mu=\rho_{2,\mu}, \quad
\zeta_\mu=\rho_{3,\mu}, \quad u_\mu=\rho_{4,\mu},\quad
v_\mu=\rho_{5,\mu},\quad w_\mu=\rho_{6,\mu}.
\label{20}
\end{eqnarray}
We also require that the following additional conditions be met:
\begin{eqnarray}
\alpha_4=\alpha_5=\alpha_6=0, \qquad \beta_4 =\beta_5= \beta_6= 0, \qquad
\zeta_4=\zeta_5=\,\zeta_6\,=0,
\nonumber\\
u_1=u_2=u_3=0,\qquad
v_1=v_2=v_3=0,\qquad
w_1=w_2=w_3=0.
 \label{21}
\end{eqnarray}

Using (\ref{19a}), (\ref{20}) and conditions (\ref{21}) from the equation (\ref{12})
we can obtain two independent  systems of algebraic equations:
\begin{eqnarray}
\alpha_1^2+\beta_1^2+\gamma^{33}\zeta_1^2\,=\,\breve{g}(\{\bar{\rho}\}),\qquad
\alpha_1\alpha_2+\beta_1\beta_2+\gamma^{33}\zeta_1\zeta_2=0,
\nonumber\\
\alpha_2^2+\beta_2^2+\gamma^{33}\zeta_2^2\,=\,\breve{g}(\{\bar{\rho}\}),\qquad
\alpha_1\alpha_3+\beta_1\beta_3+\gamma^{33}\zeta_1\zeta_3=0,
\nonumber\\
\alpha_3^2+\beta_3^2+\gamma^{33}\zeta_3^2\,=\,\breve{g}(\{\bar{\rho}\}),\qquad
\alpha_2\alpha_3+\beta_2\beta_3+\gamma^{33}\zeta_2\zeta_3=0,
\label{22}
\end{eqnarray}
 and, correspondingly:
\begin{eqnarray}
\gamma^{44}u_4^2+
\gamma^{55}v_4^2+\gamma^{66}w_4^2+2(\gamma^{45}u_4v_4
+\gamma^{46}u_4w_4+\gamma^{56}v_4w_4)=\breve{g}(\{\bar{\rho}\}),
\nonumber\\
\gamma^{44}u_5^2+
\gamma^{55}v_5^2+\gamma^{66}w_5^2+2(\gamma^{45}u_5v_5
+\gamma^{46}u_5w_5+\gamma^{56}v_5w_5)=\breve{g}(\{\bar{\rho}\}),
\nonumber\\
\gamma^{44}u_6^2+
\gamma^{55}v_6^2+\gamma^{66}w_6^2+2(\gamma^{45}u_6v_6
+\gamma^{46}u_6w_6+\gamma^{56}v_6w_6)=\breve{g}(\{\bar{\rho}\}),
\nonumber\\
a_4u_4+a_5v_4+a_6w_4=0,
\nonumber\\
b_4u_5+b_5v_5+b_6w_5=0,
\nonumber\\
c_4u_6+c_5v_6+c_6w_6=0.
 \label{23}
\end{eqnarray}
In equations (\ref{23}) the following notations are made:
$$a_{i}=\gamma^{i4}u_5+\gamma^{i5}v_5+\gamma^{i6}w_5,\quad
b_{j}=\gamma^{j4}u_6+\gamma^{j5}v_6+\gamma^{j6}w_6, \quad
c_{k}=\gamma^{k4}u_4+\gamma^{k5}v_4+\gamma^{k6}w_4,$$
where $i,j,k=\overline{4,6}.$

It should be noted that the solutions of algebraic systems (\ref{22}) and (\ref{23}) form
two different 3$D$ manifolds  $\mathfrak{S}^{(3)} $ and $\mathfrak{R}^{(3)}$,
respectively. Since the manifold $\mathfrak{S}^{(3)}$   play a key role in the  proofs
and the theoretical constructions of representation, the features of its structure
are studied  in detail  (see Appendix \textbf{B}). Note that the manifold $\mathfrak{S}^{(3)}$
is in  a one-to-one mapping on the one hand with the subspace $\mathbb{E}^3 \ni\{\bar{\rho}\}$
(where $\mathbb{E}^3\subset \mathbb{E}^6$ the \emph{internal space} in the hyperspherical
coordinate system), and on the other hand with the submanifold $\mathcal{M}^{(3)}_{\{\bar{J}\}}$
(see FIG. 2). Note that this statement follows from the fact that all points of the submanifold
$\mathcal{M}^{(3)}_{\{\bar{J}\}}$ and the subspace $\mathbb{E}^3 \subset\mathbb{R}^3$, are
pairwise connected through the corresponding derivatives (see (\ref{12})),
which, as unknown variables, enter the algebraic equations (\ref{22}), and, in addition, as shown
there exist also inverse coordinate transformations (see Appendix \textbf{C}).

Now we prove continuity of these mappings.
Recall that the unknowns in the equations (\ref{22}) are in fact functions of coordinates
$\{\bar{\rho}\}$. By making infinitely small coordinate shifts $\{\bar{\rho}\}\to
 \{\bar{\rho}\}+\{\delta\bar{\rho}\}$ in  (\ref{22}), we  get the following system of equations:
\begin{eqnarray}
\bar{\alpha}_1^2+\bar{\beta}_1^2+\bar{\gamma}^{33}\bar{\zeta}_1^2=\bar{g}(\{\bar{\rho}\}),\qquad
\bar{\alpha}_1\bar{\alpha}_2+\bar{\beta}_1\bar{\beta}_2+\bar{\gamma}^{33}\bar{\zeta}_1\bar{\zeta}_2=0,
\nonumber\\
\bar{\alpha}_2^2+\bar{\beta}_2^2+\bar{\gamma}^{33}\bar{\zeta}_2^2=\bar{g}(\{\bar{\rho}\}),\qquad
\bar{\alpha}_1\bar{\alpha}_3+\bar{\beta}_1\bar{\beta}_3+\bar{\gamma}^{33}\bar{\zeta}_1\bar{\zeta}_3=0,
\nonumber\\
\bar{\alpha}_3^2+\bar{\beta}_3^2+\bar{\gamma}^{33}\bar{\zeta}_3^2=\bar{g}(\{\bar{\rho}\}),\qquad
\bar{\alpha}_2\bar{\alpha}_3+\bar{\beta}_2\bar{\beta}_3+\bar{\gamma}^{33}\bar{\zeta}_2\bar{\zeta}_3=0,
\label{22k}
\end{eqnarray}
where
  $$\bar{g} (\{\bar{\rho}\})=
\breve{g}\bigl(\{\bar{\rho}\}+\{\delta\bar{\rho}\}\bigr), \qquad
 \{\delta\bar{\rho}\}=(\delta\rho^1,\delta\rho^2,\delta\rho^3).$$
Assuming that the displacement  $||\,\delta\{\bar{\rho}\}||\ll 1$, in the equations
(\ref{22k}), we can expand the functions in a Taylor series, and  further, taking into
account the equations system (\ref{22}), we can get:
\begin{eqnarray}
\delta\rho^i\bigl\{2(\alpha_1\alpha_{1\,i}+ \beta_1\beta_{1\,i}+ \gamma^{33} \zeta_1\zeta_{1 \,i})+
\gamma^{33}_{,\,i} \zeta_1^2-\breve{g}_{,\,i}(\{\bar{\rho}\})\bigr\}+O(||\,\delta\{\bar{\rho}\}||^2)=0,
\,\,\,
\nonumber\\
\delta\rho^i\bigl\{2(\alpha_2\alpha_{2\,i}+\beta_2\beta_{2\,i}+\gamma^{33} \zeta_2\zeta_{2\,i})+
\gamma^{33}_{,\,i}\zeta_2^2 -\breve{g}_{,\,i}(\{\bar{\rho}\})\bigr\}+O(||\delta\{\bar{\rho}\}||^2)=0,
\,\,\,
\nonumber\\
\delta\rho^i\bigl\{2(\alpha_3\alpha_{3\,i}+\beta_3\beta_{3\,i}+\gamma^{33}\zeta_3\zeta_{3,i})+
\gamma^{33}_{,\,i}\zeta_3^2 -\breve{g}_{,\,i}(\{\bar{\rho}\})\bigr\}\,+\,O(||\delta\{\bar{\rho}\}||^2)=0,
\,\,\,
\nonumber\\
\delta\rho^i\bigl\{{\alpha}_1{\alpha}_{2\,i}+{\alpha}_2{\alpha}_{1 \,i}+
{\beta}_1{\beta}_{2 \,i}+{\beta}_2{\beta}_{1\,i}+\gamma^{33}({\zeta}_1
{\zeta}_{2\,i}+{\zeta}_2{\zeta}_{1\,i})+
{\gamma}^{33}_{,\,i}{\zeta}_1{\zeta}_2\bigl\}+O(||\,\delta\{\bar{\rho}\}||^2)=0,\,\,\,
\nonumber\\
\delta\rho^i\bigl\{{\alpha}_1{\alpha}_{3 \,i}+{\alpha}_3{\alpha}_{1 \,i}+
{\beta}_1{\beta}_{3\,i}+{\beta}_3{\beta}_{1 \,i}+\gamma^{33}({\zeta}_1{\zeta}_{3\,i}+{\zeta}_3{\zeta}_{1\,i})+
{\gamma}^{33}_{,\,i}{\zeta}_1{\zeta}_3\bigl\}+O(||\,\delta\{\bar{\rho}\}||^2)=0,\,\,\,
\nonumber\\
\delta\rho^i\bigl\{{\alpha}_2{\alpha}_{3,\,i}+{\alpha}_3{\alpha}_{2\,i}+
\beta_2{\beta}_{3\,i}+ {\beta}_3{\beta}_{2\,i}+\gamma^{33}({\zeta}_2\zeta_{3\,i}+{\zeta}_3{\zeta}_{2\,i})+
\gamma^{33}_{,\,i}{\zeta}_2{\zeta}_3\bigl\}+O(||\,\delta\{\bar{\rho}\}||^2)=0,
\,\,\,
\label{22t}
\end{eqnarray}
where $i=\overline{1,3}$  and, in addition, summation is performed  by dummy indices.\\
If we require that the expressions with the same increments be equal to zero,
then from (\ref{22t}) one can obtain an underdetermined system of algebraic
equations, i.e.  18 equations for finding 27 unknowns variables:
$$
 2(\alpha_1\alpha_{1\,i}  + \beta_1\beta_{1\,i}+ \gamma^{33} \zeta_1\zeta_{1\,i})+
\gamma^{33}_{,\,i}\zeta_1^2 - \breve{g}_{,\,i}(\{\bar{\rho}\})=0,
$$
$$
2(\alpha_2\alpha_{2\,i}  + \beta_2\beta_{2\,i}+ \gamma^{33}\zeta_2\zeta_{2\,i})+
\gamma^{33}_{,\,i} \zeta_2^2 - \breve{g}_{,\,i}(\{\bar{\rho}\})=0,
$$
\begin{eqnarray}
2(\alpha_3\alpha_{3\,i} + \beta_3\beta_{3\,i}+ \gamma^{33} \zeta_3\zeta_{3\,i})+
\gamma^{33}_{,\,i} \zeta_3^2 - \breve{g}_{,\,i}(\{\bar{\rho}\})=0,
\nonumber\\
{\alpha}_2{\alpha}_{1\,i}+{\alpha}_1{\alpha}_{2\,i}+{\beta}_2{\beta}_{1\,i}+
{\beta}_1{\beta}_{2\,i}+\gamma^{33}({\zeta}_2{\zeta}_{1\,i}+{\zeta}_1{\zeta}_{2\,i})+
{\gamma}^{33}_{,\,i}{\zeta}_1{\zeta}_2=0,
\nonumber\\
{\alpha}_3{\alpha}_{1\,i}+{\alpha}_1{\alpha}_{3\,i}+{\beta}_3{\beta}_{1\,i}+
{\beta}_1{\beta}_{3\,i}+\gamma^{33}({{\zeta}_3{\zeta}_{1\,i}+\zeta}_1{\zeta}_{3\,i})+
{\gamma}^{33}_{,\,i}{\zeta}_1{\zeta}_3=0,
\nonumber\\
 {\alpha}_3{\alpha}_{2\,i}+{\alpha}_2{\alpha}_{3\,i}+{\beta}_3{\beta}_{2\,i}+
{\beta}_2{\beta}_{3\,i}+ \gamma^{33}({\zeta}_3{\zeta}_{2\,i}+{\zeta}_2{\zeta}_{3\,i})+
{\gamma}^{33}_{,\,i}{\zeta}_2{\zeta}_3=0.
\label{22zt}
\end{eqnarray}
 Recall that the set of coefficients
 $\{\sigma\}=(\sigma_1,...,\sigma_9)=[\alpha=(\alpha_1,\alpha_2,\alpha_3),\,
 \beta=(\beta_1,\beta_2,\beta_3),\, \zeta=(\zeta_1,\zeta_2,\zeta_3)]$
 belongs to the 3$D$ manifold $\mathfrak{S}^{(3)}$.

Now, we can require that the second derivatives be symmetric  $ \sigma_{ij} =\sigma_{ji}$,
where $\{\sigma\}=(\alpha,\beta,\zeta)$ and $i,j=\overline{1,3}$. This, as can be easily seen,
allows us to reduce the number of unknown variables and make the system of equations
definite, i.e. 18 equations for 18 unknowns variables.

The system of equations (\ref{22zt}) can be written in canonical form:
 \begin{equation}
 \mathbb{A}{\bf X}=\mathbb{B},\qquad \mathbb{A}=(d_{\mu\nu}), \qquad\mu,\nu=\overline{1,18},
  \label{22ztw}
\end{equation}
where $\mathbb{A}\in \mathbb{R}^{18\times18}$ is the basic matrix of the system,
$\mathbb{B}\in \mathbb{R}^{18}$ and $\textbf{X}\in \mathbb{R}^{18}$ are columns
of free terms and   solutions of the system, respectively (see  Appendix {\bf D}).
Note that, for an arbitrary point $\{\bar{\rho}_i\}\in\mathbb{E}^3$, the system of
equations (\ref{22}) generates sets of solutions $\{\sigma\}$
that continuously fill a region of $\mathbb{E}^3$ space, forming 3$D$ manifold
$\mathfrak{S}^{(3)}$.   As for the system of equations (\ref{22ztw}), it has a solution
if the determinant of the basic matrix $\mathbb{A}$ is nonzero:
$$
\det(d_{\mu\nu})\neq0, \qquad \mu,\nu=\overline{1,18}.
$$
On the other hand, the algebraic   system (\ref{22ztw}) does not have a
solution when $\det(d_{\mu\nu})=0$.
In this case, at each point $\{\bar{\rho}_i\}$ there exists a countable set
$\mathfrak{W}$ consisting of the coefficients $\{\sigma\}=[\alpha,\beta,\zeta]$,
on which the matrix degenerates. It is easy to verify that the
measure of this set in comparison with the measure of the
$\mathfrak{S}^{(3)}$ for which $\det(d_{\mu\nu})\neq0$, is equal to zero, i.e.
$\mathfrak{W}=\{0\}$. In other words, for the case under consideration Schwartz's
theorem holds, and $\sigma_\varsigma$, where $\varsigma=\overline{1,9}$, and
$d_{\mu\nu}$ (see (\ref{22zt})) have  the sense of the first and second derivatives,
respectively.
\begin{figure}
\includegraphics[width=75mm]{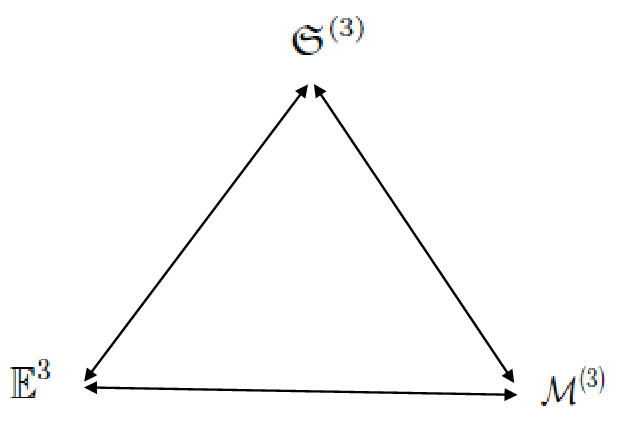}
\caption{\emph{In this diagram all spaces are homeomorphic to each other, i.e. }
$\mathbb{E}^3\simeq\mathfrak{S}^{(3)}\simeq\mathcal{M}^{(3)}.$} \label{Fig.2}
\end{figure}

The same is easy to prove for inverse mappings (see Appendix {\bf C}).\\
Let us consider the open set $\forall\,G =\cup_{\alpha} G_\alpha $, consisting of the
union of cards $G_\alpha $ arising at continuously mappings
$f:\{\bar{\rho}\}\mapsto\{\bar{x}\} $ using algebraic equations (\ref{22}).
Proceeding from the foregoing, it is obvious that the maps can be chosen so
that the immediate neighbors have intersections comprising at least
one common point, that is a necessary condition for the continuity of the
mappings. Using the above arguments,  we assert that the atlas $G$ can be
widened up to $G\cong\mathcal{M}^{(3)}$.

Thus, all the conditions of the theorem on homeomorphism between the metric
 spaces $\mathbb{E}^3$ and $\mathcal{M}^{(3)}_{\{\bar{J}\}}$ are satisfied, and therefore
 we can say that these spaces are homeomorphic or topologically equivalent,
 which means $f:\mathbb{E}^3\mapsto\mathcal{M}^{(3)}_{\{\bar{J}\}}$  and
$f^{-1}:\mathcal{M}^{(3)}_{\{\bar{J}\}}\mapsto \mathbb{E}^3$
(see   Appendix {\bf B}).

As for the  system of algebraic equations (\ref{23}), then at each
point of the \emph{internal space} $M_k(x^1,x^2,x^3)_k\in\mathcal{M}^{(3)}$,
it generates $3D$ manifold $\mathfrak{R}^{(3)}$ that is a local
analogue of the Euler angles and, consequently,
$\cup_{k}\mathcal{S}^3_{M_k}\simeq \mathfrak{R}^{(3)}$. The layer,
$\mathfrak{R}^{(3)}$ continuously passing through all points of the
basis $\mathcal{M}^{(3)}_{\{\bar{J}\}}$, fills the subspace $\mathbb{E}^6$.

Finally, taking into account the above, we can conclude that the Euclidean subspace
 $\mathbb{E}^6\subset\mathbb{R}^6$ and the Riemannian manifold $\mathcal{M}$, \emph{are also homeomorphic}.

\emph{{\bf{Proposition 2}}  is proved.}

\subsection{The classical three-body problem and the Poincar\'{e}  conjecture}
It well known that Poincar\'{e} was  the first to attempt to study of $3D$
manifolds in connection with problems of classical Hamiltonian systems. As a result
of this study, in 1904, he formulated his famous hypothesis (the \emph{Poincar\'{e} conjecture}),
which in the framework of modern mathematical conceptions could be formulated as follows:

\emph{If a smooth compact 3D manifold $\mathfrak{K}^3$ has the property that every
simple closed curve within the manifold can be deformed continuously to a point,  it
follow that $\mathfrak{K}^3$ is homeomorphic to the sphere $\mathbb{S}^3$.}

Recall that $3D$ unit sphere $\mathbb{S}^3$, that is, the locus of all
points $(x,y,z,w)$ in $4D$ Euclidean space which have distance exactly
1 from the origin \cite{Papak}:
$$
x^2+ y^2 + z^2 + w^2=1.
$$
In 2002, Perelman proved Poincar\'{e}'s conjecture without any connection to dynamical
systems \cite{Per}. In this sense it will be interesting to understand the relationship of this
\emph{Poincar\'{e} conjecture} to the classical three-body problem.

For this, in the equation (\ref{22}) it is useful to make change of variables.
In particular, the new variables will be determined by the following formulas:
\begin{eqnarray}
\tilde{\alpha}_{1}=\frac{\alpha_1+\alpha_2}{\sqrt\sigma_1},\qquad \tilde{\beta}_{1}=
\frac{\beta_1+\beta_2}{\sqrt\sigma_1},\qquad
\tilde{\zeta}_1=\frac{\sqrt{\gamma^{33}}\zeta_1}{\sqrt\sigma_1}, \qquad \tilde{\zeta}_{2(1)}=
\frac{\sqrt{\gamma^{33}}\zeta_2}{\sqrt\sigma_1},\nonumber\\
\tilde{\alpha}_{2}=\frac{\alpha_2+\alpha_3}{\sqrt\sigma_2},\qquad \tilde{\beta}_{2}=
\frac{\beta_2+\beta_3}{\sqrt\sigma_2},\qquad \tilde{\zeta}_2=
\frac{\sqrt{\gamma^{33}}\zeta_2}{\sqrt\sigma_2},\qquad \tilde{\zeta}_{3(2)}=
\frac{\sqrt{\gamma^{33}}\zeta_3}{\sqrt\sigma_2},
\nonumber\\
\tilde{\alpha}_{3}=\frac{\alpha_3+\alpha_1}{\sqrt\sigma_3},\qquad \tilde{\beta}_{3}=
\frac{\beta_3+\beta_1}{\sqrt\sigma_3},\qquad \tilde{\zeta}_3=\frac{\sqrt{\gamma^{33}}
\zeta_3}{\sqrt\sigma_3},
\qquad \tilde{\zeta}_{1(3)}=\frac{\sqrt{\gamma^{33}}\zeta_1}{\sqrt\sigma_3},
\label{23b}
\end{eqnarray}
where
$$
\sigma_1(\{\bar{x}\})=2\bigl[g(\{\bar{x}\})-\gamma^{33}(\{\bar{x}\})\,\zeta_1\zeta_2\bigr]>0,
\quad \sigma_2(\{\bar{x}\})=2\bigl[g(\{\bar{x}\})-\gamma^{33}(\{\bar{x}\})\,\zeta_2\zeta_3\bigr]>0,
$$
\begin{equation}
\quad \sigma_3(\{\bar{x}\})=2\bigl[g(\{\bar{x}\})-\gamma^{33}(\{\bar{x}\})\,\zeta_3\zeta_1\bigr]>0.
 \label{23c}
\end{equation}

Now taking into account new notations (\ref{23b}), the system of algebraic equations
(\ref{22}) can be represented in the form:
\begin{eqnarray}
\tilde{\alpha}_1^2+\tilde{\beta}_1^2+\tilde{\zeta}_1^2+
\tilde{\zeta}_{2(1)}^2 =1,\qquad
 \sigma_1\varrho_1^2-\sigma_2\varrho_2^2-\sigma_3\varrho_3^2+\varrho_{23}=0,
\nonumber\\
\tilde{\alpha}_2^2+\tilde{\beta}_2^2+\tilde{ \zeta}_2^2+
\tilde{\zeta}_{3(2)}^2 =1,\qquad
\sigma_1\varrho_1^2+\sigma_2\varrho_2^2-\sigma_3\varrho_3^2-\varrho_{12}=0,
\nonumber\\
\tilde{\alpha}_3^2+\tilde{\beta}_3^2+\tilde{\zeta}_3^2+\tilde{\zeta}_{1(3)}^2\,=1,\qquad
\sigma_1\varrho_1^2-\sigma_2\varrho_2^2+\sigma_3\varrho_3^2-\varrho_{13}=0,
\label{22az}
\end{eqnarray}
where the following notations are made:
$$
\varrho_i^2=\tilde{\alpha}_i^2+\tilde{\beta}_i^2+\tilde{\zeta}_i^2,\qquad
\varrho_{ij}=2\sqrt{\sigma_i\sigma_j}(\tilde{\alpha}_i\tilde{\alpha}_j+\tilde{\beta}_i\tilde{\beta}_j
+\tilde{\zeta}_i\tilde{\zeta}_j).
$$
As it can be seen, the (\ref{22az}) is an underdetermined system of algebraic equations consisting
of six equations and nine unknowns.  Recall that in the set of six variables $\{\tilde{\zeta}\}=
\bigl(\tilde{\zeta}_1,\tilde{\zeta}_2,\tilde{\zeta}_3,\tilde{\zeta}_{1(2)},\tilde{\zeta}_{3(2)},
\tilde{\zeta}_{1(3)}\bigr)$ only three variables
are linearly independent.  Unlike the system of equations (\ref{22}), whose domain
of definition is limited by the condition (\ref{11a}), the domain of definition of the
system (\ref{22az}) besides is limited by additional conditions (\ref{23c}).
As a result, the algebraic system (\ref{22az}) generates a manifold in the form of
$3D$  sphere with unit radius $\mathbb{S}^3\subset  \mathfrak{S}^{(3)}$ for each group of variables
$(\tilde{\alpha}_1,\tilde{\beta}_1,\tilde{\zeta}_1)$,
$(\tilde{\alpha}_2,\tilde{\beta}_2,\tilde{\zeta}_2)$ and $(\tilde{\alpha}_3,\tilde{\beta}_3,\tilde{\zeta}_3)$.

In other words, the \emph{Poincar\'{e} conjecture} for the Hamiltonian system, more precisely
for the classical three-body problem, is a special case of the \emph{\bf {Proposition 1.}}

\section{T\lowercase{ransformations between global and local coordinate systems and features
of \emph{internal time}}}

To complete the proof of the equivalence of the developed representation (\ref{08}) -
(\ref{07a}) with the original Newtonian problem, it is necessary to determine the
coordinate transformations between the two sets of coordinates  $\{\bar{x}\}$ and
$\{\bar{\rho}\}$.

As the analysis shows, the transformations between the noted
two sets of coordinates can be represented only in differential form \cite{gev0}:
\begin{eqnarray}
 d\rho_1=\alpha_1dx^1+\alpha_2dx^2+\alpha_3dx^3,
\nonumber\\
 d\rho_2=\beta_1dx^1+\beta_2dx^2+\beta_3dx^3,
\nonumber\\
\qquad d\rho_3=\zeta_1dx^1+\zeta_2dx^2+\zeta_3dx^3,\,\,
\label{24}
\end{eqnarray}
where  the coefficients $(\alpha_1,...,\beta_1,..., \zeta_3)$  are defined from the system
of underdetermined algebraic equations (\ref{22}).

Recall that a Riemannian manifold is defined in the framework of the local coordinate
system $\{\bar{x}\}\in\mathcal{M}^{(3)}_t$. A feature of this representation is that
when choosing a local coordinate system, it is necessary to take into account the
system of algebraic equations (\ref{22}). As for the \emph{timing parameter} $``s"$
(see (\ref{05t})), it can be interpreted as some trajectory in the internal space
$\mathbb{E}^3\ni (\rho_1,\rho_2,\rho_3)$, which stretches from the initial asymptotic
subspace, where the bodies form the configuration $1+ (23)$, to one of the finite
asymptotic scattering subspaces (see Sch. 1). Note that this parameter characterizes
the measure and nature of elementary atomic-molecular processes occurring in the
system and indicates the directions of their development, that is, it is characterized
by \emph{time arrow}. As can be seen from this scheme, there are four types
of elementary processes, each of which is characterized by its
own \emph{internal time} $s_i$.
 \begin{figure}
\includegraphics[width=100mm]{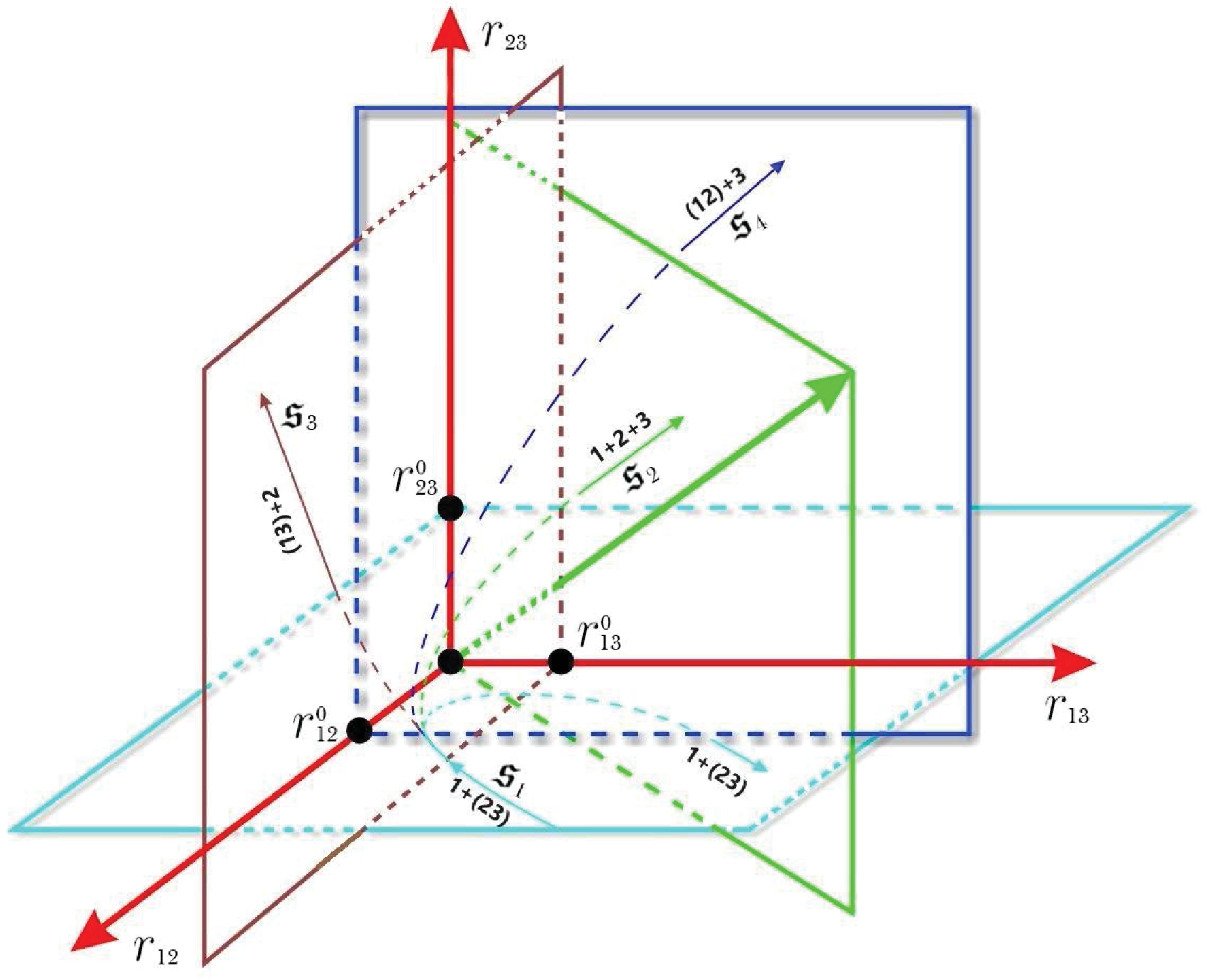}
\caption{\emph{The set of smooth curves ${\mathfrak{s}}=({\mathfrak{s}}_1,...,
{\mathfrak{s}}_4)$ connecting  $(in)$ asymptotic subspace,
where the three-body system is in a state $1+(23)$, with $(out)$ asymptotic
subspaces, where the following configurations of bodies are formed;
$1+(23)$,\, $2+(13)$,\, $3+(12)$ and $1+2+3 $, respectively.  The $r_{ij}(\{\bar{\rho}\})\,
(i,j=\overline{1,3},\, i\neq{j})$ denotes distance  between $i$ and $j$ bodies,
and $r^0_{ij}$ - the average distance between bodies in the corresponding pair.
Note that all the curves $\overline{\mathfrak{s}_{1},\mathfrak{s}_{4}}$  in
the subspace $(in)$  merges, which in the figure is shown by continuous blue.}}
\label{Fig.3}
\end{figure}
Depending on which particular elementary process is being implemented, the corresponding
\emph{internal time} $s_i$ is localized around one of the four smooth curves
$\mathfrak{s}_{i}\simeq\mathbb{R}^1 (i=\overline{1,4}\,)$ connecting two asymptotic
scattering subspaces (see FIG. 3).

When scattering between bodies occurs through the formation of a metastable
\emph{transition complex} of bodies $(123)^\ast$, the  \emph{internal time}
$s_i$ can be represented in the form of complex graph
$\mathbb{G}=(\mathbb{T},\mathbb{L})$, where $\mathbb{T}=\{\hat{s}_0,\hat{s}_1,...,
\hat{s}_n\}$ denotes a set of tops (turning points of a trajectory $s_i$)
of graph and $\mathbb{L}$ is a set of ribs of graph $\{\underline{\hat{s}_0\hat{s}_1},
\underline{\hat{s}_1\hat{s}_2},...,\underline{\hat{s}_{n-1}\hat{s}_n}\}$.
Recall that, the position of each node in the \emph{internal space}
$\mathbb{E}^3$ is determined by three coordinates $\hat{s}_i=\hat{s}_i(\rho_1,\rho_2,\rho_3)$.
Obviously, if to project the graph $\mathbb{G}_i$ of corresponding elementary process
onto the curve  $\mathfrak{s}_i$, then the
sequence of nodes will be violated, for example, as  $\mathfrak{s}_i \ni(\breve{\mathfrak{s}}_0,
...\breve{\mathfrak{s}}_9,\breve{\mathfrak{s}}_1,...\breve{\mathfrak{s}}_7,...
\breve{\mathfrak{s}}_{8},...\breve{\mathfrak{s}}_2,...\breve{\mathfrak{s}}_5,...
\breve{\mathfrak{s}}_4,...)$, where the point $\breve{\mathfrak{s}}_{i}$ denotes the
projection of the turning point (node) $\hat{s}_i$ on the curve $\mathfrak{s}_i.$
It is important to note that, depending on the initial conditions of the problem,
internal time $s_i $ very often may not have a unequivocal graph representation;
moreover, these graphs can be random.

Now, regarding the behavior of a dynamical system depending on the \emph{internal time}
$``s"$.  Formally, if in the system of equations (\ref{08}) we make the replacement $s\to-s$,
then it will not change. However, this does not mean at all that the system of equations is
 invariant with respect to this transformation and, accordingly, is invertible with respect to
the \emph{timing parameter} $``s "$.  The fact is that \emph{internal time} $``s"$ in its structure and
sense is very different from ordinary time $t$, the arrow of which is directed forward
all the time, connecting the events of the past with the future through the present.
In particular, it follows from the above that the points \emph{internal time}, generally
speaking, are not equivalent. This is due to the fact that not only the distances from
the origin, but also on which branches of the \emph{internal time} they are located are
important for their determination. Recall that the \emph{internal time} of a dynamical
system $``s"$, after leaving a region where all bodies interact strongly with each other,
is divided into four different branches  $s=(\overline{s_1,s_4})$, each of which
characterizes a specific elementary process. It should be noted that the choice
between the marked branches of further evolution of system occurs randomly, for
well-known reasons (see the system of equations (\ref{22})). In other words, with
respect to the transformation $s\to-s$, the system of equations (\ref{08}) in the
general case cannot be invariant.

Finally, to answer the question, the system of equations (\ref{08}) with respect to
the parameter $``s"$ is reversible or not, we will analyze the evolution of the
dynamical system from the point of view of the \emph{Poincar\'{e}'s  recurrence theorem}
\cite{Poinc1,Poinc2,Car1,Car2}. To do this, we
consider two possible cases $g(\{x\})>0$ and $g(\{x\})\leq 0$.

The case a. (see sec IV) $g(\{x\})>0$ or is equivalently to $\breve{g}(\{\rho\})>0$ (see sec IV), as
known corresponds to the three-body scattering problem for which the configuration
space $\mathbb{E}^3$  is unrestricted, i.e. infinite. Note that for this case,
Poincar\'{e}'s recurrence theorem is clearly not applicable.

 When $g(\{x\})\leq0$ (or $\breve{g}(\{\rho\})\leq0$), as mentioned above, we are
dealing with a restricted three-body problem. In this case, it it would be
natural to expect that the Poincar\'{e}'s theorem should be satisfied. Namely,
the system should have returned to a state arbitrarily close to its initial
state (for systems with a continuous state), after a sufficiently long but
finite time. However, even in this case, the \emph{Poincar\'{e} theorem} cannot be
is satisfied if we assume the possibility of the existence of various metastable
states characterized by distinct groupings of bodies (see Sch.1). In this case,
we can only say with some probability that the dynamical system will
return close to the  \emph{ initial state} for a long, but finite time.

Thus, analyzing the above arguments, it can be stated that irreversibility
lies in the very nature of \emph{internal time}  $s=(\overline{s_1,s_4})$,
and therefore the system of equations (\ref{08}) with respect to the
\emph{timing parameter} $``s"$, generally speaking, is \emph{irreversible}.

\section{T\lowercase{he restricted three-body problem with holonomic connections}}
An important class of solutions of the classical three-body problem describes
the bound state of three bodies $(123)$, when the motion of  bodies occurs in
a restricted space. In particular, for gravitating bodies, an exact solutions
from this class were founded by a number of outstanding researchers of the 19\emph{th}
and 20\emph{th} centuries, such as Euler \cite{Eul1,Eul2,Eul3},
Lagrange \cite{Lag}, Hill \cite{Hill,H1,H2}. In the mid-1970s, the new Brooke-Heno-Hadjidemetriu
family of orbits was discovered \cite{BrB,HC,Hen}, and in 1993 Moore
showed the existence of stable orbits, \emph{eights}, in which three bodies always
catch up with each other. In 2013, by numerical search, 13 new particular solutions
were found for the three-body problem, in which the movement of a system
of three bodies of the same mass occurs in a repeating cycle \cite{Suv}.
Finally, in 2018, more than 1800 new solutions to the restricted three-body
problem were calculated on a supercomputer \cite{Xi}.

As we will see below, the developed representation has new features and symmetries,
which allows us to obtain important information about the restricted three-body
problem by analyzing systems of algebraic equations.

Note that the state which will be spatially restricted regardless of the length
of time the interaction of bodies cannot be formed as a result of scattering (see Sch. 1)
due to the lack of a mechanism for removing energy from the system.
Nevertheless, it is clear that the character of the motions of bodies in the
states $(123)$ and $(123)^\star$ in many of features should be similar.
In any case, the solutions of the system (\ref{07a}) must satisfy the energy
conservation law (\ref{09at}) that defines  $5D$ hypersurface in the $6D$  phase space.

Some important properties of this problem can be studied by algebraic methods without
solving the equations of motion (\ref{08}) or (\ref{07a}). In particular, it is very
interesting to find solutions for which the connections between bodies remain holonomic
throughout the movement. Recall that this situation is especially interesting for three
gravitating bodies.

\textbf{Proposition 3.} \emph{The three-body system can forms a stable configuration with
holonomic connections, if in the  equations system  (\ref{07a}) all projections of geodetic
acceleration are equal to zero $\ddot{x}^i=0\,\,(i=\overline{1,3}\,)$,  and if there is
 non-empty continuous set $\mathbb{E}^3\supset\Xi\neq\oslash$,  on which the determinant
of the obtained algebraic system is equal to zero. }

\textbf{Proof.}

Let consider the case when the center of mass (\emph{imaginary point}) of
a system of bodies  moves along the manifold $\mathcal{M}^{(3)}_{\{\bar{J}\}}$
without acceleration,  i.e. $\ddot{x}^i=0\,\,(i=\overline{1,3}\,)$. This
means, we can simplify the system of equations (\ref{07a}) by writing
their in the form:
\begin{eqnarray}
 a_1 \bigl\{(\xi^1)^2-(\xi^{2})^2-(\xi^{3})^2-
\Lambda^2\bigr\}+2\xi^1\bigl\{a_2\xi^{2}+a_3\xi^{3}\bigr\}=0,
\nonumber\\
a_2\bigl\{(\xi^2)^2-(\xi^3)^2-(\xi^1)^2-
\Lambda^2 \bigr\}+ 2\xi^{2}\bigl\{a_3\xi^{3}+a_1\xi^{1}\bigr\}=0,
\nonumber\\
a_3\bigl\{(\xi^{3})^2-(\xi^{1})^2-(\xi^{2})^2-
\Lambda^2 \bigr\}+2\xi^{3}\bigl\{a_1\xi^{1}+a_2 \xi^{2}\bigr\}=0.
\label{07kt}
\end{eqnarray}
From the conditions of the absence of acceleration it follows that the projections of the
geodetic velocity $\xi^1,\,\xi^2$ and $\xi^3$ are constants and, accordingly, equations
(\ref{07kt}) can be solved with respect to three unknown coefficients:
\begin{equation}
a_i(\{\bar{x}\})=\Delta_i(\{\bar{x}\})\Delta^{-1}(\{\bar{x}\}),\qquad i=\overline{1,3},
\label{07wtb}
\end{equation}
where the determinant $\Delta(\{\bar{x}\})$ has the form:
\begin{eqnarray}
\Delta(\{\bar{x}\})=\left|\begin{array}{ccc}
K_1 &2\xi^1\xi^2 & 2\xi^1\xi^3\\
2\xi^1\xi^2& K_2 & 2\xi^2\xi^3 \\
2\xi^1\xi^3 & 2\xi^2\xi^3 & K_3
\end{array}
\right|,\qquad
\begin{array}{ccc}
K_1(\{\bar{x}\})=(\xi^1)^2-(\xi^{2})^2-(\xi^{3})^2-
\Lambda^2(\{\bar{x}\}),\\
K_2(\{\bar{x}\})=(\xi^2)^2-(\xi^3)^2-(\xi^1)^2-
\Lambda^2(\{\bar{x}\}),\\
K_3(\{\bar{x}\})=(\xi^{3})^2-(\xi^{1})^2-(\xi^{2})^2-
\Lambda^2(\{\bar{x}\}).
\end{array}
\label{07wqtk}
\end{eqnarray}
As for the determinant $\Delta_i(\{\bar{x}\})$, they can be found from the third-order determinant (\ref{07wqtk}),
 replacing the elements of the $i$-th column with zeros. In other words; $\Delta_1(\{\bar{x}\})=\Delta_2(\{\bar{x}\})
=\Delta_3(\{\bar{x}\})=0$, and, respectively, the system of equations (\ref{07kt}) will have a non-trivial solution if
the  determinant of the system (\ref{07kt})  is equal to zero too,  i.e. $\Delta(\{\bar{x}\})=0$.
More precisely, the system of equations (\ref{07kt}) will have solutions if in expressions (\ref{07wtb}),
uncertainties of the type $0/0$ can be eliminated. As the study shows, there always
exists  a non-empty  continuous set $\Xi\neq\oslash$, on which the determinant of algebraic equations (\ref {07kt})
is equal to zero and, accordingly, the above uncertainty is eliminated  (see Appendix {\bf E} for details).

\emph{ {\bf{ Proposition 3}} is proved.}

\section{D\lowercase{eviation of geodesic trajectories of one family}}
Studying the linear deviations of the geodesic trajectories of one family, one
can get valuable information about the properties of a dynamical system and, very
importantly, about the relationship between the behavior of a dynamical system and
the geometric features of a Riemannian space.

\textbf{Definition 6.}  \emph{Let $x^i=x^i(s,\eta)$ be the equation of a one-parameter
family of geodesics on the Riemannian manifold $\mathcal{M}^{(3)}_{\bar{J}}$, where $s$
is an affine parameter along geodesic the trajectory, whereas  the symbol
$\eta$ denotes the family parameter. The vector $\mathbf{j}(\{\zeta\})$
in the direction of the normal of the geodesic $\mathbf{l}(\{\bar{x}\})$ with components:
\begin{equation}
\frac{\delta x^i(s,\eta)}{\delta \eta}=\zeta^i(s,\eta),\qquad \{\zeta\}=(\zeta^1,\zeta^2,\zeta^3),
\quad i=\overline{1,3},
\label{9a}
\end{equation}
will be called the linear deviation of close geodesics.}

The components of the deviation vector $\mathbf{j}(\{\zeta\})$ satisfy the following
equations \cite{BubrNovFom}:
\begin{equation}
\frac{\mathcal{D}^2\zeta^i}{\mathcal{D}s^2}=-\mathcal{R}^i_{jkl}(\{\bar{x}\})x^j\zeta^k x^l,
\qquad i,j,k,l=\overline{1,3},
\label{9b}
\end{equation}
where $\mathcal{R}^i_{jkl}(\{\bar{x}\})$ is the Riemann tensor, which has the form:
\begin{equation}
\mathcal{R}^i_{jkl}=\Gamma^i_{lj,\,k}- \Gamma^i_{jk,\,l}+
\Gamma^i_{k\lambda} \Gamma^\lambda_{lj} -\Gamma^i_{l\lambda}
\Gamma^\lambda_{jk},\qquad \Gamma^i_{jk,\,l}(\{\bar{x}\})=\partial\Gamma^i_{jk}(\{\bar{x}\})/\partial x^l.
\label{9ab}
\end{equation}
The equation (\ref{9b}) can be written in the form of an ordinary second-order
differential equation:
\begin{equation}
\ddot{\zeta}^i+2\Gamma^i_{j\,l}\dot{x}^j\dot{\zeta}^l+\bigl(\dot{\Gamma}^i_{j\,l}\dot{x}^j-
\Gamma^i_{j\,l}\Gamma^j_{k\,p}\dot{x}^k\dot{x}^p +\Gamma^i_{j\,n}\Gamma^n_{k\,p}\,
\dot{x}^j\dot{x}^k\delta_l^p\bigr)\zeta^l
=-\mathcal{R}^i_{jkl}\,x^j\zeta^k x^l,
\label{9abc}
\end{equation}
The explicit form of specific terms of the equation (\ref{9abc}) can be found in the appendix F.
Solving equation (\ref{9abc}) together with the equations systems (\ref{08}) and (\ref{22}), we
can get a full view on deviation properties of close geodesic trajectories of a one-parameter
family, which is a very important characteristic of a dynamical system.

\section{T\lowercase{hree-body system in a random environment}}

Let us suppose that a three-body system is subject to external influences that have
regular and random components. The causes of such impacts can be different. For
example, when a system of bodies is immersed in the environment - gas, liquid, etc.
In this case, the total energy of the system of bodies changes due to random collisions.
Given the new conditions, the three-body problem can be mathematically generalized if
to assume that in the system of equations (\ref{07a}) the metric tensor $g_{ij}(\{\bar{x}\})$
is random.

When studying atomic-molecular processes even in a vacuum, it is often important to
take into account the influence of quantum fluctuations on the classical dynamics
of interacting bodies.

In the simplest case, when an external random force acts on  the dynamical system
without deformation of the metric tensor $g_{ij}(\{\bar{x}\})$, using the system
of equations (\ref{07a}), we can write the following system of stochastic differential
equations (SDE) to describe the motion of three bodies:
\begin{eqnarray}
\dot{\chi}^\mu=A^\mu(\{\chi\})+ \eta^\mu(s),\qquad
\mu=\overline{1,6},
\label{24at}
\end{eqnarray}
where the independent variables $\{\chi\}=\bigl(\{\bar{x}\},\{\bar{\xi}\}\bigr)=\overline{\chi^1,\chi^6}$
form the Euclidean $6D$ space, in addition, the following notations are made:
$$\chi^1={\xi}^1,\qquad\chi^2={\xi}^2,\qquad\chi^3={\xi}^3,\qquad
\chi^4=x^1, \qquad\chi^5={x}^2,\qquad\chi^6={x}^3.
$$
In addition, in (\ref{24at}), the coefficients $A^\mu(\{\chi\})$ are defined by the expressions:
$$
A^1\bigl(\{\chi\}\bigr)=a_1\bigl\{(\xi^1)^2-(\xi^2)^2-(\xi^3)^2-
 \Lambda^2\bigr\}+2\xi^1( a_2\xi^2+ a_3\xi^3),\qquad A^4\bigl(\{\chi\}\bigr)=\xi^1,
$$
$$
A^2\bigl(\{\chi\}\bigr)=a_2\bigl\{(\xi^2)^2-(\xi^1)^2-(\xi^3)^2
- \Lambda^2\bigr\}+2\xi^2(a_3\xi^3+a_1\xi^1), \qquad A^5\bigl(\{\chi\}\bigr)=\xi^2,
$$
$$
A^3\bigl(\{\chi\}\bigr)=a_3\bigl\{(\xi^3)^2-(\xi^2)^2-(\xi^1)^2
- \Lambda^2\bigr\}+2\xi^3(a_1\xi^1+a_2\xi^2), \qquad A^6\bigl(\{\chi\}\bigr)=\xi^3.
$$
Recall that $A^\mu(\{\chi\})$ are regular functions.

For simplicity, we assume that the stochastic functions $\eta^\mu(s)$ satisfy the
correlation relations of \emph{white noise}:
\begin{equation}
\langle\eta^\mu(s)\rangle=0, \qquad\langle\eta^\mu(s)\eta^\mu(s')\rangle=2\epsilon \delta(s-s'),
\label{24abt}
\end{equation}
where $\epsilon$ denotes the power of random fluctuations and $\delta(s-s')$
is the Dirac delta function.

Now we can move on to the problem of deriving the equation of \emph{joint probability
density} (JPD) for the independent variables $\{\chi\}$.

For further analytical study of the problem, it is convenient to present JPD in the form:
\begin{eqnarray}
P\bigl(\{\chi\},s\bigr)=\prod_{\mu=1}^6\bigl\langle\delta\bigl[\chi^\mu(s)-\chi^\mu\bigr]\bigr\rangle.
\label{24bt}
\end{eqnarray}
Using a well-known technique (see \cite{Kljat,Lif}), we can differentiate the expression
(\ref{24bt}) by \emph{internal time}  $``s"$ and taking into account (\ref{24at}) and
(\ref{24abt}) get the following second-order \emph{partial differential equation} (PDF):
\begin{eqnarray}
\frac{\partial P}{\partial s}=\sum_{\mu=1}^6\frac{\partial}{\partial
\chi^\mu}\Bigl[A^\mu\bigl(\{\chi\}\bigr) +\epsilon
 \frac{\partial}{\partial \chi^\mu }\Bigr]P.
\label{24abct}
\end{eqnarray}
It is easy to see the function (\ref{24abct}) determines the probability of the
position and momentum of \emph{ imaginary point} characterizing the three-body
system in the $6D$ phase space. In the case when $\epsilon=\hbar$, the function
$P\bigl(\{\chi\},s\bigr)$ in principle play the same role as the Wigner quasi-probability
distribution \cite{Wig,Weil}. However, unlike the Wigner function, which in some
regions of the phase space can take negative values, and therefore is not a probability
distribution, the solution of the equation (\ref{24abct}) is positive definite in the
entire phase space. In other words, the function $P\bigl(\{\chi\},s\bigr)$ really has
the meaning of a probability distribution, which describes the probabilistic evolution
of the classical three-body system in phase space  taking into account the influence
of quantum fluctuations.

Developing the same ideology, we can obtain the equation of probability distribution
of an elementary process in momentum and coordinate representations, taking into
account the influence of the environment.

In particular, for the probability current in the momentum representation
$P^{(m)}_{\{\bar{x}\}}\bigl(\{\bar{\xi}\}\bigr)$, at the point
$\{\bar{x}\}\ni \mathbb{E}^3$ we obtain the following second-order PDF:
\begin{eqnarray}
\dot{P}^{(m)}_{\{\bar{x}\}}=\sum_{i=1}^3\frac{\partial}{\partial
\xi^i}\Bigl[A^i\bigl(\{\bar{x}\},\{\bar{\xi}\}\bigr) +\epsilon
 \frac{\partial}{\partial \xi^i}\Bigr]P^{(m)}_{\{\bar{x}\}},\qquad
\dot{P}^{(m)}_{\{\bar{x}\}}= \partial{P}^{(m)}_{\{\bar{x}\}}/\partial{s}.
\label{t26a}
\end{eqnarray}
In other words, by calculating equation (\ref{t26a}) at a given point $\{\bar{x}\}$, we
can find the distribution of the velocity (momentum) $\{\bar{\xi}\}$ of the
\emph{imaginary point} depending on the \emph{internal time} $``s"$. We can also trace
the evolution of the momentum distribution along the trajectory by substituting
$\{\bar{x}\}\to\{\bar{x}(s)\}$ in the equation (\ref{t26a}). Note that in this
case the equation (\ref{t26a}) is solved in combination with the system of
equations (\ref{07a}).

Now we consider the case when the metric of the \emph{internal space} $\mathbb{E}^3$
depending on the \emph{internal time} $``s"$ is continuous, however its first derivative
is already a random function. The above task will be mathematically equivalent to
random mappings of the type:
$$
R_f:a_i\bigl(\{\bar{x}\}\bigr)\mapsto\tilde{a}_i\bigl(s,\{\bar{x}\}\bigr)=\frac{d}{dx^i}
\ln g\bigl(s,\{\bar{x}\}\bigr), \qquad i=\overline{1,3},
$$
or more detail:
\begin{eqnarray}
 \tilde{a}_i\bigl(s,\{\bar{x}\}\bigr)=\frac{\partial\ln \tilde{g}\bigl(s,\{\bar{x}\}\bigr)}{\partial x^i}+
 \frac{\partial s}{\partial x^i}\frac{\partial\ln \tilde{g}\bigl(s,\{\bar{x}\}\bigr)}{\partial s}
 = \tilde{a}_i(s,\{\bar{x}\}\bigr) +\frac{\dot{\tilde{g}}(s,\{\bar{x}\}\bigr)}{\sqrt{\tilde{g}(s,\{\bar{x}\}\bigr)}},
 \label{a25}
 \end{eqnarray}
where $\tilde{a}_i\bigl(s,\{\bar{x}\}\bigr)$  are regular functions, $R_f$ denotes the operator of
random mappings and $\tilde{\eta}\bigl(s,\{\bar{x}\}\bigr) =\dot{\tilde{g}}/\sqrt{\tilde{g}}$
is a random function, which will be defined below. Taking into account the above, the system
of equations (\ref{07a}) can be decomposed and presented in the form of stochastic Langevin
type equations:
\begin{equation}
\dot{\xi}^\mu=A^\mu\bigl(\{\chi\}\bigr)+B^{\mu}\bigl(\{\chi\}\bigr)
\eta\bigl(s,\{\bar{x}\}\bigr), \qquad \mu=\overline{1,6},
\label{25}
\end{equation}
where
$$
 B^{1}\bigl(\{\chi\}\bigr)=\bigl(\xi^1\bigr)^2-\bigl(\xi^2\bigr)^2-\bigl(\xi^3\bigr)^2
 +2\xi^1\bigl(\xi^2+\xi^3\bigr)-{\Lambda}^2(\{\bar{x}\}),\qquad B^{4}\bigl(\{\chi\}\bigr)=0,
$$
$$
B^{2}\bigl(\{\chi\}\bigr)=\bigl(\xi^2\bigr)^2-\bigl(\xi^1\bigr)^2-\bigl(\xi^3\bigr)^2+
2\xi^2\bigl(\xi^1+\xi^3\bigr)-{\Lambda}^2(\{\bar{x}\}),\qquad B^{5}\bigl(\{\chi\}\bigr)=0,
$$
$$
 B^{3}\bigl(\{\chi\}\bigr)=\bigl(\xi^3\bigr)^2-\bigl(\xi^2\bigr)^2-\bigl(\xi^1\bigr)^2+
 2\xi^3\bigl(\xi^1+\xi^2\bigr)- {\Lambda}^2(\{\bar{x}\}),\qquad B^{6}\bigl(\{\chi\}\bigr)=0.
$$
The JPD for the independent variables $\{\chi\}$ again can be represented in the form (\ref{24bt}).
For simplicity we will assume that a random generator $\tilde{\eta}\bigl(s,\{\bar{x}\}\bigr)=
\eta\bigl(s\bigr)/\sqrt{g}$ and, in addition, that it  satisfy the correlation properties of
the \emph{white noise} with fluctuation power $\epsilon$ (see (\ref{24abt})).
Further, performing calculations similar to  (\ref{24bt})-(\ref{24abct}) using the SDE
(\ref{25}), we get the following second-order PDE for JPD:
\begin{eqnarray}
\frac{\partial P}{\partial s}=\sum_{\mu=1}^6\frac{\partial }{\partial \xi^\mu}\bigl(A^\mu P\bigr)+
\epsilon g^{-1/2}\sum_{i,j=1}^3 \frac{\partial }{\partial \xi^i}\Bigl[B^{i}
\frac{\partial }{\partial \xi^j}\bigl(B^{j}P\bigr)\Bigr].
\label{25b}
\end{eqnarray}
Finally, for the probabilistic current in the momentum representation at the given point
$\{\bar{x}\}\in \mathbb{E}^3$ we get the following second-order PDF:
\begin{eqnarray}
\dot{P}^{(m)}_{\{\bar{x}\}}=\sum_{i=1}^3\frac{\partial }{\partial \xi^i}\bigl(A^i P\bigr)+
\epsilon g^{-1/2}\sum_{i,j=1}^3 \frac{\partial }{\partial \xi^i}\Bigl[B^{i}
\frac{\partial }{\partial \xi^j}\bigl(B^{j}{P}^{(m)}_{\{\bar{x}\}}\bigr)\Bigr].
\label{25bk}
\end{eqnarray}
Substituting $\{\bar{x}\}\to\{\bar{x}(s)\}$ into the equation (\ref{25bk}), we can
study the evolution of the momentum distribution along the trajectory of a dynamical system.

Thus, we have obtained equations describing geodesic flows in the phase space (\ref{24abct})
and (\ref{25b}), as well as in the momentum space (\ref{t26a}) and (\ref{25bk}), which must
be solved in combination with a system of differential equations of the first order (\ref{07a}).
Recall that the method used to obtain the noted equations can be attributed to Nelson's type
stochastic quantization \cite{Nelson}, with the only difference being that  \emph{internal time}
$``s"$ cardinally changes the sense of the developed approach. In particular, in the limit
$\epsilon \to0 $, the representation allows a continuous transition from the statistical
(see (\ref{24abct}) and (\ref{25b})) to the dynamical description (see (\ref{07a})) of the problem.

\section{A\lowercase{\,new criterion for estimating chaos in classical systems}}

When the three-body system is in an environment that has both regular and random influences on
it, then it makes sense to talk about a statistical system. In this case, the main task is to
construct the mathematical expectations of different elementary atomic-molecular
processes occurring during multichannel scattering (see  Sch. 1). Recall that the evolution
equations (\ref{24abct}) and (\ref{25b}), describing of geodesic flows depending on
 \emph{internal time} $``s"$ have an important feature. The latter circumstance makes
 it necessary to introduce new criteria for determining the measure of deviation of
 probabilistic current tubes of various elementary processes.

In particular, following the definition of \emph{Kullback-Leibler} definition of the distance  between
two continuous distributions, we can determine the criterion characterizing the deviation
 between the corresponding   tubes of probabilistic currents    \cite{Kul}.

\textbf{Definition 7.} \emph{The deviation between two different tubes of probabilistic
currents in the phase space will be defined by the expression:
\begin{equation}
d (s_a,s_b)=\int_{\mathcal{P}^6}P\bigl(\{\chi\},s_a\bigr)\ln\biggl |\frac{P\bigl(\{\chi\},s_a\bigr)}
{P\bigl(\{\chi\},s_b\bigr)}\biggr |\sqrt{g(\{\bar{x}\})}\prod_{\nu=1}^6 d\chi^\nu,
\label{26a}
\end{equation}
where $P_a\equiv P\bigl(\{\chi\},s_a\bigr)$ and $P_b\equiv P\bigl(\{\chi\},s_b\bigr)$  are two different
probabilistic currents, which at the beginning of development of elementary
processes are closely located or have an intersection.}

In the case when the distance between two flows depending on \emph{internal times} $ s\sim  s_a\sim s_b $
grows linearly, that is:
$$d(s) \sim ks,\qquad k=const>0,$$
there is reason to believe that a dynamical system exhibits chaotic behavior, i.e. it is chaotic.

\textbf{Definition 8.} \emph{Let $P_{if}(s_n)$ be the transition probability between the $(in)$
and $(out)$ asymptotic channels  with the \emph{internal time} $s_n$,  then the total mathematical
expectation of the transition between two asymptotic states $P_{ab}^{tot}$ will be defined as:
\begin{equation}
P_{if}^{tot}=\lim_{N\to\infty}\biggl[\frac{1}{N}\sum_{n=1}^N\Bigl(\lim_{s_n\to\,\infty}
P_{if}(s_n)\Bigr)\biggr],
\label{26b}
\end{equation}
where $N$ denotes the number of various solutions of the Cauchy problem for the  system (\ref{07a}).}

\section{T\lowercase{he quantum three-body problem on conformal-}E\lowercase{uclidean manifold}}
If the classical three-body problem plays a fundamental role for understanding the dynamics
of complex classical systems, then a similar problem in quantum mechanics is the key to
studying the atomic and subatomic nature of matter. In this regard, it is obvious that a
mathematically rigorous description of the system of interacting atoms is a task of primary
importance. Note that the first work on this problem was carried out by Skorniakov and
Ter-Martirosian \cite{Ter}. Recall that they derived equations for determining the wave
function of a system of three identical particles in the limiting case of zero-range forces.
The approach was generalized by Faddeev for arbitrary particles and the finite-range
forces \cite{Faddeev}. Scattering in three-particle atomic-molecular systems is
characterized by both two-particle and three-particle interactions, which makes the
Faddeev approach inaccurate for describing such processes. In this regard, subsequently,
various approaches and corresponding algorithms were developed for studying atomic-molecular
processes in the framework of the three-body scattering problem (see for example
\cite{Kosloff,Balint-Kurti}). However, on the way to the description of quantum
multichannel scattering, in our opinion, a new fundamental ideological problem arose related
to the paper of Hanney and Berry  \cite{Hannay} (see also \cite{Schuster}). Namely, as the
authors proved in this paper,  in the limit $\hbar\to 0$ there is no transition from the $Q$
system (\emph{quantum systems}) to the $P$-system (\emph{Poincar\'{e}  systems})
(see FIG. 4 ).

To solve the open problem of \emph{quantum-classical correspondence}, the three-body problem
is an ideal model, since this system very often exhibits strongly developed chaotic behavior
in the classical limit.
Recall that by \emph{strongly developed chaos}  we imply a such state of the classical system, when
the chaotic region in the $2n$ -dimensional phase space occupies a larger volume than the
volume of the \emph{quantum cell} - ${\hbar}^{n}$. Obviously, in this case the so-called
\emph{quantum  suppression of chaos}  does not occur, and we must observe chaos in the behavior
of the wave function itself.

Using the reduced classical Hamiltonian (\ref{09}), we can write the following non-stationary
quantum  for the three-body system in conformal-Euclidean space (\emph{internal space})
$\mathcal{M}^{(3)}$:
\begin{equation}
i\hbar\frac{\partial\Psi}{\partial s}= \hat{\mathcal{H}}\bigl(\{\bar{x}\};\{\bar{p}\}\bigr)\Psi,
\label{c27a}
\end{equation}
where $\hat{\mathcal{H}}$ is the Hamiltonian of the quantum problem.

By making the following substitutions in the reduced classical Hamiltonian (\ref{09}):
 $$\dot{x}^i\to -i\hbar\partial/\partial x^i\quad and\quad  J^2\to J(J+1),$$
which is equivalent to the transition to the quantum Hamiltonian  (see \cite{Zom}), we get:
\begin{equation}
 \hat{\mathcal{H}}\bigl(\{\bar{x}\};\{\bar{p}\}\bigr)=\frac{1}{2\mu_0}
\biggl\{-\hbar^2 g(\{\bar{x}\})\sum_{i=1}^3\frac{\partial^2}{(\partial x^i)^2}
+\frac{J(J+1)}{g(\{\bar{x}\})}\biggr\}.
\label{c2t7a}
\end{equation}
 In the case when the energy of the three-body system is fixed, that is,
$\mathrm{E} = const$, we can go to the stationary equation for the wave
function.
\begin{figure}
\includegraphics[width=60mm]{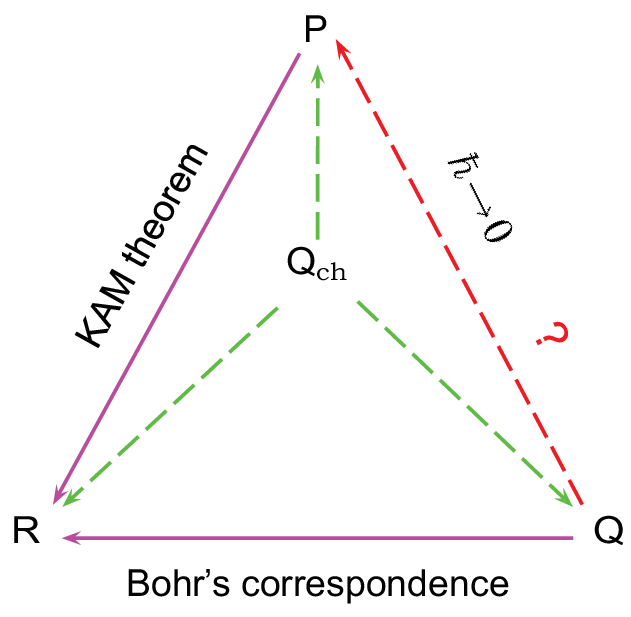}
\caption{\emph{The figure shows a diagram of the interconnections between the three
 well-known regions of matter motion ${\mathbf R}, {\mathbf P}, {\mathbf Q}$ and the
new region ${\mathbf Q}_{ch}$, which is strictly defined in this paper.
Recall that ${\mathbf R}$ denotes  classical regular systems (Newton systems),
${\mathbf P}$  denotes classical dynamically or chaotic systems (Poincar\'{e} systems),
${\mathbf Q}$  denotes regular quantum systems and ${\mathbf Q}_{ch}$ - quantum chaotic
systems. There is a possibility of passing from the  ${\mathbf P}$ system to the ${\mathbf R}$
system, which is ensured by the KAM-theorem \emph{ \cite{Poschel}}. From the system ${\mathbf Q}$, a
transition to the system $ {\mathbf R}$ is possible, but not to the system $ {\mathbf P}$,
while from the system  ${\mathbf Q}_{ch}$ there is the possibility of transition to all three
 ${\mathbf R}, {\mathbf P}$ and ${\mathbf Q}$  systems.}}
\label{fig. 4}
\end{figure}

In particular, substituting the wave function:
$$
\Psi\bigl(\{\bar{x}\},s\bigr)=\exp\bigl(-i\mathrm{E}s/{\hbar}\bigr)\bar{\Psi}\bigl(\{\bar{x}\}\bigr),
$$
 into the equation (\ref{c27a}) - (\ref{c2t7a}), we obtain the following stationary equation:
\begin{equation}
\Biggl\{\sum_{i=1}^3\frac{\partial^2}{(\partial x^i)^2}+\frac{2\mu_0}{\hbar^2 g(\{\bar{x}\})}
\biggl[\mathrm{E}-\frac{J(J+1)}{ g(\{\bar{x}\})}\biggr]\Biggr\}\bar{\Psi}(\{\bar{x}\})=0.
\label{c27b}
\end{equation}
Recall that $J^2=\sum_{i=1}^3J^2_i=const$ is the total angular momentum of the system
of bodies, which in this case is \emph{quantized}.

For any fixed $\bf J$, there is a countable number of submanifolds:
$$
\mathcal{M}^{(3)}_{\{\bf J\}}=\bigl\{\mathcal{M}^{(3)}_{\{\alpha\}}\bigr\}_{\alpha\in\mathcal{B}_{\bf J}},
$$
 on which various quantum processes flow, where $\mathcal{B}_{\bf J}$  is the family
 of sets with different projections of   $J_z$.  Recall that
these submanifolds differ by its orientations in the $6D$ manifold (space)
$\mathcal{M}$, which we can determine with two commutated quantum numbers
${\{\bf J\}}=(J,J_z)$. In other words, in the developed approach when quantizing
 a dynamical problem, a typical example of which is the three-body problem,
\emph{geometry is also quantized}.

In particular, when ${\bf J}=0$  there is only one submanifold $\mathcal{M}^{(3)}_{\{\bf{0}\}}$,
where ${\{\bf{0}\}}=(0,0)$. In the case when ${\bf J}=\bf 1$, there exists a family of three
oriented submanifolds, on each of which the Schr\"{o}dinger equation is invariant:
$$
\mathcal{M}^{(3)}_{\{\bf{1}\}}=\bigl\{\mathcal{M}^{(3)}_{\{\alpha\}}\bigr\}_{\alpha\in\mathcal{B}_{\bf 1}}, \qquad
\mathcal{B}_{\bf 1}=\bigl\{(1,+1),\,(1,0),\,(1,-1)\bigr\},
$$

We can combine submanifolds of a family with a given full rotational momentum
 $\bf J$, as  is done in the case of a family of sets:
$$
\mathcal{M}^{(3)}_{\bm1}=\bigcup_{\alpha\in\mathcal{B}_{\bm1}}\mathcal{M}^{(3)}_{\{\alpha\}}
=\bigl\{\{\bar{x}\}|\,\exists\,
\alpha\in\mathcal{B}_{\bm1},\,\{\bar{x}\}\in \mathcal{M}^{(3)}_{\alpha}\bigr\}.
$$

For further analytical constructions of the problem, it is useful to introduce a new coordinate
systems on the cards $G_\alpha$, arising at continuously mappings $f:\{\bar{\rho}\}\mapsto\{\bar{x}\}$.\\
We will consider two important cases:

 a. When three bodies form a bound state, i.e. $g(\{\bar{x}\})\leq0$,  and, accordingly,

 b. when scattering in a system occurs with a rearrangement of bodies, for example;
 $1+(23)\to (12)+3$ (see Sch. 1). Recall that in this case the scattering
 processes in the system occur under the condition  $g(\{\bar{x}\})>0.$

\subsection{The three-body coupled states}

First, consider the case  a., when three bodies form a bound state.  For this case,
it is convenient to use a  \emph{local spherical coordinate system} (LSCS) (see FIG. 5):
$$
 \{\bar{\mathrm{r}} \}=
\bigl(\mathrm{r},\theta,\varphi\bigr),\qquad
\mathrm{r}^2 =\bigl(x^1\bigr)^2+\bigl(x^2\bigr)^2+\bigl(x^3\bigr)^2, \qquad
\theta\in\bigl[0,\pi\bigr], \qquad   \varphi\in\bigl[0,2\pi\bigr].
 $$
 Note that this is firstly
due to the fact that, in a geometric sense,  bound states are localized on
2$D$ closed surfaces that are homeomorphic with isolated spheres
 having topological features (Appendix {\bf D}, family  ${\breve{\mathcal{ A}}}$
see FIG. 6 ).
 \begin{figure}
\includegraphics[width=85mm]{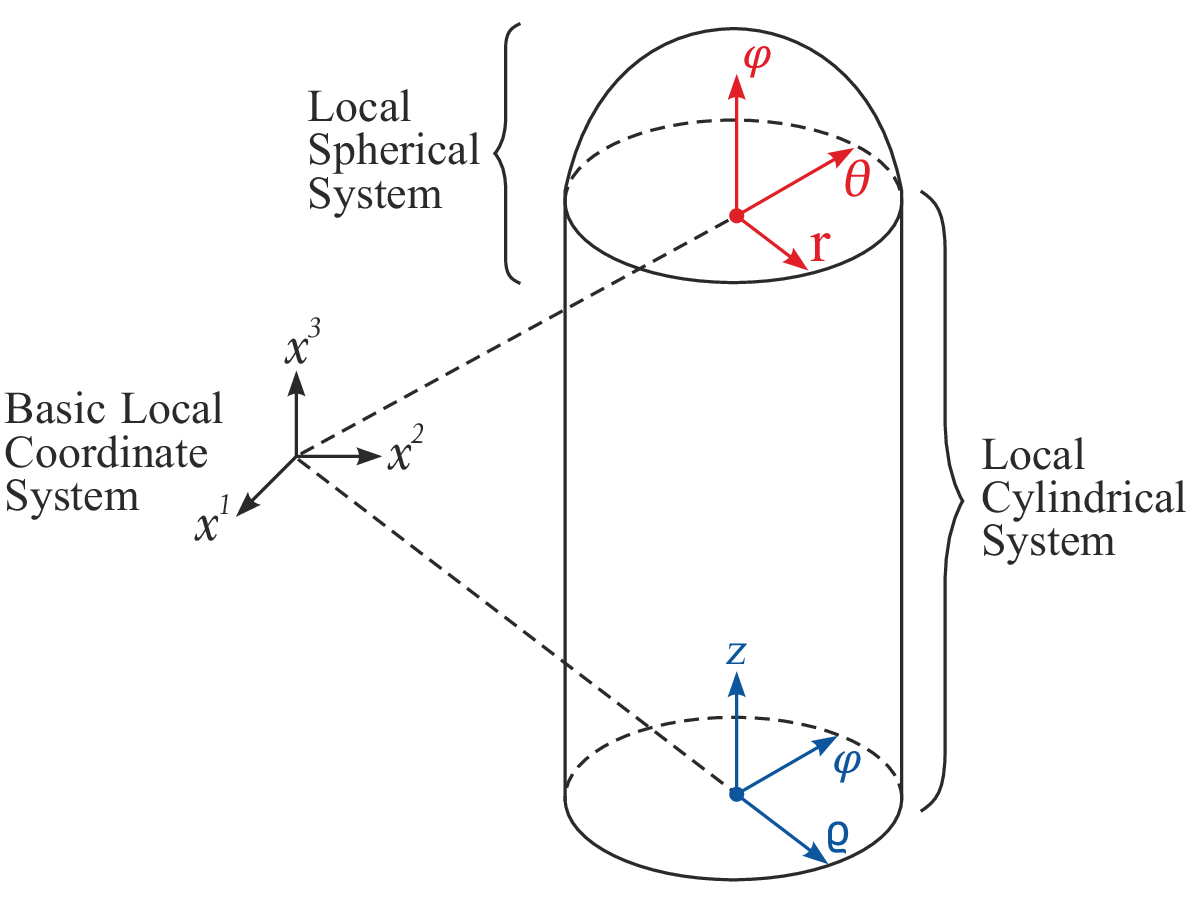}
\caption{\emph{When constructing the representation on the atlas card, a
rectangular local coordinate system (we call the basic local coordinate system)
$\{\bar{x}\}=(x^1,x^2,x^3)$ is determined. However, for further studies of the
quantum problem it is useful to use the local spherical coordinate system
$\{\bar{\mathrm{r}}\}=(\mathrm{r},\theta,\varphi)$
to describe the bound quantum state, and the local cylindrical coordinate system
$\{\bar{\mathrm{\varrho}}\}=(\mathrm{\varrho},z,\varphi)$, respectively, to
describe multichannel quantum scattering. }}
\label{fig. 5}
\end{figure}

Within the framework of LSCS, the equation (\ref{c27b}) can be written as:
\begin{equation}
\Biggl\{\Delta
+\frac{2\mu_0}{\hbar^2 g_{\varepsilon}\bigl(\{\bar{\mathrm{r}}\}\bigr)}
\biggl[\mathrm{E}-\frac{J(J+1)}{ g_{\varepsilon}\bigl(\{\bar{\mathrm{r}} \}
\bigr)}\biggr]\Biggr\}\bar{\Psi} =0.
\label{c2k7b}
\end{equation}
where $\Delta$ denotes Laplace operator in the LSCS, in addition, $f:g(\{\bar{x}\})\mapsto
g_\varepsilon(\{\bar{\mathrm{r}}\})$:
$$
\Delta=\frac{1}{\mathrm{r}^2} \frac{\partial}{\partial\mathrm{r}}
\biggl(\mathrm{r}^2 \frac{\partial}{\partial\mathrm{r}}\biggr)
+\frac{1}{\mathrm{r}^2 \sin\theta}\frac{\partial}{\partial\theta}
\biggl(\sin\theta\frac{\partial}{\partial\theta}\biggr)
+\frac{1}{\mathrm{r}^2 \sin^2\theta }\frac{\partial^2}{\partial\varphi^2 }.
$$
Recall that the function $g_{\varepsilon}\bigl(\{\bar{\mathrm{r}}\}\bigr)$
is obtained from $g(\{\bar{x}\};\varepsilon)=\bigl[\mathrm{E}+i\varepsilon
-U(\{\bar{x}\})\bigr]U^{-1}_0\neq0$, where $\varepsilon\ll1$ (see (\ref{03ab})),
after transition into the LSCS. Note that the small parameter $\varepsilon$ has a  physical meaning,
 namely, it characterizes the width of the energy level of the quantum state. Since the Laplace
spherical harmonics $Y_l^m(\theta,\varphi)$ form an orthonormal basis of the
 Hilbert space of quadratically integrable functions  \cite{Arfken}, we can use this property and
 write equation (\ref{c2k7b}) in the form:
\begin{equation}
\Biggl\{\Delta
+\frac{2\mu_0}{\hbar^2}\sum_{\bar{l}=0}^{\infty}\sum_{\bar{m}=-\bar{l}}^{\bar{l}}
\Omega_{\bar{l}\bar{m}}({\mathrm{r}} ;\, \mathrm{E}, J,\varepsilon)
Y_{\bar{l}}^{\bar{m}}(\theta,\varphi)\Biggr\}\bar{\Psi} =0,
\label{c0l7b}
\end{equation}
where
$$
\Omega_{\bar{l}\bar{m}}({\mathrm{r}} ;\, \mathrm{E}, J,\varepsilon)=
\bigl[\mathrm{E}\,\mathfrak{g}^{(1)}_{\bar{l}\bar{m}}(\mathrm{r};\varepsilon)
- J(J+1) {g}^{(2)}_{\bar{l}\bar{m}}(\mathrm{r};\varepsilon)\bigr],
$$
$$
{g^{-k}_{\varepsilon}\bigl(\{\bar{\mathrm{r}} \}\bigr)}=
\sum_{\bar{l}=0}^{\infty}\sum_{\bar{m}=-\bar{l}}^{\bar{l}}
\mathfrak{g}^{(k)}_{\bar{l}\bar{m}}(\mathrm{r} ;\,\varepsilon)
\,Y_{\bar{l}}^{\bar{m}}(\theta,\varphi),
\qquad
k=1,2.
$$
It is easy to find the functions $\mathfrak{g}^{(1)}_{\bar{l}\bar{m}}(\mathrm{r};\,\varepsilon)$
and $\mathfrak{g}^{(2)}_{\bar{l}\bar{m}}(\mathrm{r};\,\varepsilon)$. For this
we need to multiply the corresponding expressions for the functions
$g^{-1}_{\varepsilon}\bigl(\{\bar{\mathrm{r}}\}\bigr)$ and
$g^{-2}_{\varepsilon}\bigl(\{\bar{\mathrm{r}}\}\bigr)$
on the complex conjugation of a spherical function
$Y_{\bar{l}'}^{\bar{m}'\ast}(\theta,\varphi)$, and then to integrate  over the
 sphere of unit  radius:
$$
\mathfrak{g}^{(k)}_{\bar{l}\bar{m}}(\mathrm{r};\,\varepsilon)=
\int_0^{2\pi}\int_0^{\pi}g^{-k}_{\varepsilon}\bigl(\{\bar{\mathrm{r}}\}\bigr)
Y_{\bar{l}}^{\bar{m}\ast}(\theta,\varphi)\sin\theta{d\theta d\varphi},
\qquad k=1,2.
$$

We can consider the problem of finding solutions in the form:
\begin{equation}
\bar{\Psi}\bigl(\mathrm{r},\theta,\varphi;\,\varepsilon\bigr)=
\Upsilon(\mathrm{r};\varepsilon)Y_{l}^{m}(\theta,\varphi).
\label{c2lzb}
\end{equation}
where  $\Upsilon(\mathrm{r};\varepsilon)$ describes a radial wave function.

Substituting (\ref{c2lzb}) into the equation (\ref{c0l7b}) and performing
simple calculations, we can find the following \emph{ordinary differential
equation} (ODE) (see  Appendix  {\bf G}):
 \begin{eqnarray}
 \biggl\{\frac{1}{{\mathrm{r}}^2}\frac{\mathrm{d}}{\mathrm{d}\mathrm{r}}
\biggl({\mathrm{r}}^2\frac{\mathrm{d}}{\mathrm{d}{{\mathrm{r}}}}\biggr)
- \frac{l(l+1)}{{\mathrm{r}}^2}  +
  \frac{2\mu_0}{\hbar^2}\sum_{\bar{l}=0}^{2l}\sum_{\bar{m}=\,0}^{\bar{l}}
\overline{\mathcal{W}}_{m,\bar{m}\,;\,l,\bar{l}}\,
 \Omega_{\bar{l}\bar{m}}\bigl({\mathrm{r}};\,\mathrm{E},J,\varepsilon\bigr)\biggr\}\Upsilon=0,
 \label{c2lb}
\end{eqnarray}
where  $l=0,1,2,...$  is the quantum number of angular momentum in the \emph{internal
space} $\mathcal{M}^{(3)}$, in addition:
\begin{eqnarray}
\overline{\mathcal{W}}_{m,\bar{m};\,l,\bar{l}}=\sqrt{\frac{2\bar{l}+1}{\pi}}
\biggl(l+\frac{1}{2}\biggr)\biggl(\begin{array}{ccc}
l & l&\bar{l} \\
0 & 0 & 0
\end{array}\biggr)
\biggl(\begin{array}{ccc}
l & l&\bar{l} \\
-|m|&-|m|&|\bar{m}|
\end{array}\biggr).
 \label{c2lt7b}
\end{eqnarray}
Thus, we have obtained a one-dimensional equation for the radial wave function
of the coupled three-body system. It is easy to see that this equation is a bit like
 a hydrogen-like atom and can be quantized for certain energy values.
If we solve this equation taking into account the system of algebraic equations
(\ref{22}) and coordinate transformations (\ref{24}), then we obtain  the
full wave function of the system of bodies as; in global $\{\bar{\rho}\}$ (see (\ref{02})),
as well as in local $\{\bar{x}\}$ coordinate systems.

\subsection{Quantum multichannel  scattering in a three-body system}
In this section, we will consider the case b., i.e. quantum scattering with
particles rearrangement  (see Sch. 1). Recall that all coupled pairs in this scheme
are described by two quantum numbers $n$- (\emph{vibrational quantum number}), $j$-
(\emph{rotational quantum number}) and $K$- ($z$-\emph{projection of the total
angular momentum ${\bf J}$ in space-fixed coordinate system}). The regrouping process, obviously,
will occur through manifolds of the family $ \breve{\mathcal{C}}$  (see FIG. 10), which
have cylindrical symmetry. This fact dictates us to use  \emph{local cylindrical
coordinates} (LCC) (see FIG. 5):
\begin{equation}
 \{\bar{\mathrm{\varrho}}\}=
\bigl(\mathrm{\varrho},z,\varphi\bigr),\qquad \mathrm{r}=\sqrt{z^2+
\mathrm{\varrho}^2},\quad \mathrm{\varrho}\leq L,\quad
z\in\bigl(-\infty,+\infty\bigr), \quad \varphi\in\bigl[0,\pi\bigr],
\label{ag1}
 \end{equation}
where $x^1=\varrho\sin\varphi$ and $x^2=\varrho\cos\varphi$, in addition,
$L>0$ is some finite length.

In these coordinates, the quantum motion of bodies is described by the following PDE:
\begin{equation}
\biggl\{\frac{1}{\mathrm{\varrho} }\frac{\partial}{\partial\mathrm{\varrho}}
\biggl(\mathrm{\varrho}\frac{\partial}{\partial\mathrm{\varrho}}\biggr)+
\frac{\partial^2}{\partial z^2}+\frac{1}{\mathrm{\varrho}^2}
\frac{\partial^2}{\partial \varphi^2}+\frac{2\mu_0}{\hbar^2 \widetilde{g}
\bigl(\{\bar{\mathrm{\varrho}}\}\bigr)}\biggl[\mathrm{E}-\frac{J(J+1)}
{\widetilde{g}\bigl(\{\bar{\mathrm{\varrho}}\}\bigr)}\biggr]\biggr\}
\widetilde{\Psi}^J_K=0,
\label{ckz07b}
\end{equation}
where $f:g\bigl(\{\bar{x}\})\mapsto \widetilde{g}\bigl(\{\bar{\mathrm{\varrho}}\}\bigr)$.

For further study of the problem, it is convenient to represent the function;
$\widetilde{g}^{\,-k}\bigl(\{\bar{\mathrm{\varrho}}\}\bigr),\,\,(k=1,2)$  in
the form of expansion in the orthogonal Legendre functions:
\begin{equation}
\widetilde{g}^{\,-k}\bigl(\{\bar{\mathrm{\varrho}}\}\bigr)=
\sum_{m=0}^\infty\widetilde{\mathfrak{g}}^{\,(k)}\bigl(\varrho,z\bigr)
P_m\bigl(\zeta\bigr),\qquad \zeta=\cos\varphi,
\label{Qz7b}
\end{equation}
and, correspondingly;
$$
\widetilde{\mathfrak{g}}^{\,(k)}_m\bigl(\varrho,z\bigr)=\Bigl(m+\frac{1}{2}\Bigr)\int^{1}_{-1}
\widetilde{g}^{\,-k}\bigl(\{\bar{\mathrm{\varrho}}\}\bigr)P_m\bigl(\zeta\bigr)d\zeta.
$$
Representing the solution of the equation  (\ref{ckz07b}) in the form:
\begin{equation}
\widetilde{\Psi}^J_K\bigl(\{\bar{\mathrm{\varrho}}\}\bigr)=\widetilde{\Upsilon}
\bigl(\varrho,z\big)\Theta^j_{K}\bigl(\zeta\bigr),
\label{Wz2b}
\end{equation}
with consideration (\ref{Qz7b}), we get the following second-order PDE:
\begin{equation}
\biggl\{\Theta^j_{K} \biggl[\frac{1}{\mathrm{\varrho}}\frac{\partial}{\partial\mathrm{\varrho}}
\biggl(\mathrm{\varrho}\frac{\partial}{\partial\mathrm{\varrho}}\biggr)+
\frac{\partial^2}{\partial z^2}+\frac{2\mu_0}{\hbar^2}
\widetilde{\Omega}\bigl(\{\bar{\varrho}\}\big)\biggr]
 +\frac{1}{\mathrm{\varrho}^2}\frac{\partial^2\Theta^j_{K}}{\partial \varphi^2}\biggr\}
\widetilde{\Upsilon}=0,
\label{ckz7b}
\end{equation}
where
$
\widetilde{\Omega}\bigl(\{\bar{\varrho}\}\big)=\sum_{m=0}^\infty
\bigl[\mathrm{E}\widetilde{\mathfrak{g}}^{\,(1)}_m-J(J+1)
\widetilde{\mathfrak{g}}^{\,(2)}_m\bigr]\Theta^j_m\bigl(\zeta\bigr),
$
in addition,  $\Theta^j_{K}\bigl(\zeta\bigr)$ denotes the associated
Legendre functions \cite{Arfken}.

Now, having performed simple calculations, we finally obtain the following ODE for the
 wave function (seel Appendix  {\bf H}):
\begin{equation}
\biggl\{ \biggl[\frac{1}{\mathrm{\varrho}}\frac{\partial}{\partial\mathrm{\varrho}}
\biggl(\mathrm{\varrho}\frac{\partial}{\partial\mathrm{\varrho}}\biggr)
+ \frac{\partial^2}{\partial z^2}\biggr]
+ \frac{\mathbf{Q}_{jK}}{\mathrm{\varrho}^2}+ \frac{2\mu_0}{\hbar^2}
\widetilde{\mathbf{\Omega}}_{jK}\bigl(\varrho,z\big) \biggr\}\widetilde{\Upsilon}=0,
\label{c0z7b}
\end{equation}
where the following notations are made:
$$
\mathbf{Q}_{jK} =\widetilde{\Omega}\bigl(\{\bar{\varrho}\}\big)=
\bigl(\Theta^j_k(1)\bigr)^2+ \frac{(K+j)!}{(K-j)!}
\biggl[\frac{K^2}{j}-\frac{1+2j(j+1)}{2K+1}\biggr],\qquad  j\neq 0,
$$
$$
\widetilde{\mathbf{\Omega}}_{jK}\bigl(\varrho,z\big)=\sum_{m=0}^\infty
I_{mK}^j\bigl[\mathrm{E}\,\widetilde{\mathfrak{g}}^{\,(1)}_m-J(J+1)
\widetilde{\mathfrak{g}}^{\,(2)}_m\bigr], \quad\,I_{mK}^j=
\int_{-1}^1\bigl[\Theta^j_{K}\bigl(\zeta\bigr)\bigr]^2
\Theta^0_{m}\bigl(\zeta\bigr)d\zeta.
$$
The term $I_{mK}^j\equiv I_{mKK}^j$  exactly  is calculated (see Appendix  {\bf H}).

It is obvious that in the limit of $z\to -\infty$ or in the $(in)$ asymptotic state
$\lim_{z\to-\infty}\widetilde{\mathbf{\Omega}}_{jK}\bigl(\varrho,z\big)=
\widetilde{\mathbf{\Omega}}^-_{jK}\bigl(\varrho\big)$, the motion of  the three-body
quantum system breaks up into vibrational-rotational and translational components.
This means that we can write the following representation for an asymptotic wave function:
\begin{equation}
\widetilde{\Psi}^{+(J)}_{njK}
\bigl(\{\bar{\mathrm{\varrho}}\}\bigr)\quad{}_{\overrightarrow{\,\,z\to-\infty\,\,}}\quad
\widetilde{\Psi}^{(in)J}_{njK}\bigl(\{\bar{\mathrm{\varrho}}\}\bigr)=
\frac{1}{\sqrt{2\pi}}\exp\Bigl\{-\frac{i}{\hbar}p^-_{n(jK)}\,z\Bigr\}\Theta^j_K\bigl(\varphi\bigr)
\widetilde{\Upsilon}_{njK}^{(in)}\bigl(\mathrm{\varrho}\bigr),
\label{t01b}
\end{equation}
where $p_{n(jK)}^-=\sqrt{2\mu_0\bigl[\mathrm{E}-\mathcal{E}^{(in)}_{n(j,K)}\bigr]}$  is
the momentum of the \emph{imaginary point} in the  $(in)$ asymptotic subspace of scattering,
and  the wave function $\widetilde{\Upsilon}_{njK}^{(in)}\bigl(\mathrm{\varrho}\bigr)$
denotes the bound state of a three-body system that satisfies the following equation:
\begin{equation}
\biggl\{\frac{1}{\mathrm{\varrho}}\frac{\mathrm{d}}{\mathrm{d}\mathrm{\varrho}}
\biggl(\mathrm{\varrho}\frac{\mathrm{d}}{\mathrm{d}\mathrm{\varrho}}\biggr)
+ \frac{\mathbf{Q}_{jK}}{\mathrm{\varrho}^2}+ \frac{2\mu_0}{\hbar^2}
\widetilde{\mathbf{\Omega}}^-_{jK}\bigl(\varrho\big) \biggr\}\widetilde{\Upsilon}^{(in)}=0,
\label{c0z2b}
\end{equation}
where $\mathcal{E}^{(in)}_{n(j,K)}$ is the quantized energy of the coupled system
$(23)_{njK}$, which takes into account the influence of the vibrational-rotational motion
of the system. The spectrum of the energy $\mathcal{E}_{n(j,K)}$ can be
calculate by solving the equation (\ref{c0z2b}).

The total wave function $\widetilde{\Psi}^{+(J)}_{njK}$ in the limit $z\to +\infty$ goes
 into the $(out)$  asymptotic state, where it can be represented as:
\begin{equation}
\widetilde{\Psi}^{+(J)}_{njK}\bigl(\{\bar{\mathrm{\varrho}}\}\bigr) \quad
{}_{\overrightarrow{\,\,z\to+\infty\,\,}}\quad\sum_{n'j'K'}
\mathbf{ S}^J_{njK\,\rightarrow\,n'j'K'}\bigl(\mathrm{E}_c\bigr)\widetilde{\Psi}^{(out)J}_{n'j'K'}
\bigl(\{\bar{\mathrm{\varrho}}\}\bigr),
\label{t02sb}
\end{equation}
where $\mathbf{ S}^J_{n'j'K'\,\leftarrow\,njK}\bigl(\mathrm{E}_c\bigr)$ is the
 $\mathbf{ S}$ - matrix element of the rearrangement process, which depends
 on the collision energy  $\mathrm{E}_c=\bigl[\mathrm{E}-\mathcal{E}^{(in)}_{n(j,K)}\bigr]$ of
 particles and the  quantum numbers of asymptotic states.
The total wave function of the system of bodies also satisfies the following boundary conditions:
\begin{equation}
\lim_{|\varrho\,|\to\,\infty}\widetilde{\Psi}^{+(J)}_{njK}\bigl(\{\bar{\mathrm{\varrho}}\}\bigr)
=\lim_{|\varrho\,|\to\,\infty}\frac{\partial}{\partial\varrho}
\widetilde{\Psi}^{+(J)}_{njK}\bigl(\{\bar{\mathrm{\varrho}}\}\bigr)=0.
\label{tz01}
\end{equation}

As is known, the main goal of quantum scattering theory is to construct
$\mathbf{ S}$ - matrix elements of different quantum transitions.  In the body-fixed LCC system,
 we can write the following exact representation connecting two different
representations of the full wave function \cite{New}:
\begin{equation}
\widetilde{\Psi}^{+(J)}_{njK}\bigl(\{\bar{\mathrm{\varrho}}\}\bigr)=\sum_{n'j'K'}
\mathbf{ S}^J_{njK\,\rightarrow\,n'j'K'} \widetilde{\Psi}^{-(J)}_{n'j'K'}
\bigl(\{\bar{\mathrm{\varrho}}\}\bigr),
\label{ck0w7b}
\end{equation}
where $\widetilde{\Psi}^{+(J)}_{njK}\bigl(\{\bar{\mathrm{\varrho}}\}\bigr)$ and
$\widetilde{\Psi}^{-(J)}_{njK}\bigl(\{\bar{\mathrm{\varrho}}\}\bigr)$ are total
stationary wave functions that develop, respectively, from pure $(in)$ and $(out)$
 asymptotic states. Recall that this case the coordinate $z$ plays role of \emph{timing parametr}.

As for asymptotic wave functions, it is convenient to represent them in  global
coordinates $\{\bar{\rho}\}\in \mathbb{E}^3$, and then display them on a
manifold $\mathcal{M}^{(3)}_t\ni \{\bar{x}\}$. In order to implement the
 mapping  $f:\Psi^{(in)J}_{njK}\bigl(\{\bar{\rho}\}\bigr)\mapsto
\widetilde{\Psi}^{(in)J}_{njK}\bigl(\{\bar{\mathrm{\varrho}}\}\bigr)$,
 in the function $\Psi^{(in)J}_{njK}\bigl(\{\bar{\rho}\}\bigr)$, we need to
perform a coordinate transformation using the expressions (\ref{24}) and (\ref{ag1}).
Recall that for the asymptotic state $1+(23)_{njK}$ the wave function in
global system $\{\bar{\rho}\}\in \mathbb{E}^3$ can be represented as:
\begin{eqnarray}
\Psi^{(in)J}_{njK}\bigl(\{\bar{\rho}\}\bigr)
=\frac{1}{\sqrt{2\pi}}\exp\Bigl\{-\frac{i}{\hbar}p^-_{nj}\rho_1\Bigr\}
\Pi^{(in)}_{n(j)}(\rho_2)\Theta^j_{K}(\rho_3),\quad  p^-_{n(j)}=
\sqrt{2\mu_0\bigl[\mathrm{E}-\mathcal{E}^{(in)}_{n(j)}\bigr]},
\label{zagb}
\end{eqnarray}
where $\mathcal{E}^{(in)}_{n(j)}$ is the vibration-rotational energy
of the coupled state $(23)_{njK}$, and the function $\Pi^{(in)}_{n(j)}(\rho_2)$,
which describes the wave state satisfying the following ODE \cite{GBN}:
$$
\biggl[-\frac{\hbar^2}{2\mu_0}\frac{\mathrm{d}^2}{\mathrm{d}\rho_2^2}
+U^{(in)}(\rho_2)
+\frac{\hbar^2j(j+1)}{2\mu_0\rho_2^2}\biggr]\Pi^{(in)}_{n(j)}=
\mathcal{E}^-_{n(j)}\Pi^{(in)}_{n(j)}.
$$
Note that in the $(in)$ asymptotic state:
$\lim_{\rho_1\to\,\infty}\mathbb{V}(\textbf{r})=U^{(in)}(\rho_2)$ (see expression (\ref{01c})).

It is easy to verify that the asymptotic wave functions (\ref{t01b}) and
 (\ref{zagb}), despite being represented in different coordinate systems,
however, consist of similar functions.

Finally, based on the foregoing, we can construct the full stationary wave function
of the scattering process on the $6D$ manifold
$\{x\}\sim\bigl(\{\bar{\varrho}\};\{\underline{x}\}\bigr)\in \mathcal{M}$:
\begin{equation}
{\bf{\Psi}}^{+}\bigl(\{\bar{\varrho}\};\{\underline{x}\}\bigr)=\sum^J_{K=-J}
\widetilde{\Psi}^{+(J)}_K\bigl(\{\bar{\mathrm{\varrho}}\}\bigr)
\emph{D}^J_{KM}\bigl(\{\underline{x}\}\bigr), \qquad
\{\underline{x}\}=(x^4,x^5,x^6),
\label{ckw7b}
\end{equation}
where $\emph{D}^J_{KM}$ is the Wigner $D$\,-matrix \cite{Ed,Zar}, in addition,
$K$ and $M$ are space-fixed and body-fixed $z$ projections of the angular
momentum $\bf J$.
 \begin{figure}
\includegraphics[width=95mm]{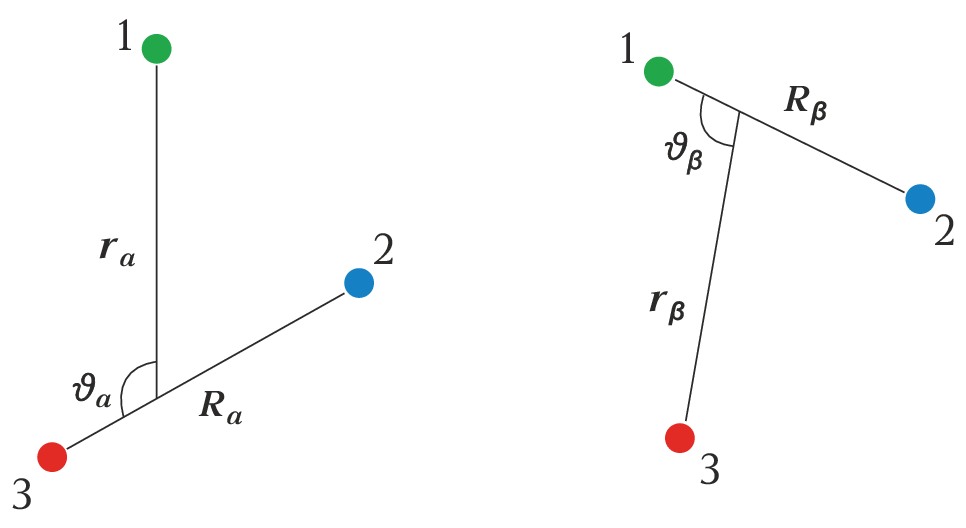}
\caption{\emph{The set of Jacobi coordinates $(\textbf{R}_\alpha,\textbf{r}_\alpha,\vartheta_\alpha)$
is convenient for describing the asymptotic states $1+(23)_{njK}$, whereas another set of
Jacobi coordinates $(\textbf{R}_\beta,\textbf{r}_\beta,\vartheta_\beta)$ is convenient for
describing the asymptotic states $(12)_{n'j'K'}+ 3.$}}
\label{fig. 6}
\end{figure}

Returning to the problem of constructing of $S$-matrix elements, it
should be noted that each of the scattering channels in the global
coordinate system is conveniently described by its own coordinate system.
In other words, it is convenient to describe quantum states in the
\emph{initial} $(in)$ and \emph{final} $(out)$ channels by various
Jacobi coordinate systems. In this regard, it is obvious that local
systems associated with the corresponding global systems must also be different.
For example, if the wave function $\widetilde{\Psi}^{+(J)}_{njK}$ is conveniently
described using the coordinate system $\{\bar{\varrho}_\alpha\}\in
\mathcal{M}^{(3)}_\alpha\simeq\mathbb{E}^3_\alpha\ni\{\bar{\rho}_\alpha\}$,
then the wave function $\widetilde{\Psi}^{-(J)}_{njK}$ will
naturally be described using the coordinate system
$\{\bar{\varrho}_\beta\}\in \mathcal{M}^{(3)}_\beta\simeq\mathbb{E}^3_\beta
\ni\{\bar{\rho}_\beta\}$  (see FIG. 5).

The correspondence conditions between the asymptotic wave functions
written in two various  global coordinate systems $\{\bar{\rho}_\alpha\}$ and
$\{\bar{\rho}_\beta\}$  can be specified using the equation  \cite{Ed,Zar}:
\begin{equation}
 {\Psi}^{(out)J}_{K'}\bigl(\{\bar{\mathrm{\rho}_\beta}\}\bigr)=
\sum_{\bar{K}}d^J_{K'\bar{K}}(\vartheta) {\Psi}^{(out)J}_{K'}
\bigl(\{\bar{\mathrm{\rho}_\alpha}\}\bigr),
\label{c0w7b}
\end{equation}
where $d^J_{K'\bar{K}}(\vartheta)$  is the
Wigner's small matrix, which has the following form \cite{Wig0}:
$$
d^J_{K'\bar{K}}(\vartheta)=D^J_{KK'}(0,\vartheta,0)=\bigl[(J+K')!(J-K')!(J+K')!(J-K')!\bigr]^{1/2}
\times\,
$$
$$
\quad\qquad\qquad\sum_{s}\Biggl[\frac{(-1)^{K'-K+s}\bigl[\cos(\vartheta/2)\bigr]^{K-K'+2(J-s)}
\bigl[\sin(\vartheta/2)\bigr]^{K'-K+2s}}{(J+K-s)!s!(K'-K+s)!(J-K'-s)!}\Biggr],
$$
where the sum over $``s"$  exceeds such values that factorials are non-negative, in addition,
$\vartheta$ is the angle between the vectors $\textbf{r}_\alpha$ and
$\textbf{r}_\beta$, that is $\textbf{r}_\alpha\textbf{r}_\beta=
r_\alpha r_\beta\cos\vartheta$, which are distances of free particle from the center of mass
of coupled pair in the Jacobi coordinates of the initial $(in)$ and
final $(out)$ channels, respectively.

Now we have all the necessary mathematical objects for constructing of
the $\mathbf{S}$- matrix elements  of a quantum reactive process.

Taking into account the fact that the coordinate $z$ is the \emph{timing parameter}
of the problem, we can obtain a new  exact representation for the transition
$\mathbf{S}$ - matrix elements in terms of stationary wave functions (this idea
was first implemented for the collinear model \cite{ASG,GaKG}):
\begin{eqnarray}
\mathbf{S}^J_{njK\,\rightarrow\,n'j'K'}\bigl(\mathrm{E}_c\bigr)=
\lim_{z\,\,\to\,\,+\infty}
\Bigl\langle\widetilde{\Psi}^{+(J)}_{njK}\bigl(\{\bar{\mathrm{\varrho}_\alpha}\}\bigr)
{\widetilde{\Psi}^{(out)J}_{n'j'K'}\bigl(\{\bar{\mathrm{\varrho}_\beta}\}\bigr)}^\ast\Bigr\rangle
=
\nonumber\\
\sum_{\bar{K}}\Bigl\langle d^J_{K'\bar{K}}(\vartheta)
\widetilde{\Psi}^{+(J)}_{njK}\bigl(\{\bar{\mathrm{\varrho}_\alpha}\}\bigr)
  {\widetilde{\Psi}^{(out)J}_{n'j'K'}
\bigl(\{\bar{\mathrm{\varrho}_\alpha}\}\bigr)}^\ast\Bigr\rangle,
\label{c0w1b}
\end{eqnarray}
where is the sign $``\,{}^\ast"$ denotes the complex conjugation of a function, in addition:
$$
 f:{{\Psi}^{(out)J}_{n'j'K'} \bigl(\{\bar{\mathrm{\rho}_\alpha}\}\bigr)}
\mapsto {\widetilde{\Psi}^{(out)J}_{n'j'K'} \bigl(\{\bar{\mathrm{\varrho}_\alpha}\}\bigr)},
\qquad
\bigl\langle\cdot\cdot\cdot\bigr\rangle=\int_0^\pi d\vartheta\int_0^\infty\int_0^\pi
\sqrt{\widetilde{g}\bigl(\{\bar{\mathrm{\varrho}}_\alpha\}\bigr)}
\varrho_\alpha d\varrho_\alpha\mathrm{d}\varphi_\alpha.
$$
Note that in the limit $z\to - \infty$ as the initial asymptotic condition for
$\widetilde{\Psi}^{+(J)}_{njK}\bigl(\{\bar{\mathrm{\varrho}_\alpha}\}\bigr) $,
 we must choose an asymptotic wave function in the global system
$ {{\Psi}^{(in)J}_{njK} \bigl(\{\bar{\mathrm{\rho}_\alpha}\}\bigr)}$.
In other words, we have to do a mapping
$ f: {\widetilde{\Psi}^{(in)J}_{njK} \bigl(\{\bar{\mathrm{\varrho}_\alpha}\}\bigr)}
\mapsto {{\Psi}^{(in)J}_{njK} \bigl(\{\bar{\mathrm{\rho}_\alpha}\}\bigr)}$,
which we can implement using coordinate transformations (\ref{24}) and (\ref{ag1}).

It is often convenient to obtain equations for $\mathbf{S}$ - matrix elements.
Let us consider the following representation for a complete wave function that
uses the \emph{time-independent  coupled-channel approach} \cite{Wal}:
\begin{equation}
\widetilde{\Psi}^{+(J)}_{\bar{K},[\mathcal{K}]}\bigl(\{\bar{\mathrm{\varrho}}\}\bigr)
=\sum_{\bar{n}\bar{j}}\mathbf{\Xi}^{+(J)}_{[\mathcal{K}]\,[\bar{\mathcal{K}}]}(z)
\Pi_{\bar{n}(\bar{j}\bar{K})}(\varrho;z)\Theta_{\bar{K}}^{\bar{j}}(\zeta),
\qquad [\mathcal{K}]=(n,j,K).
\label{t07sb}
\end{equation}
Substituting (\ref{t07sb}) into the equation (\ref{ckz07b}) and performing
 not  complicated calculations, we obtain:
\begin{equation}
\biggl\{\frac{\mathrm{d}^2}{\mathrm{d} z^2}+
\overline {\mathcal{E}}_{n'(j^{\,'}K')}(z)
\biggr\}\mathbf{\Xi}^{+(J)}_{[{\mathcal{K}}]\,[\mathcal{K}']}(z)=0,
\label{a5ckz07b}
\end{equation}
where $\overline {\mathcal{E}}_{n'(j^{\,'}K')}(z)\equiv\overline
{\mathcal{E}}_{n'(j^{\,'}K')}^{\,n'(j^{\,'}K')}(z)$ is a regular function
(for more details see Appendix  {\bf H}).

It is easy to verify that the solutions of equation (\ref{a5ckz07b}) in the
limit $z\to+\infty$  go over to the corresponding $\mathbf{S}$ - matrix elements:
\begin{equation}
 \lim_{z\to+\infty}\mathbf{\Xi}^{+(J)}_{[{\mathcal{K}}]\,[\mathcal{K}']}(z)=
\mathbf{S}^J_{[{\mathcal{K}}]\,\rightarrow\,[{\mathcal{K}}']}\bigl(\mathrm{E}_c\bigr),\qquad
[{\mathcal{K}}]=(njK).
\label{a6ckz07b}
\end{equation}

Returning to the quantum equations, both non-stationary (\ref{c27a}) and stationary
(\ref{c27b}), we note that they are solved together with the classical equations
(\ref{07a}) taking into account coordinate transformations (\ref{24}) and (\ref{ag1}). It is
important to note that the meaning of additional classical equations and coordinate
transformations is that they generate trajectory  tubes with various geometric and
topological features, which are quantized using equations (\ref{c27a}) and
(\ref{c27b}). In view of the foregoing, it is obvious that \emph{non-integrability} and,
 moreover, the \emph{randomness} in behavior of the classical problem will affect the
quantum problem. In the case of \emph{strongly developed chaos}, this can lead to chaos
generation and, in the \emph{main object of quantum mechanics},  in the \emph{wave function}.
Recall that this significantly distinguishes our understanding of quantum chaos from
the interpretation of this phenomenon by other authors (see for example \cite{Gut}).
This means that in the limit $\hbar\to0$ the dynamical quantum system (conditionally
${\mathbf{Q}_{ch}}$ -  \emph{quantum chaotic system}) will be goes over to the
\emph{classical dynamical system} (${\mathbf{P}}$ - system),  without violating the
\emph{quantum generalization of Arnold's theorem} \cite{Hannay} (see FIG. 4). In
other words, in connection with the statement of M. Gutswiller that  \emph{"the concept
of quantum chaos is a mystery, not a well-formulated problem"}, we argue that
quantum chaos - $Q_{ch}$ a separate,  more general and well-defined area-of-motion
is represented.

Recent studies by the authors have shown  that quantum chaotic behavior even
manifests itself in a low-dimensional model problem, such as a collinear collision of
three bodies  \cite{AshG}, on  the example of the \emph{bimolecular chemical reaction} with
the rearrangement $Li+(FH)\to (LiF) +H$. In particular, as shown by numerical
calculations, the total wave function for the system under study exhibits strongly
chaotic behavior, which also affects the amplitude of quantum transitions
${\bf \mathcal{A}}^J_{[{\mathcal{K}}]\,\rightarrow\,[{\mathcal{K}}']}=
\bigl|\mathbf{S}^J_{[{\mathcal{K}}]\,\rightarrow\,[{\mathcal{K}}']}
\bigl(\mathrm{E}_c\bigr)\bigr|^2$. In other words, to calculate the mathematical
expectation of the amplitude  of the quantum transition, it is necessary to carry
out additional averaging,  which is done using formula (\ref{26b}) based on the idea
of {\bf Definition 8}.

In the end, we note that, as the study showed, not all bimolecular reactions show
chaotic behavior.  For example, as shown by numerical simulation of the reacting
systems $ N +N_2, \, \, O+ O_2, \, \, N +O_2 $ in the framework of the collinear model
\cite{ASG}, these systems are generally regular in the behavior of wave functions and,
accordingly, in transition amplitudes, which indicates \emph{insufficient development
of chaos} in the  corresponding classical counterparts.

\section{C\lowercase{onclusion}}
The study of the classical three-body problem with the aim of revealing new
regularities of both celestial mechanics and elementary atomic-molecular processes,
is still of great interest. In addition, it is very important to answer the
fundamental question for quantum foundations, namely: \emph{is irreversibility
fundamental for describing the classical world} \cite{Brig}? Recall that the
answer to this question on the example of the three-body problem can significantly
deepen our understanding regarding the type and nature of complexities
that arise in dynamical systems.

Note that if the main task for celestial mechanics is finding stable trajectories,
for atomic-molecular collisions the studying of multichannel scattering processes
are of primary importance.

Following the Krylov's idea, we considered the general classical three-body
problem on a conformal-Euclidean-\emph{Riemann manifold}. The new formulation
of the known problem made it possible to identify a number of important and still
unknown fundamental features of the dynamical system. Below we list only
the four most important ones:
\begin{itemize}
\item
The Riemannian geometry with its local coordinate system in the most general
case allows us to reveal additional hidden symmetries of the internal motion
of a dynamical system. This circumstance makes it possible to reduce the dynamical
system from the 18\emph{th} to the 6\emph{th} order (see Eqs. (\ref{07a})) instead
of the generally accepted 8\emph{th} order. In case when the energy of the
system is fixed, the dynamical problem is reduced to a 5\emph{th}-order system.
Obviously, the fact of a more complete reduction of the equations system is very
useful for creating efficient algorithms for numerical simulation. Note that
the obtained system of differential equations differs in principle from the
Newtonian equations in that it is symmetric with respect to all variables and is
non-linear since it  includes quadratic terms of the velocity projections. These
equations play a crucial role in deriving equations for a probability distributions
of geodesic flows both in the phase and configuration spaces.
\item
The equivalence between the Newtonian three-body problem (\ref{19f}) and the problem
of geodesic flows on the Riemannian manifold (\ref{07a}) provides the coordinate
transformations (\ref{24}) together with  the system of algebraic equations (\ref{22}).
Note that due to the algebraic system, which is absent in Krylov's representation, the
chronological parameter of the $s$ dynamical system, conventionally called \emph{internal
time} $s$ (see FIG. 3), can branch and fluctuate. Moreover, in some intervals it may
show a chaotic character that essentially distinguishes it from usual time $t$.
As the analysis shows, the \emph{internal time} in this microscopic classical problem
has the same non-trivial behavior as the \emph{time's arrow}  of more complex systems
\cite{Misra}. Obviously, \emph{internal time} $``s"$  makes the system of equations
(\ref{07a}) \emph{irreversible}, because it has a structure and an arrow of development,
which significantly distinguishes it from ordinary time $t$. The latter radically changes our
understanding of \emph{time as a trivial parameter that chronologizing events} in a dynamical
system and connects the past with the future through the present. And, in spite of the
pessimistic statements of Bergson and Prigogine \cite{Eric,Henri,Ilya}, a new approach,
in our view, will allow classical mechanics to describe the whole spectrum of various
phenomena, including the \emph{irreversibility} inherent of elementary atomic-molecular
processes.
\item
The developed representation allows taking into account external regular and random
forces on the evolution of the dynamical system without using perturbation theory methods.
In particular, equations have been obtained that describe the propagation of probabilistic
flows of geodesic trajectories in both the phase space (\ref{24abct}) and the configuration
space (\ref{25b}). Note that this makes it possible to calculate the probabilities of
elementary transitions between different asymptotic subspaces taking into account the
multichannel character of scattering with all its complexities.
\item
The quantization of the reduced Hamiltonian (\ref{09}), taking into account algebraic
equations (\ref{22}) and coordinate transformations (\ref{24}) makes the quantum-mechanical
equations (\ref{c27a}) and (\ref{c27b}) irreversible. This circumstance is a necessary
condition for generating chaos in the wave function. The latter without violating the
quantum generalization of Arnold's theorem, in the limit $\hbar\to 0$ allows us to make
the transition from the quantum region to the region of classical chaotic motion, that
solves an important open problem of the \emph{quantum-classical correspondence}
(see \cite{Schuster,Hannay}).
\end{itemize}

Lastly, it is important to note that, despite Poincar\'{e}'s pessimism regarding the
usefulness of using non-Euclidean geometry in physics, this study rather shows the
truthfulness of his other statement. Namely, Poincar\'{e} believed that geometry and
physics are closely related, and therefore the choice of geometry to solve the problem
should be made based on the convenience of describing the problem under consideration.

We are confident that the ideas discussed will be useful and promising for study,
especially for more complex dynamical problems, both classical and quantum.

 \section{a\lowercase{cknowledgment }}
The author is grateful to Profs. L. Beklaryan and A. A. Saharian for detailed discussions of various
aspects of the considered problem and for useful comments.

\section{A\lowercase{ppendix}}
\subsection{ }
Let us consider vector product of vectors encountered in the expression of the kinetic energy (\ref{16a}).
Taking into account the fact that the direction $\textbf{k}=\textbf{\emph{R}}||\textbf{\emph{R}}||^{-1}$
coincides with the axis $z$ we get:
\begin{equation}
[\bm\omega\times \textbf{k}]=(\hat{x}\omega_x+\hat{y}\omega_y+\hat{z}\omega_z)\times(\hat{x}\cdot 0
+\hat{y}\cdot 0+\hat{z}\cdot k_z)
=\hat{x}\omega_y-\hat{y}\omega_x,\quad \textbf{k}=\hat{z}\cdot k_z,
\label{1b}
\end{equation}
and respectively,
\begin{equation}
[\bm\omega\times \textbf{k}]^2=\omega_x^2+\omega_y^2,\qquad ||\hat{x}||=||\hat{y}||=||\hat{z}||=1.
\label{2b}
\end{equation}
Similarly, we can calculate the second term:
\begin{equation}
[\bm\omega\times {\bf{r}}]=\hat{x}\omega_yr\cos\theta  +\hat{y}r(\omega_z\sin\theta-\omega_x\cos\theta)
-\hat{z}r\omega_y\sin\theta,\quad {\bf{r}}=||{\bf{r}}||{\bm{\gamma}}=r\bm{\gamma},
\label{3b}
\end{equation}
using which we can get:
\begin{eqnarray}
[\bm\omega\times{\bf{r}}]^2=r^2\bigr\{\omega_y^2+(\omega_z\sin\theta-\omega_x\cos\theta)^2\bigl\},
\quad  {\bf{\dot{r}}}^2=\bigl(||\bf{r}||\dot{\bm\gamma}+||\bf{\dot{r}}||{\bm\gamma}\bigr)^2=
 r^2\dot{\bm\gamma}^2+
\nonumber\\
 2r\dot{r}{\bm\gamma}\dot{\bm\gamma}
+\dot{r}^2{\bm\gamma}^2 =r^2\dot{\theta}^2+\dot{r}^2,\quad \bm\gamma=(\sin\theta,0,\cos\theta),
\quad  {\bm\gamma}\dot{\bm\gamma}=0,\qquad {\bf\dot{r}}\cdot[\bm\omega\times {\bf{r}}]=
\nonumber\\
 (r\dot{\bm\gamma}+\dot{r}{\bm\gamma})\cdot[\bm\omega\times {\bf{r}}]
=r\dot{r}\omega_y\sin\theta\cos\theta-r\dot{r}\omega_y\sin\theta\cos\theta=0.\qquad\qquad\qquad
\label{4b}
\end{eqnarray}
Taking into account (\ref{1b})-(\ref{4b}),  the expression of the kinetic energy (\ref{16a})
 can be written in the form  (\ref{17a}).

Now it is important to calculate the terms $A$ and $B$ that enter in the expression (\ref{17a}).
Taking into account the equations system (\ref{18a}), it is easy to calculate:
 \begin{eqnarray}
A=\omega_x^2+\omega_y^2=(\dot{\Phi}\sin\Theta\sin\Psi+\dot{\Theta}\cos\Psi)^2+(\dot{\Phi}\sin\Theta\cos\Psi
-\dot{\Theta}\sin\Psi)^2=
\nonumber\\
\dot{\Phi}^2\sin^2\Theta\sin^2\Psi+2\dot{\Phi}\dot{\Theta}\sin\Theta\sin\Psi\cos\Psi+\dot{\Theta}^2\cos^2\Psi+
\dot{\Phi}^2\sin^2\Theta\cos\Psi^2
\nonumber\\
 -2\dot{\Phi}\dot{\Theta}\sin\Theta\cos\Psi\sin\Psi+\dot{\Theta}^2\sin^2\Psi=\dot{\Phi}^2\sin^2\Theta
 +\dot{\Theta}^2,
\label{5b}
\end{eqnarray}
and
 \begin{eqnarray}
 B=\omega^2_y+\bigl(\omega_x\cos\theta-\omega_z\sin\theta\bigr)^2=(\dot{\Phi}\sin\Theta\cos\Psi-
 \dot{\Theta}\sin\Psi)^2 +(\dot{\Phi}\sin\Theta\sin\Psi+
 \nonumber\\
 \dot{\Theta}\cos\Psi)^2\cos^2\theta-2(\dot{\Phi}\sin\Theta\sin\Psi+
\dot{\Theta}\cos\Psi)(\dot{\Phi}\cos\Theta-\dot{\Psi})\sin\theta\cos\theta+
\nonumber\\
(\dot{\Phi}\cos\Theta-\dot{\Psi})^2\sin^2\theta=\dot{\Phi}^2\sin^2\Theta\cos^2\Psi-
\dot{\Phi}\dot{\Theta}\sin\Theta\sin2\Psi+\dot{\Theta}^2\sin^2\Psi
\nonumber\\
+ \,\dot{\Phi}^2\sin^2\Theta\sin^2\Psi\cos^2\theta\,+ \dot{\Phi}\dot{\Theta}\sin\Theta\sin2\Psi
\cos^2\theta+\dot{\Theta}^2\cos^2\Psi\cos^2\theta\,-
\nonumber\\
 \frac{1}{2}\,\dot{\Phi}^2\sin2\Theta\sin\Psi\sin2\theta+\dot{\Phi}\dot{\Psi}\sin\Theta
\sin\Psi\sin2\theta-\dot{\Phi}\dot{\Theta}\cos\Theta\cos\Psi\sin2\theta
\nonumber\\
 +\,\, \dot{\Theta}\dot{\Psi}\cos\Psi\sin2\theta\, +\,
\dot{\Phi}^2\cos^2\Theta\sin^2\theta\,-\,2\dot{\Phi}\dot{\Psi}\,\cos\Theta\sin^2\theta\,+
\,\dot{\Psi}^2\sin^2\theta.
\label{6b}
\end{eqnarray}
Finally, taking into account the calculations (\ref{5b}) and (\ref{6b}), it is easy to
calculate the components of the tensor $\gamma^{\alpha\beta}$ (see expression (\ref{19a})).

 \subsection{}
As we saw in section IV, the manifold $\mathfrak{S}^{(3)}$ plays a key role at proofing
direct one-to-one transformation  between the manifolds $\mathcal{M}^{(3)}$ and
$\mathbb{E}^3$. In particular, a set of nine unknown parameters $(\alpha_1,...,\zeta_3)$
 forms $9D$ space $\mathbb{R}^9$. In the case when we impose additional restrictions on
 these variables in the form of a system of six algebraic equations (see Eqs. (\ref{22})),
we are thereby isolate the set of 3$D$ manifolds $\mathfrak{S}^{(3)}$ in $\mathbb{R}^9$ space.

Now let us see how these 3$D$ manifolds are formed and what their geometric and topological
features are. Using simple  notations, we  can rewrite the system of equations (\ref{22})
in a universal form:
 \begin{eqnarray}
\tilde{\alpha}_1^2+\tilde{\beta}_1^2+ \tilde{\zeta}_1^2\,=1,\qquad\quad
\tilde{\alpha}_1\tilde{\alpha}_2+\tilde{\beta}_1\tilde{\beta}_2+\tilde{\zeta}_1\tilde{\zeta}_2=0,
\nonumber\\
\tilde{\alpha}_2^2+\tilde{\beta}_2^2+\tilde{\zeta}_2^2\,=1,\qquad\quad
\tilde{\alpha}_1\tilde{\alpha}_3+\tilde{\beta}_1\tilde{\beta}_3+\tilde{\zeta}_1\tilde{\zeta}_3=0,
\nonumber\\
\tilde{\alpha}_3^2+\tilde{\beta}_3^2+\tilde{\zeta}_3^2\,=\,1,\qquad\quad
\tilde{\alpha}_2\tilde{\alpha}_3+\tilde{\beta}_2\tilde{\beta}_3+\tilde{\zeta}_2\tilde{\zeta}_3=0,
\label{22f}
\end{eqnarray}
where $\tilde{\alpha}_i=\alpha_i/\sqrt{\breve{g}(\{\bar{\rho}\}}),\,\,
\tilde{\beta}_i=\beta_i/\sqrt{\breve{g}(\{\bar{\rho}\}})$ and
$\tilde{\zeta}_i=\zeta_i \sqrt{{\gamma}^{33}(\{\bar{\rho}\}})/\sqrt{\breve{g}(\{\bar{\rho}\}})$.
It is well known that the number of combinations $C_n^k$ from the $n$-elements
in $k$ is determined by the expression $C^k_n=\frac{n!}{k!(n-k)!}$.
In our case, if we take into account the fact that the number of algebraic equations
is 6 and the number of  unknowns is 9, then it is obvious that the system of equations
(\ref{22f}) will generate $C^6_9=84$ oriented smooth $3D$ -manifolds
$\mathfrak{S}^{(3)}_{\{\alpha\}}$, which are immersed in the
space $\mathbb{R}^9$. Note that $\{\alpha\}$ denotes the certain family
of manifolds. Recall that the symmetry of the equations (\ref{22f}) suggests
that only four families of manifolds are possible
$\{\alpha\}\in\bigl(\mathcal{\breve{A}},\mathcal{\breve{B}},\mathcal{\breve{C}},
\mathcal{\breve{D}}\bigr)$,  where in each family there is a different number of manifolds.
\begin{figure}
\includegraphics[width=70mm]{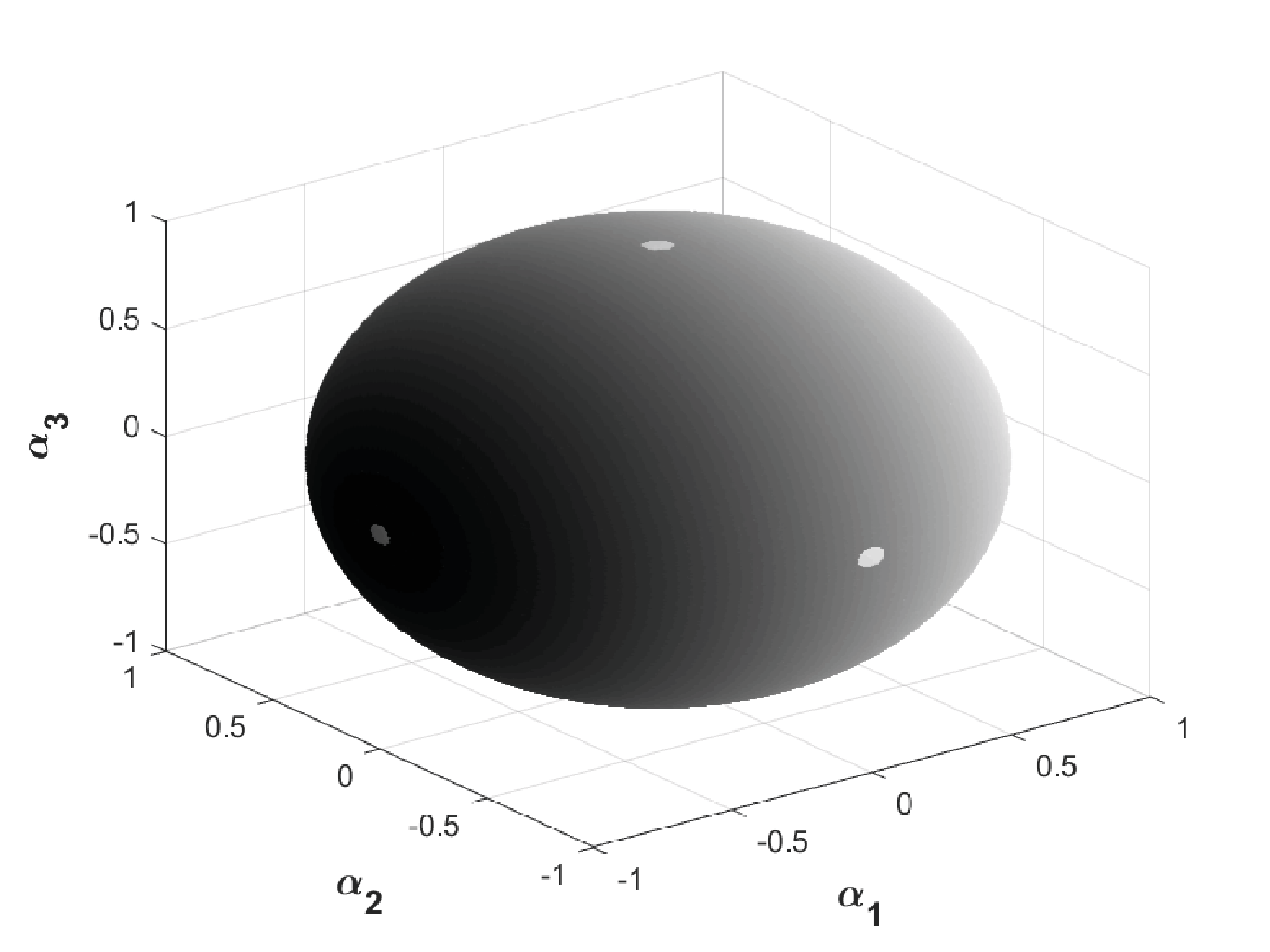}\quad
\includegraphics[width=50mm]{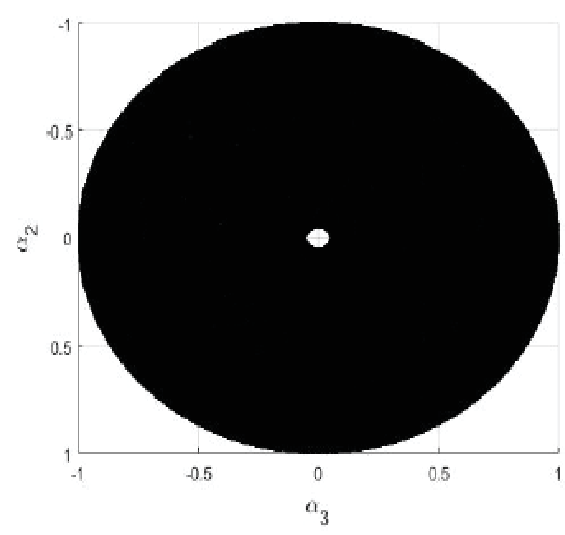}
\caption{\emph{The left figure shows $3D$ submanifold typical of the $\breve{\mathcal{A}}$
family with six topological features. The right figure shows the projection
of this submanifold onto the plane $(\alpha_2,\alpha_3)$. Recall that similar
pictures arise when we projecting manifold on the plane $(\alpha_1,\alpha_2)$
and $(\alpha_1,\alpha_3)$.}}
\label{Fig. 8}
\end{figure}
\begin{figure}
\includegraphics[width=70mm]{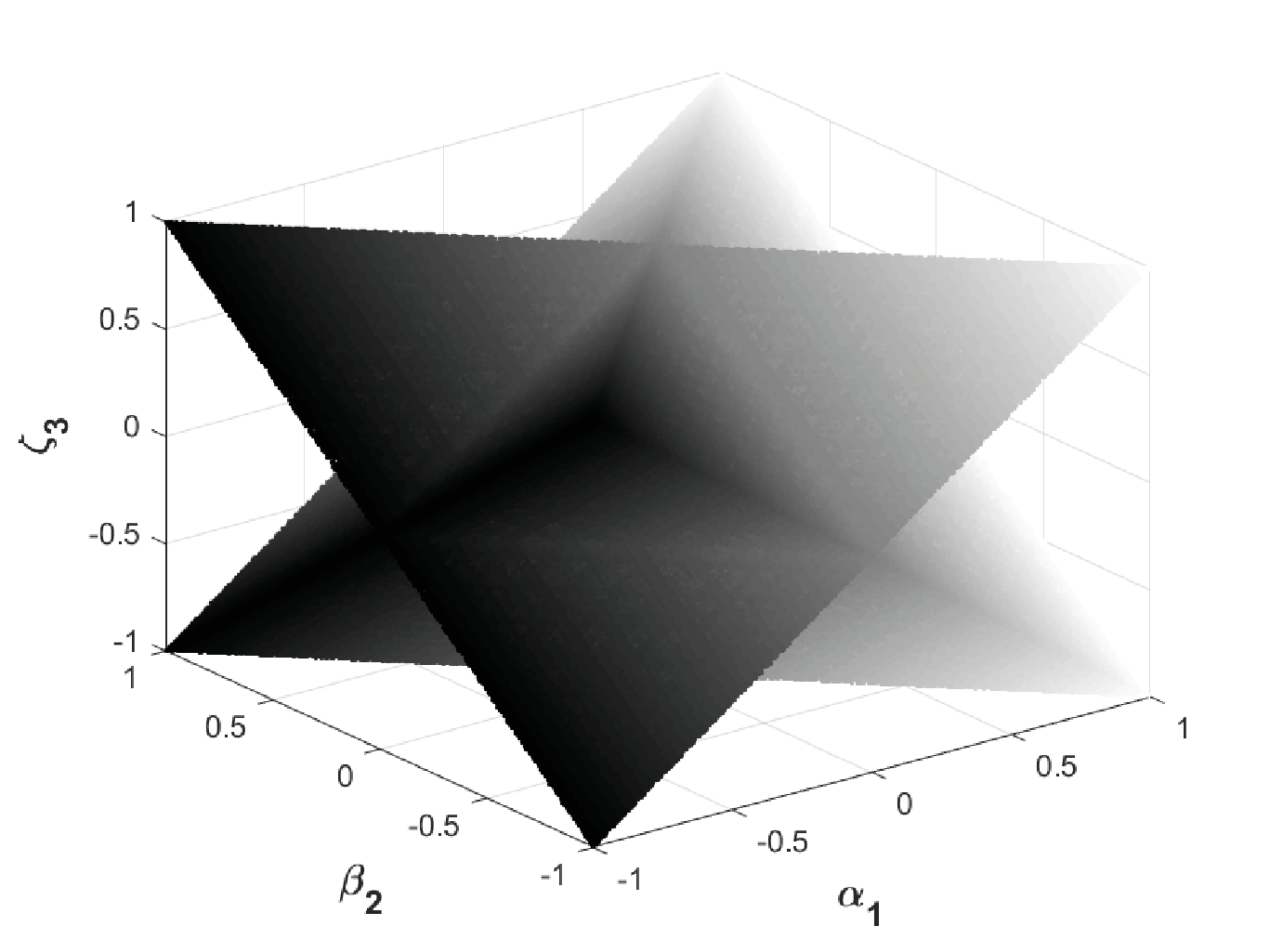}\quad
\includegraphics[width=60mm]{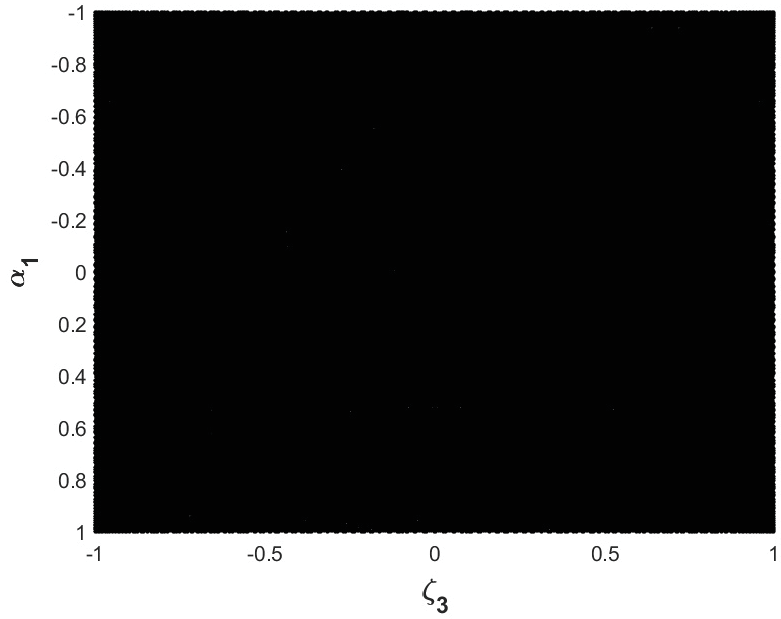}
\caption{\emph{The left image shows a typical $3D$ submanifold of the $\breve{\mathcal{B}}$ family.
As seen, a submanifold is smooth and has no topological features.
The right figure shows the projection of this manifold on the plane $(\alpha_1,\zeta_3)$.
Recall that similar pictures arise when we projecting manifold on the plane
$(\alpha_1,\beta_2)$ and $(\beta_2,\zeta_3)$.}}
\label{Fig.7}
\end{figure}

The first family  $\breve{\mathcal{A}}$ consists of six submanifolds
$\breve{\mathcal{A}}=\overline{\breve{\mathcal{A}}_1,\,\breve{\mathcal{A}}_6}$
 (see FIG. 7).
\begin{figure}
\includegraphics[width=70mm]{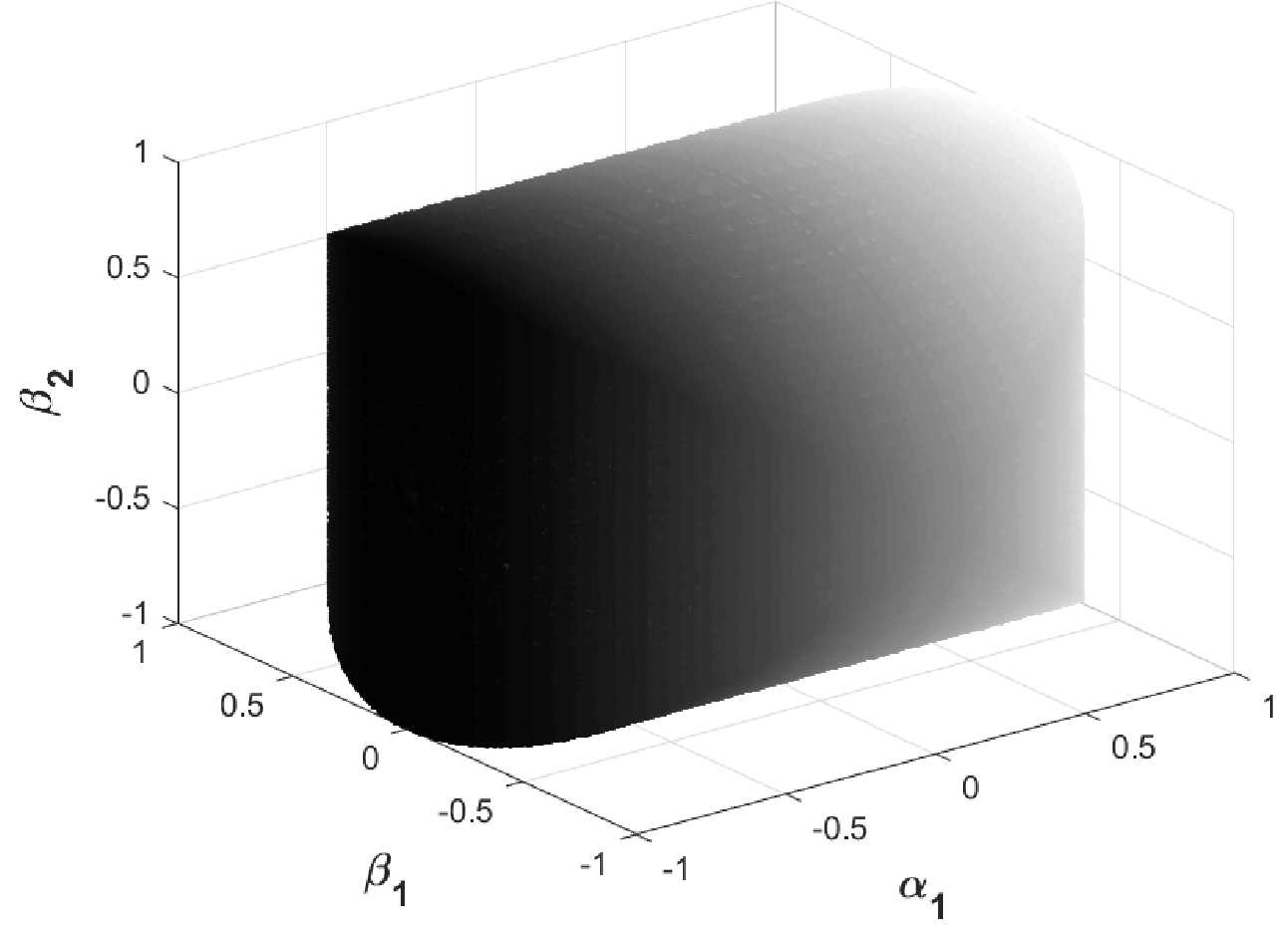}\quad
\includegraphics[width=50mm]{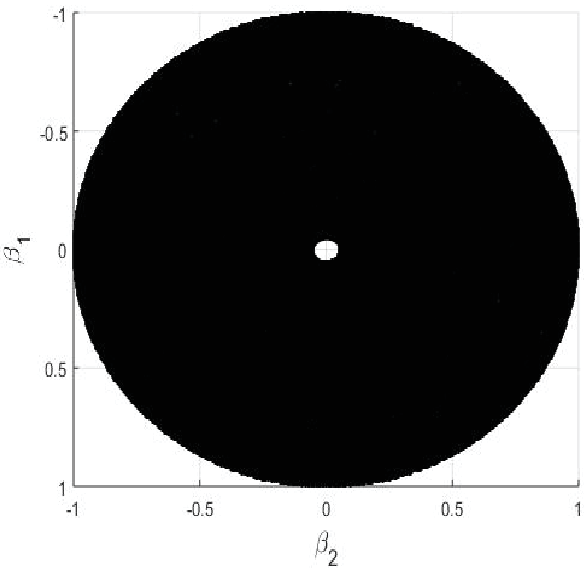}
\caption{\emph{The left image shows a typical $3D$ submanifold of
the $\breve{\mathcal{C}}$ family that has a topology. The right figure shows the
projection of this submanifold on the plane $(\alpha_1,\beta_1)$.
 Recall that the projections of the submanifold on the plane $(\beta_1,
\beta_2)$ and $(\alpha_1,\beta_2)$ do not contain topologies. }}
\label{Fig.8}
\end{figure}
\begin{figure}
\includegraphics[width=70mm]{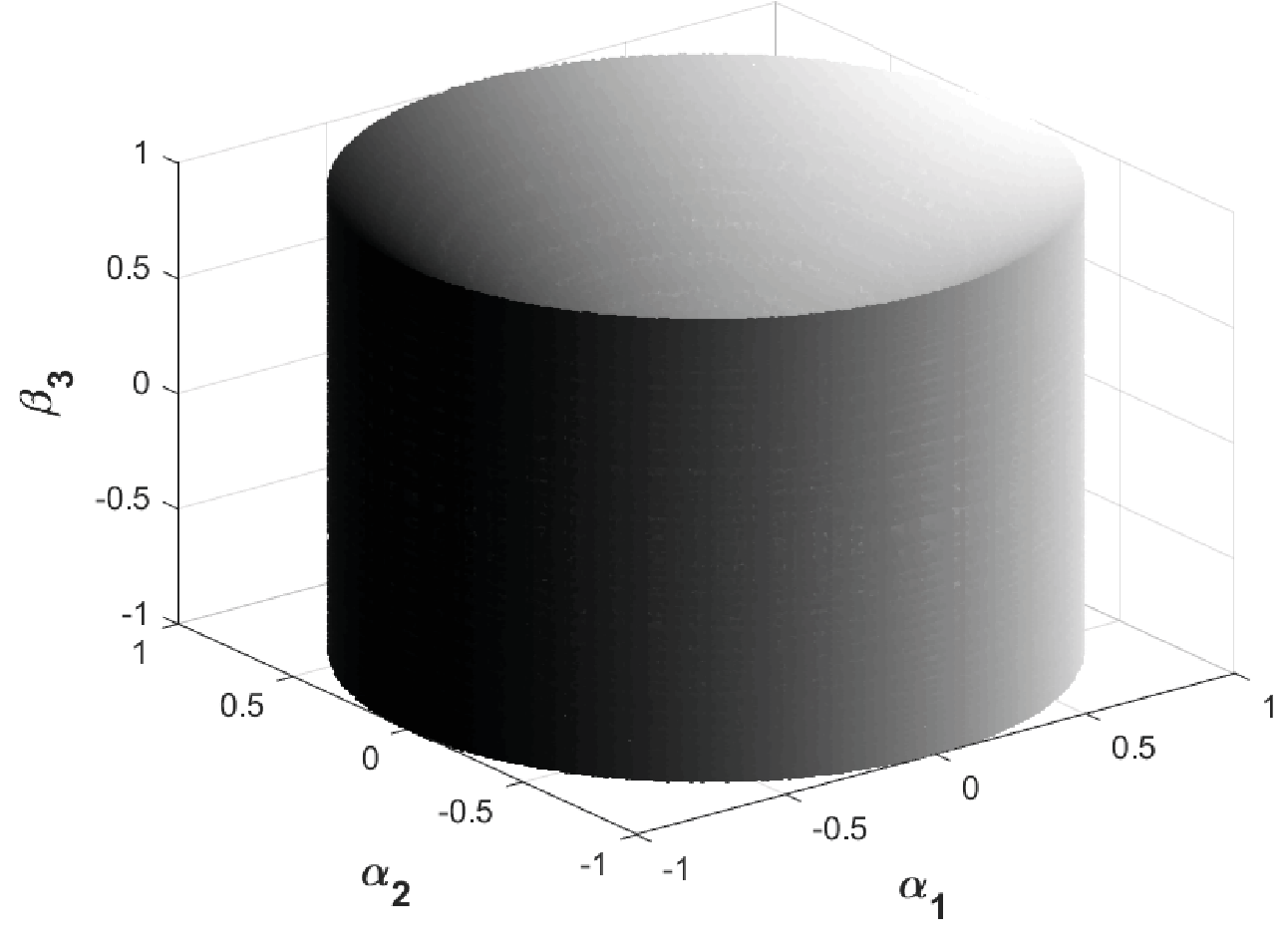}\quad
\includegraphics[width=50mm]{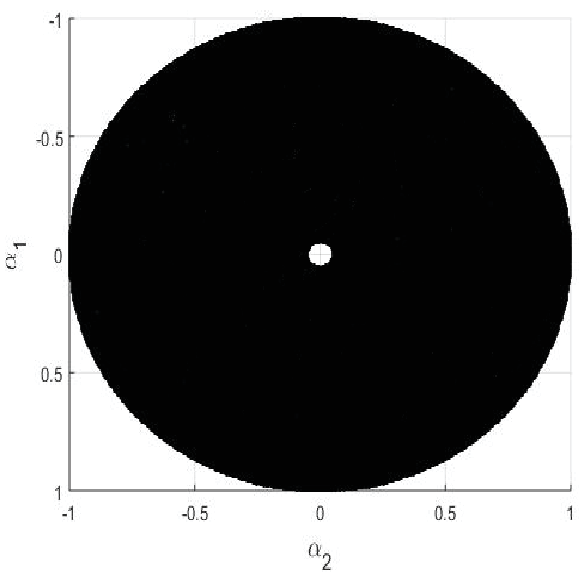}
\caption{\emph{The left image shows a typical $3D$ submanifold of
the $\breve{\mathcal{D}}$ family that has a topology. The right figure shows the
projection of this submanifold on the plane $(\alpha_1,\alpha_2)$.
 Recall that the projections of the submanifold on the plane $(\alpha_2,
\beta_3)$ and $(\alpha_1,\beta_3)$ do not contain topologies.}}
\label{Fig.9}
\end{figure}
We can combine the submanifolds of this family similarly to the family
of sets and form $3D$ - manifold immersed in the space $\mathbb{R}^9$:
\begin{eqnarray}
\mathfrak{S}^{(3)}_{\breve{\mathcal{A}}}=\bigcup_{\alpha\in {\breve{\mathcal{A}}}}\mathfrak{S}^{(3)}_\alpha
=\bigl\{\{\mathcal{x}\}|\,\exists\,\alpha\in {\breve{\mathcal{A}}},\,\{\mathfrak{x}\}\in {\breve{\mathcal{A}}}\bigr\},
\label{ap01}
 \end{eqnarray}
where
$
\{\mathfrak{x}\}=\bigl[(\alpha_1,\alpha_2,\alpha_3),(\beta_1,\beta_2,\beta_3), (\zeta_1,\zeta_2,\zeta_3),
(\alpha_1,\beta_1,\zeta_1),(\alpha_2,\beta_2,\zeta_2),(\alpha_3,\beta_3,\zeta_3)\bigr].
$

The second family of $\breve{\mathcal{B}}$ also consists of six submanifolds
$\breve{\mathcal{B}}=\overline{\breve{\mathcal{B}}_1,\,\breve{\mathcal{B}}_6}$ (see FIG. 8).
The united manifold in this case has the form:
\begin{eqnarray}
\mathfrak{S}^{(3)}_{\breve{\mathcal{B}}}=\bigcup_{\alpha\in{\breve{\mathcal{B}}}}\mathfrak{S}^{(3)}_\alpha
=\bigl\{\{\mathfrak{y}\}|\,\exists\,\alpha\in{\breve{\mathcal{B}}},\,\{\mathfrak{y}\}\in{\breve{\mathcal{B}}}\bigr\},
\label{ap02}
 \end{eqnarray}
where
$
\{\mathfrak{y}\}=\bigl[(\alpha_1,\beta_2,\zeta_3),(\alpha_1,\beta_3,\zeta_2),(\alpha_2,\beta_3,\zeta_1),
(\alpha_2,\beta_1,\zeta_3),(\alpha_3,\beta_1,\zeta_2),(\alpha_3,\beta_2,\zeta_1)\bigr].
$

The third $\breve{\mathcal{C}}=\overline{\breve{\mathcal{C}}_1,\,\breve{\mathcal{C}}_{36}}$
and  fourth
$\breve{\mathcal{D}}=\overline{\breve{\mathcal{D}}_1,\,\breve{\mathcal{D}}_{36}}$
families (see FIG. 9 and FIG. 10), each of which individually consists of 36
submanifolds, can be combined similarly to the previous cases. In particular:
\begin{eqnarray}
\mathfrak{S}^{(3)}_{\breve{\mathcal{G}}}=\bigcup_{\alpha\in {\breve{\mathcal{G}}}}
\mathfrak{S}^{(3)}_\alpha=\bigl\{\{\mathfrak{t}\}|\,\exists\,
\alpha\in{\breve{\mathcal{G}}},\,\{\mathfrak{t}\}\in{\breve{\mathcal{G}}}\bigr\},
\label{ap01}
 \end{eqnarray}
where ${\breve{\mathcal{G}}}=({\breve{\mathcal{C}}},{\breve{\mathcal{D}}})$ and $\{\mathfrak{t}\}
=\bigl(\{\mathfrak{u}\},\{\mathfrak{v}\}\bigr)$, in addition:
$$
\{\mathfrak{u}\}=\bigl[(\alpha_1,\alpha_2,\beta_3),(\beta_3,\zeta_1,\zeta_2),(\beta_2,\zeta_1,\zeta_3),
(\beta_2,\beta_3,\zeta_1),(\beta_1,\zeta_2,\zeta_3),(\beta_1,\beta_3,\zeta_2),(\beta_1,\beta_2,\zeta_3),
$$
$$
\quad\qquad(\alpha_3,\zeta_1,\zeta_2),(\alpha_3,\beta_3,\zeta_2),(\alpha_3,\beta_3,\zeta_1),(\alpha_3,\beta_2,\zeta_3),
(\alpha_3,\beta_2,\zeta_2),(\alpha_3,\beta_1,\zeta_3),(\alpha_3,\beta_1,\zeta_1),
$$
$$
\qquad\quad(\alpha_3,\beta_1,\zeta_2),(\alpha_2,\zeta_1,\zeta_3),(\alpha_2,\beta_3,\zeta_3),(\alpha_2,\beta_3,\zeta_2),
(\alpha_2,\beta_2,\zeta_3),(\alpha_2,\beta_2,\zeta_1),(\alpha_2,\beta_1,\zeta_2),
$$
$$
\qquad\quad(\alpha_2,\beta_1,\zeta_1),(\alpha_2,\beta_1,\beta_3),(\alpha_2,\alpha_3,\zeta_1),(\alpha_2,\alpha_3,\beta_1),
(\alpha_1,\zeta_2,\zeta_3),(\alpha_1,\beta_3,\zeta_3),(\alpha_1,\beta_3,\zeta_1),
$$
$$
\quad\qquad(\alpha_1,\beta_2,\zeta_2),(\alpha_1,\beta_2,\zeta_1),(\alpha_1,\beta_2,\beta_3),(\alpha_1,\beta_1,\zeta_3),
(\alpha_1,\beta_1,\zeta_2),(\alpha_1,\alpha_3,\zeta_2),(\alpha_1,\alpha_3,\beta_2),
$$
$$
(\alpha_1,\alpha_2,\zeta_3)\bigr],
$$
and
$$
\{\mathfrak{v}\}=\bigl[(\alpha_1,\beta_1,\beta_2),\,(\alpha_1,\alpha_2,\beta_2),\,(\alpha_1,\alpha_2,\beta_1),
(\beta_3,\zeta_2,\zeta_3),(\beta_3,\zeta_1,\zeta_3),(\beta_2,\zeta_2,\zeta_3),(\beta_2,\zeta_1,\zeta_2),
$$
$$
\quad\qquad(\beta_2,\beta_3,\zeta_3),\,(\beta_2,\beta_3,\zeta_2),\,(\beta_1,\zeta_1,\zeta_3),\,(\beta_1,\zeta_1,\zeta_2),
\,(\beta_1,\beta_3,\zeta_3),\,(\beta_1,\beta_3,\zeta_1),(\beta_1,\beta_2,\zeta_2),
$$
$$
\qquad\quad(\beta_1,\beta_2,\zeta_1),\,(\alpha_3,\zeta_2,\zeta_3),\,(\alpha_3,\zeta_1,\zeta_3),
\,(\alpha_3,\beta_2,\beta_3),
\,(\alpha_3,\beta_1,\beta_3),(\alpha_2,\zeta_2,\zeta_3),\,(\alpha_2,\zeta_1,\zeta_2),
$$
$$
\qquad\quad(\alpha_2,\beta_2,\beta_3),(\alpha_2,\beta_1,\beta_2),(\alpha_2,\alpha_3,\zeta_3),
(\alpha_2,\alpha_3,\zeta_2),(\alpha_2,\alpha_3,\beta_3),(\alpha_2,\alpha_3,\beta_2),
(\alpha_1,\zeta_1,\zeta_3),
$$
$$
\quad\qquad(\alpha_1,\zeta_1,\zeta_2),(\alpha_1,\beta_1,\beta_3),(\alpha_1,\alpha_3,\zeta_3),
(\alpha_1,\alpha_3,\zeta_1),
(\alpha_1,\alpha_3,\beta_3),(\alpha_1,\alpha_3,\beta_1),(\alpha_1,\alpha_2,\zeta_2),
$$
$$
(\alpha_1,\alpha_2,\zeta_1)\bigr].
$$
Finally, we can combine all the manifolds and find the $3D$ manifold that
is immersed in the configuration space $9D$:
\begin{eqnarray}
\mathfrak{S}^{(3)}=\bigcup_{\alpha\in (\breve{\mathcal{A}},\breve{\mathcal{B}},\breve{\mathcal{C}},\breve{\mathcal{D}})}\mathfrak{S}^{(3)}_\alpha
=\bigl\{\{\mathfrak{l}\}|\,\exists\,
\alpha\in ({\breve{\mathcal{A}}},{\breve{\mathcal{B}}},{\breve{\mathcal{C}}},{\breve{\mathcal{D}}}),
\,\{\mathfrak{l}\}\in ({\breve{\mathcal{A}}},{\breve{\mathcal{B}}},{\breve{\mathcal{C}}},{\breve{\mathcal{D}}})\bigr\},
\label{ap04}
 \end{eqnarray}
where $\{\mathfrak{l}\}=\bigl(\{\mathfrak{x}\},\{\mathfrak{y}\},\{\mathfrak{u}\},\{\mathfrak{v}\}\bigr).$

\subsection{  }
Since the existence of \emph{inverse coordinate transformations} is very important for the
proof of the proposition, we now consider the system of algebraic equations (\ref{13}).

Let us make the following  notations:
\begin{eqnarray}
\bar{\alpha}_\mu=x^{1}_{\,\,,\,\mu},\quad \bar{\beta}_\mu=x^{2}_{\,\,,\,\mu}, \quad
\bar{\zeta}_\mu=x^{3}_{\,\,,\,\mu}, \quad \bar{u}_\mu=x^{4}_{\,\,,\,\mu},\quad
\bar{v}_\mu=x^{5}_{\,\,,\,\mu},\quad \bar{w}_\mu=x^{6}_{\,\,,\,\mu}.
\label{20a}
\end{eqnarray}
In addition, we require the following conditions to be fulfilled:
\begin{eqnarray}
\bar{\alpha}_4=\bar{\alpha}_5=\bar{\alpha}_6=0, \quad \bar{\beta}_4
 =\bar{\beta}_5= \bar{\beta}_6= 0, \quad
\bar{\zeta}_4=\bar{\zeta}_5=\bar{\zeta}_6=0,\,
\nonumber\\
\bar{u}_1=\bar{u}_2=\bar{u}_3=0,\quad
\bar{v}_1=\bar{v}_2=\bar{v}_3=0,\quad
\bar{w}_1=\bar{w}_2=\bar{w}_3=0.
 \label{21a}
\end{eqnarray}
Now, performing similar arguments and calculations, as in the case of direct coordinate
transformations, from (\ref{13}) it is easy to get the following two systems of algebraic equations:
\begin{eqnarray}
\bar{\alpha}_1^2+\bar{\beta}_1^2+ \bar{\zeta}_1^2= \frac{1}{{g}(\{\bar{x}\})},\qquad
\bar{\alpha}_1\bar{\alpha}_2+\bar{\beta}_1\bar{\beta}_2+
 \bar{\zeta}_1\bar{\zeta}_2=0,
\nonumber\\
\bar{\alpha}_2^2+\bar{\beta}_2^2+ \bar{\zeta}_2^2=\frac{1}{{g}(\{\bar{x}\})},\qquad
\bar{\alpha}_1\bar{\alpha}_3+\bar{\beta}_1\bar{\beta}_3+ \zeta_1\zeta_3=0,
\nonumber\\
\bar{\alpha}_3^2+\bar{\beta}_3^2+ \bar{\zeta}_3^2= \frac{\zeta^{33}}{{g}(\{\bar{x}\})},\qquad
\bar{\alpha}_2\bar{\alpha}_3+\bar{\beta}_2\bar{\beta}_3+
\bar{\zeta}_2\zeta_3=0,
\label{22a}
\end{eqnarray}
and, correspondingly:
\begin{eqnarray}
\bar{u}_4^2+ \bar{v}_4^2+ \bar{w}_4^2 = \gamma^{44}{g}^{-1}(\{\bar{x}\}),
\qquad \bar{u}_4\bar{u}_5+\bar{v}_4\bar{v}_5+
\bar{w}_4\bar{w}_5=\gamma^{45}{g}^{-1}(\{\bar{x}\}),
\nonumber\\
\bar{u}_5^2+\bar{v}_5^2+\bar{w}_5^2 =\gamma^{55}{g}^{-1}(\{\bar{x}\}),
\qquad \bar{u}_4\bar{u}_6+\bar{v}_4\bar{v}_6+
\bar{w}_4\bar{w}_6= \gamma^{46}{g}^{-1}(\{\bar{x}\}),
\nonumber\\
 \bar{u}_6^2+ \bar{v}_6^2+\bar{w}_6^2 =\gamma^{66}{g}^{-1}(\{\bar{x}\}),
 \qquad \bar{u}_5\bar{u}_6+\bar{v}_5\bar{v}_6+
\bar{w}_5\bar{w}_6= \gamma^{56}{g}^{-1}(\{\bar{x}\}),
 \label{23a}
\end{eqnarray}
where $f^{-1}:g(\{\bar{x}\})\mapsto \breve{g}(\{\bar{\rho}\})$.\\
In particular, systems of algebraic equations (\ref{22a}) and (\ref{23a}), as in the
case direct coordinate transformations (see (\ref{22}) and (\ref{23})), generate two
3$D$ manifolds $\bar{\mathfrak{S}}^{(3)}$ and $\bar{\mathfrak{R}}^{(3)}$, respectively.

Thus, we have proved that there are also inverse coordinate transformations.

\subsection{ }
As mentioned (see  (\ref{22ztw})), the vector $\textbf{X}$ consists of 18
independent components. Its transposed form looks like this:
$$
\textbf{X}^T=\bigl(\alpha_{11}, \alpha_{12}, \alpha_{13}, \alpha_{22}, \alpha_{23}, \alpha_{33}, \beta_{11}, \beta_{12}, \beta_{13},
 \beta_{22}, \beta_{23}, \beta_{33}, \zeta_{11}, \zeta_{12},  \zeta_{13}, \zeta_{22}, \zeta_{23}, \zeta_{33}
 \bigr).
 $$
 Taking into account the form of the vector $\textbf{X}$, we can write the explicit form of the basic matrix:
\begin{equation}
\mathbb{A}=\left(
  \begin{array}{cccccccccccccccccc}
d_{\,1}^{\,1} & 0 &0 & 0 & 0 & 0&d_{\,1}^{\,7} & 0 & 0 & 0 & 0 &0 &d_{\,1}^{13} & 0 & 0 & 0 &0 & 0 \\
0 &d_{\,2}^{\,2} & 0 &0 & 0 & 0 &0&d_{\,2}^{\,8} & 0 & 0 & 0 & 0 & 0 & d_{\,2}^{14} & 0 & 0 &0 & 0 \\
0 & 0 &d_{\,3}^{\,3} & 0 & 0 & 0 & 0 &0 &d_{\,3}^{\,9}&0 & 0 & 0 & 0 & 0 &d_{\,3}^{15} & 0 & 0 & 0 \\
0 &d_{\,4}^{\,2} & 0 & 0 &0 &0& 0&d_{\,4}^{\,8}& 0 & 0 & 0&0& 0 &d_{\,4}^{14} & 0 & 0 & 0 & 0 \\
0 & 0 & 0 & d_{\,5}^{\,4} & 0 & 0 & 0 & 0 & 0 & d_{\,5}^{10} & 0 & 0 & 0 & 0 & 0 &d_{\,5}^{16} &0 &0 \\
0 & 0 & 0 & 0 & d_{\,6}^{\,5} & 0 & 0 & 0 & 0 & 0 &d_{\,6}^{11} & 0 & 0 &0 &0 & 0 &d_{\,6}^{17} & 0\\
0& 0&d_{\,7}^{\,3} & 0 & 0 &0 & 0 & 0 &d_{\,7}^{\,9} & 0 & 0 & 0 & 0 & 0 &d_{\,7}^{15} & 0 &0 & 0 \\
0 & 0 & 0 & 0 & d_{\,8}^{\,5} & 0 & 0 & 0 & 0 & 0 & d_{\,8}^{11} & 0 & 0 & 0 & 0 & 0 &d_{\,8}^{17} & 0 \\
0 & 0 & 0 & 0 & 0 & d_9^{\,6} & 0 & 0 & 0 &0 & 0 &d_{\,9}^{12}& 0 & 0 & 0 & 0 & 0 & d_{\,9}^{18} \\
d_{10}^{\,1} & d_{10}^{\,2} & 0 & 0 &0 &0&d_{10}^{\,7} &d_{10}^{\,8}& 0 & 0 & 0 & 0 &d_{10}^{13}
 & 0 & 0 & 0 & d_{10}^{17} & 0 \\
0 & d_{11}^{\,2} & 0 &d_{11}^{\,4} & 0 & 0 & 0 & d_{11}^{\,8}& 0 & d_{11}^{10} & 0 & 0 & 0
 &d_{11}^{14} & 0 & d_{11}^{16} &0 &0 \\
0 & 0 &d_{12}^{\,3} & 0 & d_{12}^{\,5}& 0 & 0 & 0 & d_{12}^{\,9} & 0 &d_{12}^{11} & 0 & 0
&0 &d_{12}^{15} & 0&d_{12}^{17} & 0\\
d_{13}^{\,1}& 0&d_{13}^{\,3}& 0 & 0 & 0 &d_{13}^{\,7} & 0 &d_{13}^{\,9} & 0 & 0 & 0 &
d_{13}^{13} & 0
& d_{13}^{15} & 0 &0 & 0 \\
0 &d_{14}^{\,2} & 0 &0&d_{14}^{5}& 0& 0 &d_{14}^{\,8} & 0 & 0 &d_{14}^{11} & 0 & 0 &
d_{14}^{14} & 0 & 0 &d_{14}^{17} & 0 \\
0 & 0 &d_{15}^{\,3}& 0 & 0 &d_{15}^{\,6}&0&0& d_{15}^{\,9} &0 & 0 &d_{15}^{12}& 0 & 0 &
d_{15}^{15}& 0 & 0 & d_{15}^{18} \\
0 &d_{16}^{\,2}&d_{16}^{\,3}& 0 &0 &0& 0 &d_{16}^{\,8}&d_{16}^{\,9}&0 &0 &0 & 0 &d_{16}^{14}&
d_{16}^{15} &0& 0 & 0 \\
0 & 0 & 0 &d_{17}^{\,4}&d_{17}^{5}& 0 & 0 & 0 & 0 &d_{17}^{10}& d_{17}^{11}& 0 & 0& 0 & 0 &
d_{17}^{16}&d_{17}^{17} &0 \\
0 & 0 & 0 & 0 &d_{18}^{\,5}&d_{18}^{\,6}& 0 & 0 & 0 & 0 &d_{18}^{11}&d_{18}^{12}& 0 &0 &0 &0
&d_{18}^{17}&d_{18}^{18}\\
\end{array}
\right),
 \label{19at}
\end{equation}
where the superscript indicates the column number, while the subscript indicates the
line number. As for the explicit form of elements $d^{\,\nu}_\mu=d_{\mu\nu}$, where
$\mu,\nu=\overline{1,18}$, then we can find they by multiplying the basic matrix
$\mathbb{A}$ with the vector $\textbf{X}$ (see equation (\ref{22ztw})) and
comparing with the system of equations (\ref{22zt}). 

In particular, it is easy to verify these terms are equal:
$$
d^{\,1}_{\,1}\,=\,d^{\,2}_{\,2}\,=\,d^{\,3}_{\,3}\,=2d^{\,2}_{\,10}=2d^{\,4}_{11}=2d^{\,5}_{12}=
2d^{\,3}_{13}=\,2d^{\,5}_{14}=2d^{\,6}_{15}=2\alpha_1,\quad\,
$$
$$
d^{\,2}_{\,4}\,=\,d^{\,4}_{\,5}\,=\,d^{\,5}_{\,6}\,=\,2d^{\,1}_{10}=2d^{\,2}_{11}=2d^{\,3}_{12}=
2d^{\,3}_{16}=2d^{\,5}_{17}=2d_{18}^{\,6}=2\alpha_2,\quad\,
$$
$$
d^{\,3}_{\,7}\,=\,d^{\,5}_{\,8}\,=\,d^{\,6}_{\,9}\,=\,2d^{\,1}_{13}=2d^{\,2}_{14}=2d^{\,3}_{15}=
2d^{\,2}_{16}=2d^{\,4}_{17}=2d^{\,5}_{18}=2\alpha_3,\quad\,
$$
$$
 d^{\,1}_{\,7}\,=\,d^{\,2}_{\,8}\,=\,d_{\,9}^{\,3}=\,2d^{\,8}_{10}\,=2d^{10}_{11}=2d^{11}_{12}=
 2d^{\,9}_{13}=2d^{11}_{14}=2d^{12}_{15}
 =2\beta_1,\quad\,
$$
$$
 d^{\,8}_{\,4}= d^{10}_{\,5}=d_{11}^{\,6}=2d^{\,7}_{10}=2d^{\,8}_{11}=2d^{\,9}_{12}=
 2d^{\,9}_{16}\,=2d^{11}_{17}\,=2d^{12}_{18}=2\beta_2,\quad\,
$$
\begin{eqnarray}
 d^{\,9}_{\,7}\,=d^{11}_{\,6}=\, d_{12}^{\,6}=2d^{\,7}_{13}=2d^{\,8}_{14}=2d^{\,9}_{15}=
 2d^{\,8}_{16}=2d^{10}_{17}=2d^{11}_{18}= 2\beta_3,\quad\,
 \nonumber\\
 d^{\,1}_{13}=d^{\,2}_{14}=d_{15}^{\,3}=2d^{17}_{10}=2d^{16}_{11}=2d^{17}_{12}=
 2d^{15}_{13}=2d^{17}_{14}=2d^{18}_{15}=2\gamma^{33}\zeta_1,
 \nonumber\\
 d^{14}_{\,4}=d^{16}_{5}=d_{17}^{\,6}=2d^{13}_{\,10}=2d^{14}_{11}=2d^{15}_{12}=
 2d^{13}_{13}=2d^{14}_{14}=2d^{15}_{15}=2\gamma^{33}\zeta_2,
 \nonumber\\
 d^{15}_{\,7}=d^{17}_{\,8}=d_{18}^{\,9}=2d^{14}_{16}=2d^{16}_{17}=2d^{17}_{18}=
 2d^{15}_{16}=2d^{17}_{17}=2d^{18}_{18}=\, 2\gamma^{33}\zeta_3.
\label{23cat}
\end{eqnarray}

As is known, the algebraic system (\ref{22zt}) or (\ref{22ztw}) does not have a
solution in the case when the determinant of the matrix is zero
$\det(\mathbb{A})=\det(d_{\mu\nu})=0.$  A class consisting of sets of coefficients
 $\{\sigma\} $ for which the determinant is zero can be countable and the
 measure, respectively,  will be equal to zero $\mathfrak{W}= \oslash$.

\subsection{}
Let us consider third-order matrices $\Delta_i(\{\bar{x}\})\,\,(i=\overline{1,3}\,)$,
that are included in the solutions of the system of  algebraic equations (\ref{07kt}):
\begin{eqnarray}
\Delta_1=\left|\begin{array}{ccc}
\delta &2\xi^1\xi^2 & 2\xi^1\xi^3\\
\delta& K_2 & 2\xi^2\xi^3 \\
\delta & 2\xi^2\xi^3 & K_3
\end{array}
\right|,\,\,
\Delta_2=
\left|\begin{array}{ccc}
K_1 &\delta & 2\xi^1\xi^3\\
2\xi^1\xi^2&\delta& 2\xi^2\xi^3 \\
2\xi^1\xi^3 & \delta & K_3
\end{array}\right|,\,\,
\Delta_3=
\left|\begin{array}{ccc}
K_1 &2\xi^1\xi^2 & \delta\\
2\xi^1\xi^2& K_2 & \delta\\
2\xi^1\xi^3 & 2\xi^2\xi^3 & \delta
\end{array}\right|.
\label{7wtk}
\end{eqnarray}
By calculating these determinants we get:
\begin{eqnarray}
\Delta_1(\{\bar{x}\})= \delta\cdot\bigl\{K_2K_3-2\xi^1[\xi^2K_3+\xi^3K_2]+
4\xi^2\xi^3[\xi^1(\xi^2+\xi^3)-\xi^2\xi^3]\bigr\},
\nonumber\\
\Delta_2(\{\bar{x}\})=\delta\cdot\bigl\{K_1K_3-2\xi^2[\xi^1K_3+\xi^3K_1]+
4\xi^1\xi^3[\xi^2(\xi^1+\xi^3)-\xi^1\xi^3]\bigr\},
\nonumber\\
\Delta_3(\{\bar{x}\})=\delta\cdot\bigl\{K_1K_2-2\xi^3[\xi^2K_1+\xi^1K_2]+
4\xi^1\xi^2[\xi^3(\xi^1+\xi^2)-\xi^1\xi^2]\bigr\}.
\label{07wtk}
\end{eqnarray}
The main determinant $\Delta(\{\bar{x}\})$ (see (\ref{07wqtk})) is easy to to calculate:
\begin{eqnarray}
\Delta(\{\bar{x}\})=K_1 K_2 K_3-4\bigl[(\xi^2\xi^3)^2K_1+(\xi^1\xi^3)^2K_2+
(\xi^1\xi^2)^2K_3\bigl]+16(\xi^1\xi^2\xi^3)^2.
\label{07al}
\end{eqnarray}
In a coupled system,  given the conditions  $\ddot{x}^i=0\,\,(i=\overline{1,3}\,)$,
bodies can have different constant velocities $\xi^i=const_i\,\,(i=\overline{1,3})$
depending on their mass. To simplify  the determinant $\Delta(\{\bar{x}\})$, it is useful to introduce
 two  new parameters;  $\alpha=(const_2)^2= (\xi^2/\xi^1)^2$ and $\beta=(const_3)^2=(\xi^3/\xi^1)^2$,  and
also notation $(\xi^1)^2=(const_1)^2=y>0$. In addition, we  assume that;  $(\xi^1)^2\geq[(\xi^2)^2, (\xi^3)^2]$,
from which follows that parameters $\alpha,\beta\in[0,1]$.

Using these notations,  we can represent the expression (\ref{07al}) in the form of a third-order polynomial:
 \begin{eqnarray}
\Delta(\{\bar{x}\})=\mathcal{A}y^3
 +\mathcal{B} y^2+\mathcal{C}y-\Lambda^6,
\label{07alb}
\end{eqnarray}
where
$$
\mathcal{A}=\bigl\{12\alpha^2\beta^2+(1-\alpha^2-\beta^2)(1+\alpha^2-\beta^2)(1-\alpha^2+\beta^2)+
4(\alpha^2+\beta^2)(1+\alpha^2\beta^2)+ 4(\alpha^2-\beta^2)^2 \bigr\},
$$
$$
\mathcal{B}=\bigl\{1+2(\alpha^2+\beta^2)+(\alpha^2+\beta^2)^2\bigr\}\Lambda^2,\qquad \mathcal{C}=
 -(1+\alpha^2+\beta^2)\Lambda^4.
$$
Now to eliminate  uncertainties like $0/0$  in expressions (\ref{07wtb}), we need to find the conditions, that is,
the parameters $\alpha$ and $\beta$, for which  $\Delta(\{\bar{x}\})\sim\delta$, and later $\delta\to 0$.

Let us consider  the cubic equation:
 \begin{equation}
 \Delta(\{\bar{x}\})=0.
 \label{06alb}
\end{equation}
To find the roots of the cubic equation (\ref{06alb}), it is convenient to use the   Vieta trigonometric formula.
Recall that the determinant  of the equation (\ref{06alb}) has the following form:
 $$\mathcal{D} = \mathcal{Q}^3-\mathcal{R}^2,$$
where
 $\mathcal{Q}=\bigl([\mathcal{B}/\mathcal{A}]^3 -3[\mathcal{C}/\mathcal{A}]\bigr)/9 $
and $\mathcal{R}=\bigl(2[\mathcal{B}/\mathcal{A}]^2-9[\mathcal{B}\mathcal{C}/\mathcal{A}^2]
 -27\Lambda^6/\mathcal{A}\bigr)/54 $.\\
According to the analysis, depending on the values of the parameters $\alpha$ and
 $\beta$, three cases are possible for determinant $\mathcal{D}$.

Case 1: When $\mathcal{D}>0$,  there are three real solutions:
\begin{eqnarray}
 y_1=-2\sqrt{\mathcal{Q}}\cos(\phi)-\mathcal{B}/(3\mathcal{A}), \qquad\qquad\qquad\qquad\qquad\qquad\qquad\quad\,\,\,
 \nonumber\\
 y_2=-2\sqrt{\mathcal{Q}}\cos(\phi+2\pi/3)-\mathcal{B}/(3\mathcal{A}),\qquad\qquad\qquad\qquad\qquad\qquad\,
 \nonumber\\
 y_3=-2\sqrt{\mathcal{Q}}\cos(\phi-2\pi/3)-\mathcal{B}/(3\mathcal{A}),\quad \phi=
 \bigl[\arccos(\mathcal{R}/\mathcal{Q}^{3/2})\bigr]/3.\,\,
 \label{06blt}
\end{eqnarray}

Case 2:  When $\mathcal{D}<0$, depending on the sign of the parameter $\mathcal{Q}$, there are three
possible solutions.

$\bullet\,\, \mathcal{Q}>0$, there is one real solution:
 \begin{equation}
 y=-2\mathrm{sgn}(\mathcal{R})|\mathcal{Q}|^{1/2}\cosh(\phi)-\mathcal{B}/(3\mathcal{A}),
 \quad \phi=\bigr[\mathrm{Arch}\bigl(|\mathcal{R}|/|\mathcal{Q}|^{3/2}\bigr)\bigr]/3.
 \label{06altk}
\end{equation}

$\bullet\,\,\mathcal{Q}<0$, in this case, the real solution is:
 \begin{equation}
 y=-2\mathrm{sgn}(\mathcal{R})|\mathcal{Q}|^{1/2} \sinh(\phi)-\mathcal{B}/(3\mathcal{A}),
 \quad \phi=\bigr[\mathrm{Arsh}\bigl(|\mathcal{R}|/|\mathcal{Q}|^{3/2}\bigr)\bigr]/3.
 \label{06altk}
\end{equation}

$\bullet\,\,\mathcal{Q}=0$, in this case, the real solution, accordingly, has the form:
\begin{equation}
 y=\Bigl( {\Lambda^6}/{\mathcal{A}}+\bigl[{\mathcal{B}/}{3\mathcal{A}}\bigr]^3\Bigr)^{1/3}.
 \label{06altk}
\end{equation}

 Case 3: When $\mathcal{D}=0$, there are three real solutions, however, two of them coincide:
\begin{equation}
 y_1=-2\mathcal{R}^{1/3}-\mathcal{B}/(3\mathcal{A}),\qquad  y_2= y_3=\mathcal{R}^{1/3}
 -\mathcal{B}/(3\mathcal{A}).
 \label{06bltz}
\end{equation}

Below, as an example, we will analyze case 1, i.e. when $\mathcal{D}>0$.\\
Taking into account the solutions (\ref{06blt}), the determinant $\Delta(\{\bar{x}\})$
can be represented as:
\begin{equation}
 \Delta(\{\bar{x}\})=(y-y_1)(y-y_2)(y-y_3).
 \label{6blt}
\end{equation}
Consider solutions (\ref{07wqtk}) near the value:
\begin{equation}
y=y_1\pm\delta.
 \label{6ablt}
\end{equation}
Using (\ref{6ablt}) and (\ref{07wtk})-(\ref{07al}) for solutions (\ref{07wtb}),   we  obtain the
following expressions:
\begin{eqnarray}
 a_1(\{\bar{x}\})=\pm \frac{K_2K_3-2(y_1\pm\delta)^2[\alpha K_3+\beta K_2]+4\alpha\beta(y_1\pm\delta)^4[\alpha
 +\beta-\alpha\beta]}{(y_2-y_1\pm\delta)(y_3-y_1\pm\delta)},
 \nonumber\\
 a_2(\{\bar{x}\})=\, \pm\, \frac{K_1K_3-2\alpha(y_1\pm\delta)^2[ K_3\,+\beta K_1]+4\beta(y_1\pm\delta)^4[\alpha-\beta
+\alpha\beta]}{(y_2-y_1\pm\delta)(y_3-y_1\pm\delta)},
 \nonumber\\
 a_3(\{\bar{x}\})=\pm\, \frac{K_1K_2-2\beta(y_1\pm\delta)^2[\alpha  K_1\,+ K_2]\,+\,4\alpha(y_1\pm\delta)^4[\beta-\alpha
 +\alpha\beta]}{(y_2-y_1\pm\delta)(y_3-y_1\pm\delta)}.
 \label{6abt}
\end{eqnarray}
Now,  making the transition to the limit $\delta\to0 $ in the expressions (\ref{6abt}) for the coefficients
(\ref{07wqtk}), we get clearly defined regular expressions. Assuming that $y_1(\{\bar{x}\})=\lambda_1=const$, we can
generate by this equation 2$D$ surface  in the internal space
$\mathbb{E}^3$,  on which the system of equations (\ref{07kt}) has a solution. Similarly, we can find
solutions of the system of algebraic equations  (\ref{07wqtk}) on $2D$ manifolds generated by equations
$y_2(\{\bar{x}\})=\lambda_2=const$ and $y_3(\{\bar{x}\})=\lambda_3=const$,
respectively.

To analyze the problem, of particular interest is the case when all the masses are the same.
In this case, obviously, $\alpha=\beta=1$, using which  from the equation (\ref{07alb}),
taking into account  (\ref{06alb}), it is easy to find the following cubic equation:
\begin{equation}
27y^3+9\Lambda^2y^2-3\Lambda^4y-\Lambda^6=0,
 \label{a01}
\end{equation}
which can be written as:
\begin{equation}
(3y+\Lambda^2)^2(3y-\Lambda^2)=0.
 \label{a02}
\end{equation}
From the equation (\ref{a02}) it follows that there is only one real solution:
\begin{equation}
y=\Lambda^2(\{\bar{x}\})/3, \quad or\quad \xi^1=const_1=\Lambda(\{\bar{x}\})/\sqrt{3}.
 \label{a03}
\end{equation}
Finally, using (\ref{07wtb}), (\ref{07wtk})-(\ref{07al})   and  (\ref{a03}), we can find
the coefficients of algebraic equation (\ref{07kt}):
$$
a_1(\{\bar{x}\})=a_2(\{\bar{x}\})=a_3(\{\bar{x}\})=\biggl(\frac{K-2y}{\Lambda^2+3y}\biggr)^2=1.
$$
Solving the second equation in (\ref{a03}) for a specific value of $\xi^1=const_1$, we can find
a 2$D$ surface $\Xi$ on which a restricted three-body system with holonomic connections is localized.

For other cases, also using similar reasoning, we can find surfaces on which
configurations with holonomic connections are localized.

\subsection{}
The equation for the covariant derivative (\ref{9abc}) can be written as:
\begin{equation}
\frac{\mathcal{D} F^i}{\mathcal{D}s}=\dot{F}^i+Y^i, \qquad
Y^i=\Gamma^i_{j\,l}(\{\bar{x}\})\dot{x}^jF^l,\qquad \dot{q}=\frac {d q}{ds},\quad i,j,l=\overline{1,3},
 \label{24a}
\end{equation}
where $Y^i\in \mathcal{M}^{(3)}$ is a component of the 3$D$ vector.

Using (\ref{24a}), we can calculate the covariant derivative of the second order:
\begin{eqnarray}
\frac{\mathcal{D}^2 \zeta^i}{\mathcal{D}s^2}\,=\,\ddot{\zeta}^i\,+\,\Gamma^i_{j\,l}\dot{x}^j\dot{\zeta}^l
+\dot{Y}^i\,+\,\Gamma^i_{j\,l}\dot{x}^jY^l\,=\,\ddot{\zeta}^i\,+
\Gamma^i_{j\,l}\dot{x}^j\dot{\zeta}^l\,+\,\frac{d}{ds}\bigl(\Gamma^i_{j\,l}\dot{x}^j\zeta^l\bigr)\,+\qquad
\nonumber\\
\Gamma^i_{j\,l}  \Gamma^l_{k\,p}\dot{x}^j\dot{x}^k\zeta^p
=\ddot{\zeta}^i+2\Gamma^i_{j\,l}\dot{x}^j\dot{\zeta}^l+\dot{\Gamma}^i_{j\,l}\dot{x}^j\zeta^l
+\Gamma^i_{j\,l}\ddot{x}^j\zeta^l+\Gamma^i_{j\,l}\Gamma^l_{k\,p}\dot{x}^j\dot{x}^k\zeta^p
\,\quad
\nonumber\\
=\ddot{\zeta}^i+2\Gamma^i_{j\,l}\dot{x}^j\dot{\zeta}^l
+(\dot{\Gamma}^i_{j\,l}\dot{x}^j\zeta^l-
\Gamma^i_{j\,l}\Gamma^j_{k\,p}\dot{x}^k\dot{x}^p\zeta^l+\Gamma^i_{j\,n}\Gamma^n_{k\,p}
\dot{x}^j\dot{x}^k\zeta^p),\quad
 \label{24b}
\end{eqnarray}
where $k,n,p=\overline{1,3}$. In addition:
\begin{eqnarray}
\Gamma^i_{j\,l} =\frac{1}{2}g^{ip}
\bigl(\partial_{l}g_{p j}+\partial_{j}g_{lp}-\partial_{p}g_{jl}\bigr)=-\delta_l^i a_j-
\delta_j^i a_l+\delta^{ip}\delta_{jl}a_p,\quad a_k=-\frac{1}{2}\partial_{x_k}\ln g,
\label{24ab}
\end{eqnarray}
\begin{eqnarray}
\dot{\Gamma}^i_{j\,l}=\frac{d\Gamma^i_{j\,l}}{ds}=\frac{1}{2}\dot{g}^{ip}
\bigl(\partial_{l}g_{p j}+\partial_{j}g_{lp}-\partial_{p}g_{jl}\bigr)+
\frac{1}{2}{g}^{ip}\bigl(\partial_{l}\dot{g}_{p j}+\partial_{j}\dot{g}_{lp}
-\partial_{p}\dot{g}_{jl}\bigr)\,
\nonumber\\
=\frac{1}{g}\Bigl(\sum_{k=1}^3a_k\dot{x}^k\Bigr)\Bigl[\bigl(\delta^i_{j}a_l+\delta^i_{p}a_j-
\delta^{ip}\delta_{j\,l}a_p\bigr)
-\bigl(\delta^i_{j}b_l+\delta^i_{p}b_j-\delta^{ip}\delta_{jl}b_p\bigr)\Bigl]\,
\nonumber\\
=\frac{1}{g}\Bigl(\sum_{k=1}^3a_k\dot{x}^k\Bigr)\Bigl[\bigl(\delta^i_{j}(a_l-b_l)+
\delta^i_{p}(a_j-b_j)-\delta^{ip}\delta_{jl}(a_p-b_p)\Bigl],
\label{24abk}
\end{eqnarray}
where $b_k=-(1/2)\partial_{x_k}\ln\bigl|\sum_{i=1}^3g_{;i}\dot{x}^i\bigr|$ and
$g_{;\,k}=\partial g/\partial x^k.$

\subsection{}
Substituting (\ref{c2lzb}) into the equation (\ref{c0l7b}), we get:
\begin{equation}
\Biggl\{\frac{1}{{\mathrm{r}}^2}\frac{\mathrm{d}}{\mathrm{d}\mathrm{r}}
\biggl({\mathrm{r}}^2\frac{\mathrm{d}}{\mathrm{d}{{\mathrm{r}}}}\biggr)
-\frac{{l(l+1)}}{{\mathrm{r}}^2}+\frac{2}{\hbar^2}\sum_{\bar{l}=0}^{\infty}
\sum_{\bar{m}=-\bar{l}}^{\bar{l}}\Omega_{\bar{l}\bar{m}}({\mathrm{r}}\,;\,\mathrm{E},J,\varepsilon)
Y_{\bar{l}}^{\bar{m}}(\theta,\varphi)\Biggr\}\bar{\Psi}=0,
\label{G01}
\end{equation}
where  $\Omega_{\bar{l}\bar{m}}({\mathrm{r}}\,;\,\mathrm{E},J,\varepsilon)
=\bigl[\mathrm{E}\,\mathfrak{g}^{(1)}_{\bar{l}\bar{m}}(\mathrm{r};\varepsilon)
-J(J+1)\mathfrak{g}^{(2)}_{\bar{l}\bar{m}}(\mathrm{r};\varepsilon)\bigr]$.

To simplify the equation (\ref{G01}), we first multiply it by the complex
conjugate of the spherical function, that is  $Y_{l'}^{m'\ast}(\theta,\varphi)$
then using the well-known orthogonal properties of the spherical functions \cite{Hob}:
$$
\int_0^{2\pi}\int_0^{\pi}{Y_{l}^{ m}(\theta,\varphi){Y_{l'}^{ m'}}^\ast(\theta,\varphi)}
\sin\theta d\theta d\varphi=\delta_{mm'}\delta_{ll'},
$$
we  obtain the following \emph{ordinary differential equation} (ODE)
for the radial component of the wave function:
\begin{eqnarray}
 \Biggl\{\frac{1}{{\mathrm{r}} ^2}\frac{\mathrm{d}}{\mathrm{d} \mathrm{r}}
\biggl({\mathrm{r}} ^2\frac{\mathrm{d}}{\mathrm{d} {{\mathrm{r}} }}\biggr)
- \frac{l(l+1)}{{\mathrm{r}} ^2}\Biggr\}\delta_{mm'}\delta_{ll'}\,\Upsilon =
- \frac{2}{\hbar^2}\sum_{\bar{l}=0}^{\infty} \sum_{\bar{m}=
-\bar{l}}^{\bar{l}}\mathcal{W}_{m,m',\bar{m}\,;\,l,l',\bar{l}}\,
\Omega_{\bar{l}\bar{m}}({\mathrm{r}} ;\, \mathrm{E}, J)\Upsilon,
\label{c2l7b}
\end{eqnarray}
where
$$
\mathcal{W}_{m_1,m_2, m_3\,;\,l_1,l_2,l_3}=\int_0^{2\pi}\int_0^{\pi}
 Y_{l_1}^{ m_1}(\theta,\varphi)Y_{l_2}^{ m_2\ast}(\theta,\varphi)
Y_{l_3}^{ m_3}(\theta,\varphi) \sin\theta{d\theta}d\varphi.
$$
For calculation the integral of the product of three spherical harmonics
$\mathcal{W}_{m_1,m_2,m_3\,;\,l_1,l_2,l_3}$  we will use the following formula
 \cite{Zar}:
\begin{eqnarray}
\int_0^{2\pi}\int_0^{\pi} Y_{l_1 m_1}(\theta,\varphi)Y_{l_2 m_2}(\theta,\varphi)
Y_{l_3 m_3}(\theta,\varphi) \sin\theta{d\theta}d\varphi=\qquad
\nonumber\\
 \sqrt{\frac{(2l_1+1)(2l_2+1)(2l_3+1)}{4\pi}}\,
\biggl(\begin{array}{ccc}
 l_1 & l_2 & l_3 \\
0 & 0 & 0
\end{array}\biggr)
\biggl(\begin{array}{ccc}
 l_1 & l_2 & l_3 \\
 m_1 & m_2 & m_3
\end{array}\biggr),
\label{c2ld7b}
\end{eqnarray}
where $Y_{l m}(\theta,\varphi)$ is the real spherical function,
which can be represented by a complex spherical function $Y_{l}^{m}(\theta,\varphi)$
(see \cite{Hob}) and
$
\biggl(\begin{array}{ccc}
l_1 & l_2 & l_3 \\
m_1 & m_2& m_3
\end{array}\biggr)
$
denotes the Wigner $3j$ symbol (see \cite{Wigner}).
Using the transform:
$$
Y_{l}^{m}(\theta,\varphi)=\left\{
  \begin{array}{ll}
 \,\,\, \frac{1}{\sqrt{2}}\,\bigl(Y_{l|m|}\,-\,iY_{l,-|m|}\bigr), \qquad& m<0, \\
 \,\,\,Y_{0m}, \qquad& m=0, \\
 \frac{(-1)^m}{\sqrt{2}}\bigl(Y_{l|m|}+iY_{l,-|m|}\bigr),\qquad & m>0,
  \end{array}
\right.
$$
we can calculate the function $\mathcal{W}_{m_1,m_2,m_3\,;\,l_1,l_2,l_3}$.

As follows from (\ref{c2l7b}), this equation, depending on the ratios of the quantum
numbers $ m, m ', l $ and $ l' $, can go over into two different equations:

1. Into the algebraic equation:
\begin{equation}
\sum_{\bar{l}=0}^{\infty} \sum_{\bar{m}=
-\bar{l}}^{\bar{l}}\mathcal{W}_{m,m',\bar{m}\,;\,l,l',\bar{l}}\,
\Omega_{\bar{l}\bar{m}}({\mathrm{r}} ;\, \mathrm{E}, J,\varepsilon)=0,
\label{c2ld1b}
\end{equation}
when one of the inequalities holds; $m\neq m'$ or $l\neq l'$, or when take
place of both inequalities $m\neq m'$ and $l\neq l'$, and, accordingly,

2.  into the ODE for the radial wave function of bodies system (see   (\ref{c2lb}) ),
 if $m=m'$ and $l=l'$.

Note that the algebraic equation (\ref{c2ld1b}) generates the discrete set of points
$\mathcal{Y}$ at which the wave function is not defined. However, the cardinality
 of the set $\mathcal{Y}$ with respect to the cardinality of the set that forms the
\emph{internal space} $\mathcal{M}^{(3)}$  is equal to zero. The latter means
 that the wave function of a dynamical system is defined in the space
$\mathcal{M}^{(3)}\setminus\mathcal{Y}. $

 Based on this, below we will calculate only those $3j$ symbols that will
be needed to determine the ODE for the quantum motion (see (\ref{c2lb})).

Case 1. Assuming that $m=m'<0$ and $l=l'$, as well as taking into
account the selection rules for the Wigner $3j$ symbol, we obtain:
\begin{eqnarray}
\mathcal{W}_{m,m,\bar{m};\,l,\,l,\,\bar{l}}=(-1)^{\bar{m}}\frac{2l+1}{4}\sqrt{\frac{2\bar{l}+1}{\pi}}
\biggl(\begin{array}{ccc}
l & l&\bar{l} \\
0 & 0 & 0
\end{array}\biggr)\biggl(\begin{array}{ccc}
l & l&\bar{l} \\
-|m|&-|m|&|\bar{m}|
\end{array}\biggr),\quad \bar{m}>0,
\nonumber\\
\mathcal{W}_{m,m,\bar{m};\,l,\,l,\,\bar{l}}=\frac{2l+1}{4}\sqrt{\frac{2\bar{l}+1}{\pi}}
\biggl(\begin{array}{ccc}
l & l&\bar{l} \\
0 & 0 & 0
\end{array}\biggr)\biggl(\begin{array}{ccc}
l & l&\bar{l} \\
-|m|&-|m|&|\bar{m}|
\end{array}\biggr),\quad    \bar{m}<0.
\label{k01g}
\end{eqnarray}
It is easy to see that the second $3j$ symbol in (\ref{k01g}) is not equal to zero only
if the equality $\bar{m}=2m$ holds. Recall that it follows directly from selection rules.
From this condition, in particular, it follows that the first and second expressions
in (\ref{k01g}) are equal.

Case 2. When $m=m'>0$ and $l=l'$, the Wigner $3j$ symbol is calculated  in the
same way and gives the result similarly (\ref{k01g}).

Case 3. When $\bar{m}=0$, in addition, $m=m'$ and $l=l'$. For this case we obtain:
\begin{eqnarray}
\mathcal{W}_{0,0,0;\,l,l,\bar{l}}=\frac{2l+1}{2}\sqrt{\frac{2\bar{l}+1}{\pi}}
\biggl(\begin{array}{ccc}
l & l&\bar{l} \\
0 & 0 & 0
\end{array}\biggr)^2.
\label{k01tg}
\end{eqnarray}

To calculate the  $3j$ symbol, we turn to the well-known general representation \cite{Ed}:
\begin{eqnarray}
\biggl(\begin{array}{ccc}
 l_1 & l_2 & l_3 \\
 m_1 & m_2 & m_3
\end{array}
\biggr)=\biggl[\frac{(l_1+l_2-l_3)!(l_1-l_2+l_3)!(-l_1+l_2+l_3)!}{(l_1+l_2+l_3+1)!}\biggr]^{1/2}\times
\nonumber\\
\bigl[(l_1+m_1)!(l_1-m_1)!(l_2+m_2)!(l_2-m_2)!(l_3+m_3)!(l_3-m_3)!\bigr]^{1/2}\,\,\times
\nonumber\\
\sum_{\nu}\Bigl\{\bigl(-1\bigr)^{\nu+l_1-l_2-m_1}\,\bigl[\nu!(l_1+l_2-l_3-\nu)!
(l_1-m_1-\nu)!\,\times
\nonumber\\
(l_2+m_2-\nu)!(l_3-l_1-m_2+\nu)!(l_3-l_2+m_1+\nu)!\bigr]^{-1}\Bigr\},
\label{01g}
\end{eqnarray}
where summation over $\nu$ is carried out over all integers.

Using (\ref{01g}) and the selection rules for the Wigner $3j$ symbol,
 we can calculate the following specific $3j$ symbols:
\begin{eqnarray}
\Biggl(\begin{array}{ccc}
 l & l & \bar{l} \\
- |m| &-|m| &|\bar{m}|
\end{array}
\Biggr)\,=\, \biggl[\frac{(2l-\bar{l}\,)!(\bar{l}+
|\bar{m}|)!(\bar{l}-|\bar{m}|)!}{(2l+\bar{l}+1)!}\biggr]^{1/2}\,\bar{l}!\,(l+|m|)!\,(l-|m|)!
\times
\nonumber\\
\sum_{\nu}\frac{\bigl(-1\bigr)^{\nu+|m|}}{\nu!(2l -\bar{l}-\nu)!(l+|m|-\nu)!(l-|m|-\nu)
!(\bar{l}-l+|m|+\nu)!(\bar{l}-l-|m|+\nu)!},
\label{02g}
\end{eqnarray}
and correspondingly:
\begin{eqnarray}
\biggl(\begin{array}{ccc}
 l & l&\bar{l}\\
 0 &0 &0
\end{array}
\biggr)=  \biggl[\frac{(2l-\bar{l}\,)!}{(2l+\bar{l}+1)!}\biggr]^{1/2}\bigl(\bar{l}!l!\bigr)^2
\sum_{\nu}\frac{\bigl(-1\bigr)^{\nu}}{\nu!(2l -\bar{l}-\nu)!\bigl[(l-\nu)!(\bar{l}-l+\nu)!\bigr]^{2}}.
\label{03g}
\end{eqnarray}
 Based on the above  analysis and selection rules for $3j$ symbols, the quantum
equation  (\ref{c2l7b}) can be written as:
\begin{eqnarray}
 \Biggl\{\frac{1}{{\mathrm{r}}^2}\frac{\mathrm{d}}{\mathrm{d} \mathrm{r} }
\biggl({\mathrm{r}} ^2\frac{\mathrm{d}}{\mathrm{d} {{\mathrm{r}} }}\biggr)
- \frac{l(l+1)}{{\mathrm{r}} ^2}\Biggr\}\Upsilon=
-\frac{2l+1}{\hbar^2}\sum_{\bar{l}=\,0}^{2l} \sum_{\bar{m}=\,0}^{\bar{l}}
\sqrt{\frac{2\bar{l}+1}{\pi}}\times
\nonumber\\
\biggl(\begin{array}{ccc}
l & l&\bar{l} \\
0 & 0 & 0
\end{array}\biggr)\biggl(\begin{array}{ccc}
l & l&\bar{l} \\
-|m|&-|m|&|\bar{m}|
\end{array}\biggr)
\Omega_{\bar{l}\bar{m}}({\mathrm{r}} ;\, \mathrm{E}, J)\Upsilon.
\label{z2l7b}
\end{eqnarray}
Note that the upper limit of summation over $\bar{l}$ is  the value $2l$.
Recall that this fact is related to the selection rules, according to which
the symbol $3j$ is not equal to zero, in particular, if  $|l-l'|\leq\bar{l}\leq l+l'$.
Since in the case under consideration $l =l'$,  therefore, $0\leq\bar{l}\leq 2l$.

\subsection{}
If we assume that $ \zeta = \cos\varphi$, then the second-order derivative
$\mathrm{d}^2\Theta_K^j/(\mathrm{d}\varphi)^2$ will have the following form:
\begin{equation}
\frac{\mathrm{d}^2\Theta^j_{K}}{\mathrm{d}\varphi^2}=-\zeta\frac{\mathrm{d}
\Theta^j_{K}}{\mathrm{d}\zeta}
+\bigl(1-\zeta^2\bigr)\frac{\mathrm{d}^2\Theta^j_{K}}{\mathrm{d}\zeta^2}=
\zeta\frac{\mathrm{d}\Theta^j_{K}}{\mathrm{d}\zeta}-\biggl[j(j+1)-
\frac{K^2}{1-\zeta^2}\biggr]\Theta^j_{K}.
\label{H1}
\end{equation}
Using (\ref{H1}), we can calculate the following integral, which will play an
important role in further calculations:
\begin{equation}
\mathbf{Q}_{jKK'}=\int_{-1}^1\,\Theta^j_{K'}\frac{\mathrm{d}^2\Theta^j_{K}}{\mathrm{d}\varphi^2}
\,\mathrm{d}\zeta=\int_{-1}^1\,\zeta\Theta^j_{K'}\frac{\mathrm{d}\Theta^j_{K}}{\mathrm{d}\zeta}\,
\mathrm{d}\zeta - \int_{-1}^1\biggl[j(j+1)-
\frac{K^2}{1-\zeta^2}\biggr] \Theta^j_{K'}\Theta^j_{K}\,d\zeta.
\label{H2}
\end{equation}
Multiplying the equation (\ref{ckz7b}) by the associated Legendre function
$\Theta^j_{K'}\bigl(\zeta\bigr)$ and integrating it over the variable $\zeta$
 in the range $ [1, -1] $ we get:
\begin{equation}
\biggl\{\delta_{KK'}\biggl[\frac{1}{\mathrm{\varrho}}\frac{\partial}{\partial\mathrm{\varrho}}
\biggl(\mathrm{\varrho}\frac{\partial}{\partial\mathrm{\varrho}}\biggr)
+ \frac{\partial^2}{\partial z^2}\biggr]
+ \frac{\mathbf{Q}_{jKK'}}{\mathrm{\varrho}^2}+ \frac{2\mu_0}{\hbar^2}
\widetilde{\mathbf{\Omega}}_{jKK'}\bigl(\varrho,z\big) \biggr\}\widetilde{\Upsilon}=0,
\label{H3}
\end{equation}
where
\begin{eqnarray}
{\mathbf{Q}}_{jKK'}=\sum_{m=0}^\infty I^j_{mKK'},\qquad
 I^j_{mKK'}=\int_{-1}^1\Theta^j_{K}\bigl(\zeta\bigr)
\Theta^j_{K'}\bigl(\zeta\bigr)\Theta^0_{m}\bigl(\zeta\bigr)d\zeta,
\nonumber\\
 \widetilde{\mathbf{\Omega}}_{jKK'}\bigl(\varrho,z\big)=
\bigl[\mathrm{E}\widetilde{\mathfrak{g}}^{\,(1)}_m-J(J+1)\widetilde{\mathfrak{g}}^{\,(2)}_m\bigr].
\qquad\qquad
 \label{H4}
\end{eqnarray}
To calculate the term $I(j,K;j,K';0,m)\equiv I^j_{mKK'}$, we can use the following general formula \cite{Mav,Shi}:
\begin{eqnarray}
I(m_1,j_1;m_2,j_2;m_3,j_3)=
\qquad\qquad\qquad \qquad\qquad\qquad\qquad \qquad
\nonumber\\
\int_{-1}^{+1}\Theta^{m_1}_{j_1}(x)\Theta^{m_2}_{j_2}(x)\Theta^{m_3}_{j_3}(x)dx=
\sqrt{\frac{(j_2+m_2)!(j_1+m_1)!}{(j_2-m_2)!(j_1-m_1)!}}
 \sum_{n}\biggl[(-1)^{m_1+m_2}(2n+1)\times\quad
\nonumber\\
\biggl(\begin{array}{ccc}
j_1& j_2&n\\
0 & 0 & 0
\end{array}\biggr)\biggl(\begin{array}{ccc}
j_1& j_2&n \\
m_1&m_2&-m_1-m_2
\end{array}\biggr)
\sqrt{\frac{(n-m_1-m_2)!}{(n+m_1+m_2)!}}
\int_{-1}^{+1} \Theta^{m_3}_{j_3}(x)\Theta^{m_1+m_2}_{n}(x)dx\biggr],\quad
\label{H5}
\end{eqnarray}
where it is assumed that;
$
j_1+m_1+j_2+m_2+j_3+j_3,
$
is \emph{even} in addition,  also are \emph{even} $ |j_1-j_2|\leq n\leq j_1+j_2,\, j_1+j_2+n$
and $n+m_1+m_2+m_3+j_3$. As for the integral from two associated Legendre
polynomials, it is calculated exactly for an arbitrary case:
$$
\int_{-1}^{+1}\Theta^{m_1}_{j_1}(x)\Theta^{m_2}_{j_2}(x)dx=
\frac{(-1)^{m_2}2^{-2|m_2-m_1|-1}\pi}{ \Gamma(1/2+|m_2-m_1|/2)
\Gamma(3/2+|m_2-m_1|/2)}\,\times
$$
$$
\sqrt{\frac{(j_1+m_1)!(m_2+j_2)!}{(j_1-m_1)!(m_2-j_2)!}}\sum_k
G_{\{\bullet\}}\bigl(1+(-1)^{k+|m_2-m_1|}\bigr)\sqrt{\frac{(k+|m_2-m_1|)!}{(k-|m_2-m_1|)!}}\,\times
$$
$$
\frac{\Gamma(1/2)\Gamma(k/2)\Gamma(|m_2-m_1|+1)\Gamma(-[k+1]/2)}
{\Gamma([|m_2-m_1|+1-k]/2)\Gamma(|m_2-m_1|/2)\Gamma([|m_2-m_1|+k]/2+1)
\Gamma(-[|m_2-m_1|-1]/2) },
$$
where again $|j_2-j_1|\leq k\leq j_2+j_1$ and $k+j_1+j_2$ are even.
Additionally one requires that the integrand is even, i.e.
$j_1+m_1+j_2+m_2=even.$ As for the  function $G_{\{\bullet\}}$,
 then it is defined by the help of $3j$ symbols as:
$$
G_{\{\bullet\}}=(-1)^{-m_1+m_2}(2k+1)\biggl(\begin{array}{ccc}
j_1& j_2&j_3\\
0 & 0 & 0
\end{array}\biggr)\biggl(\begin{array}{ccc}
j_1& j_2&k \\
-m_1&m_2&m_1-m_2
\end{array}\biggr).
$$

From the equation (\ref{H3}) in the case $K\neq K'$ we obtain the following
algebraic equation:
\begin{equation}
 \frac{\mathbf{Q}_{jKK'}}{\mathrm{\varrho}^2}+\frac{2\mu_0}{\hbar^2}
\widetilde{\mathbf{\Omega}}_{jKK'}\bigl(\varrho,z\big)=0.
\label{H4}
\end{equation}
The set of points $\mathcal{Z}$ that generates the equation (\ref{H4}) with respect
to  the set of points forming the \emph{internal space} $\mathcal{M}^{(3)}$ has power zero.
Recall that the wave function is not uniquely determined on the set of points $\mathcal{Z}$,
 i.e.  it  can be defined in the space $\mathcal{M}^{(3)}\setminus\mathcal{Z}$.

We now turn to the question of obtaining an equation whose solution in the
limit $z\to+\infty$  goes over to the $\mathbf{S}$ -matrix elements.
For this, we substitute the full wave function of the  three-body  system (\ref{t07sb})
 into the Schr\"{o}dinger equation (\ref{ckz07b}):
\begin{equation}
\sum_{\bar{n}\bar{j}} \biggl\{\frac{1}{\mathrm{\varrho}}
\frac{\partial}{\partial\mathrm{\varrho}}
\biggl(\mathrm{\varrho}\frac{\partial}{\partial\mathrm{\varrho}}\biggr)+
\frac{\partial^2}{\partial z^2}+\frac{1}{\mathrm{\varrho}^2}
\frac{\partial^2}{\partial \varphi^2}+\frac{2\mu_0}{\hbar^2 }
\widetilde{\Omega}\bigl(\{\bar{\varrho}\}\bigr)\biggr\}
\mathbf{\Xi}^{+(J)}_{[\mathcal{K}]\,[\bar{\mathcal{K}}]} (z)
\Pi_{\bar{n}(\bar{j}\bar{K})}(\varrho;z)\Theta_{\bar{K}}^{\bar{j}}(\zeta)=0,
\label{a2ckz07b}
\end{equation}
where $\mathbf{\Xi}^{+(J)}_{[\mathcal{K}]\,[\bar{\mathcal{K}}]} (z)$ and
$\Pi_{\bar{n}(\bar{j}\bar{K})}(\varrho;z)$ functions that still need to be defined.

Multiplying the equation (\ref{a2ckz07b}) by the associated Legendre function
$\Theta^j_{K'}\bigl(\zeta\bigr) $ and integrating it over the variable $\zeta$
in the range $[1, -1]$, taking into account the condition of orthogonality of
these functions, we obtain:
 \begin{equation}
\sum_{\bar{n}\bar{j}} \biggl\{\delta_{j^{\,'}\bar{j}}\,\delta_{K'\bar{K}}
 \biggl[ \frac{1}{\mathrm{\varrho} }\frac{\partial}{\partial\mathrm{\varrho}}
\biggl(\mathrm{\varrho}\frac{\partial}{\partial\mathrm{\varrho}}\biggr)+
\frac{\partial^2}{\partial z^2} \biggr]+
\frac{\mathbf{Q}_{j^{\,'}\bar{j}K'\bar{K}}}{\mathrm{\varrho}^2}+\frac{2\mu_0}{\hbar^2 }
\widetilde{\mathbf{\Omega}}_{j^{\,'}\bar{j}K'\bar{K}}\bigl(\varrho,z\big)
\biggr\}\mathbf{\Xi}^{+(J)}_{[\mathcal{K}]\,[\bar{\mathcal{K}}]} (z)
\Pi_{\bar{n}(\bar{j}\bar{K})}(\varrho;z)=0.
\label{a3ckz07b}
\end{equation}
Let us consider the following \emph{reference} equation:
\begin{equation}
\biggl\{\frac{1}{\mathrm{\varrho}}\frac{\partial}{\partial\mathrm{\varrho}}
\biggl(\mathrm{\varrho}\frac{\partial}{\partial\mathrm{\varrho}}\biggr)
+\frac{\mathbf{Q}_{j^{\,'}\bar{j}K'\bar{K}}}{\mathrm{\varrho}^2}+ \frac{2\mu_0}{\hbar^2}
\widetilde{\mathbf{\Omega}}_{j^{\,'}\bar{j}K'\bar{K}}\bigl(\varrho,z\big) \biggr\}
\Pi_{\bar{n}(\bar{j}\bar{K})}(\varrho;z)
=\mathcal{E}^{(j^{\,'}K')}_{\bar{n}(\bar{j}\bar{K})}(z)\Pi_{\bar{n}(\bar{j}\bar{K})}(\varrho;z),
\label{c1z7tb}
\end{equation}
which is actually a parametric, second-order ODE.

Based on the fact that the localization of the quantum current occurs
near the coordinate $z$ by the coordinate $\varrho $, it can be assumed
that the solution $\Pi_{\bar{n}(\bar{j}\bar{K})}(\varrho;z)$ is quantized.
In other words, the solutions $\Pi_{\bar{n}(\bar{j}\bar{K})}(\varrho;z)$
form an orthonormal basis in a Hilbert space,  and we can write the
following condition of orthonormality:
$$
\int_0^\infty \Pi_{n(jK)}(\varrho;z){\Pi^\ast_{\bar{n}(\bar{j}\bar{K})}(\varrho;z)}
d\varrho=\delta_{n\bar{n}}.
$$
Finally, multiplying the equation (\ref{c1z7tb}) by the solution
$ {\Pi_ {n'(j ^ {\,'} K ')} (\varrho; z)}^\ast $ and integrating, we obtain the following ODE :
\begin{equation}
\sum_{\bar{n}\bar{j}} \delta_{n'\bar{n}}
\biggl\{\delta_{j^{\,'}\bar{j}}\,\delta_{K'\bar{K}} \frac{\mathrm{d}^2}{\mathrm{d} z^2}+
\overline {\mathcal{E}}^{\,n'(j^{\,'}K')}_{\bar{n}(\bar{j}\bar{K})}(z)
\biggr\}\mathbf{\Xi}^{+(J)}_{[\mathcal{K}]\,[\bar{\mathcal{K}}]}(z)=0,
\label{a4ckz07b}
\end{equation}
where $\overline{\mathcal{E}}^{\,n'(j^{\,'}K')}_{\bar{n}(\bar{j}\bar{K})}(z)=
\int_0^\infty\Pi_{n'(j^{\,'}K')}(\varrho;z)\mathcal{E}^{(j^{\,'}K')}_{\bar{n}(\bar{j}\bar{K})}(z)
{\Pi^\ast_{\bar{n}(\bar{j}\bar{K})}(\varrho;z)}d\varrho.$

The equation (\ref{a4ckz07b}) at  the $n'=\bar{n},\,j^{\,'}=\bar{j}$ and $K=K'$ takes
the simple form of the second -order ODE (see equation (\ref{a5ckz07b})).

In the case when at least one pair of quantum numbers does not coincide between
two sets $[\mathcal{K}']$ and $[\bar{\mathcal{K}}]$, from (\ref{a4ckz07b}) we
obtain the algebraic equations:
\begin{equation}
\sum_{\bar{j}}
\overline {\mathcal{E}}^{\,n'(j^{\,'}K')}_{\, \bar{n}\,(\bar{j}\,\,\bar{K})}(z)=0,
\qquad n'\neq  \bar{n} \quad or \quad K\neq \bar{K}.
\label{av07b}
\end{equation}
The algebraic equation (\ref{av07b}) generates a line on which the function
 should be equal to zero. Note that this is an additional condition imposed on the  function
 $\mathcal{E}_{\,\bar{n}\,(\bar{j}\,\,\bar{K})}(z)$.

 \end{document}